**Title**

The Target Study: A Conceptual Model and Framework for Measuring Disparity

**Author(s)**

John W. Jackson (1, 2, 3), Yea-Jen Hsu (4), Raquel C. Greer (5), Romsai T. Boonyasai (5,6), Chanelle J. Howe (7)

**Affiliation(s)**

1) Departments of Epidemiology, Biostatistics, and Mental Health, Johns Hopkins Bloomberg School of Public Health, Baltimore, MD; 2) Johns Hopkins Center for Health Equity, Baltimore, MD; 3) Johns Hopkins Center for Health Disparities Solutions, Baltimore, MD; 4) Department of Health Policy and Management, Johns Hopkins Bloomberg School of Public Health, Baltimore, MD; 5) Department of Medicine, Johns Hopkins School of Medicine, Baltimore, MD; 6) Agency for Healthcare Research and Quality, Rockville, MD; 7) Center for Epidemiologic Research, Department of Epidemiology, School of Public Health, Brown University, Providence, RI

**Correspondence**

John W. Jackson, ScD, 615 N. Wolfe St., Baltimore, MD, 21205, email: john.jackson@jhu.edu

**Author Contributions**

Dr. Jackson conceived of the work, developed the formal results, carried out the data application, and drafted the initial and revised manuscripts. Dr. Hsu constructed the analytic cohort for the data application. Drs. Jackson, Hsu, Greer, and Boonyasai oversaw the construction of the analytic cohort and data application. Drs. Hsu, Greer, Boonyasai, and Howe critically edited the initial and revised manuscripts for scientific content.

**Conflict of Interest Statement**

Dr. Jackson was supported by a grant from the National Heart, Lung, and Blood Institute (K01HL145320). Dr. Howe has received funding via a grant from Sanofi Pasteur administered directly to Brown University (unrelated to the current work). This work does not necessarily represent the views or opinions of the Agency for Healthcare Research and Quality.

**Abstract**

We present a conceptual model to measure disparity—the target study—where social groups may be similarly situated (i.e., balanced) on allowable covariates. Our model, based on a sampling design, does not intervene to assign social group membership or alter allowable covariates. To address non-random sample selection, we extend our model to generalize or transport disparity or to assess disparity after an intervention on eligibility-related variables that eliminates forms of collider-stratification. To avoid bias from differential timing of enrollment, we aggregate time-specific study results by balancing calendar time of enrollment across social groups. To provide a framework for emulating our model, we discuss study designs, data structures, and G-computation and weighting estimators. We compare our sampling-based model to prominent decomposition-based models used in healthcare and algorithmic fairness. We provide R code for all estimators and apply our methods to measure health system disparities in hypertension control using electronic medical records.

## 1. Introduction

Measuring disparity is a key step in making progress towards health equity. Disparity measures underlie descriptive reports and trends and serve as benchmarks for evaluating the effects of interventions and policies. (Cooper, Hill and Powe 2002) Although the measurement of disparity is critical and there has been much discussion and debate about what constitutes a disparity, (Institute of Medicine Committee on Understanding and Eliminating Racial Ethnic Disparities in Healthcare 2003, Braveman 2006, Duran and Pérez-Stable 2019) there has been limited discussion about best practices and principles for measurement of disparity, especially when using secondary data not collected for research purposes.

Conceptual models serve as important guides for the analysis and interpretation of secondary data. For example, consider the target trial framework, (Hernán and Robins 2016) which lays out the hypothetical randomized controlled trial one would conduct if the goal were to estimate the effect of a treatment strategy to inform clinical decision-making. The elements of the trial (eligibility, treatment strategies, outcome follow-up) guide the design and analysis of a study based on secondary data help ensure that the measure of association has a causal interpretation that applies to (a) the population of interest, (b) treatment strategies of interest, and (c) outcomes of interest, all of which are critical for informing treatment policy decisions.

The target trial framework cannot guide a descriptive measurement of disparity where there is no intervention. Still, without a conceptual guide, the population of interest and the follow-up period that pertain to unjust processes or outcomes may be unclear which can impede appropriate policymaking. Without a conceptual model it is difficult to justify and interpret covariate adjustment in health disparities research. (Kaufman 2017) Causal models have been used to define disparities, (Duan, Meng, Lin et al. 2008) but they have stringent assumptions and abstract away important realities. Meanwhile, there are intense discussions about non-random sample selection and its impact on related concepts such as discrimination. (Knox, Lowe and Mummolo 2020, Gaebler, Cai, Basse et al. 2022) Outlining the hypothetical study one could do in the real world to measure disparity will provide clarity on these issues.

We present a novel conceptual model—the target study—to address these issues and provide a framework for emulating it. The paper is organized as follows. Section 2 introduces a motivating example. Section 3 reviews key issues in disparity measurement. Section 4 presents our model under the case where investigators wish to capture all effects of non-random sample selection. Section 5 expands the model to address non-random sample selection by generalizing to a broader population, transporting to a different population, or estimating disparity in a counterfactual population where certain consequences of non-random sample selection are absent. Section 6 proposes data structures and estimators to emulate the target study. Section 7 outlines our contributions and compares our model to others widely used to study disparity in healthcare and algorithmic fairness. Section 8 discusses strengths and limitations. To aid readability, we use modular sections with ample cross-referencing so readers may skip directly to sections of interest.

## 2. Motivating Example

Consider the measurement of racial disparities in hypertension outcomes of primary care patients diagnosed with hypertension who receive care at a large regional health system in the United States. The outcomes of interest are a health-related quantity $Y$ (e.g., hypertension control) or a healthcare decision $D$ made by a clinician (e.g., to intensify hypertension treatment). We are concerned with average outcomes across a categorical social grouping, such as race $R$, where a socially disadvantaged (henceforth referred to as marginalized) group (e.g., Black persons) is denoted as $R = 1$ and a socially advantaged (henceforth referred to as privileged) group (e.g., White persons) is denoted as $R = 0$. The available data are electronic medical records stamped at time $k$ (e.g., in minutes, hours, seconds) from office-based primary care visits over multiple years that include measures of prior hypertension, demographics $\boldsymbol{X}_k$ (e.g., age and sex assigned at birth), comorbidity and socioeconomic status (SES) $\boldsymbol{L}_k$, hypertension control as of that visit, denoted by $Y_k$ (1: yes, 0: no), antihypertensive treatment intensification (i.e., initiation, change in dose, or change in class) within the 14 days after the visit, denoted by $D_k$ (1: yes, 0: no), and time-stamped enrollment in an electronic patient portal program (EPPP) for care management. The notation $k$ refers to the timing of a variable's measurement rather than the time at which its value is realized. Finally, note that persons may have multiple visits per day or week.

## 3. Conceptual Issues in Measuring Disparity

### *3.1 Defining Disparity*

In medicine and public health, the definition of disparity depends on whether the outcome is a health status (e.g., hypertension control $Y$) or a healthcare commodity (e.g., treatment intensification $D$). For health status outcomes, the Healthy People 2020 report committee defined disparities as "systematic, plausibly avoidable health differences adversely affecting socially disadvantaged groups." (Braveman, Kumanyika, Fielding et al. 2011) This builds upon a World Health Organization definition (Whitehead 1992) and relates to the National Institute of Minority Health and Health Disparities (NIMHD) definition (Duran and Pérez-Stable 2019): "a health difference that adversely affects disadvantaged populations, based on one or more health outcomes" where outcomes range from health behaviors, to diagnosis- or stage-specific- clinical endpoints or self-reported measures, to overall mortality. For healthcare outcomes (e.g. treatment), the Institute of Medicine (IOM) report "Unequal Treatment" (Institute of Medicine Committee on Understanding and Eliminating Racial Ethnic Disparities in Healthcare 2003) defines disparities as "differences in the quality of care that are *not due to* access-related factors or clinical needs, preferences, and appropriateness of intervention…[where] analysis is focused at two levels: 1) the operation of the health systems and the legal and regulatory climate…; 2) discrimination at the individual, patient-provider level" (emphasis added). Disparity reflects society's failure to achieve equity in health, defined as "everyone having a fair and just opportunity to be as healthy as possible." (Whitehead 1992, Braveman et al. 2011)[1]

*3.2 Temporal Framing*

To aid decision-makers, community members, and other stakeholders, disparity refers not to a universal, general phenomenon, but to outcomes among people nested in a particular context at a point in (or span of) calendar time. For example, we could describe the disparity in prevalent uncontrolled hypertension for primary care visits during each month during the peak of the COVID-19 pandemic in 2020-2022. If we include one care episode per person per month, we can meaningfully summarize disparity over the entire period by averaging over the month-specific estimates of disparity. Such a summary measure would be interpreted as an average disparity over populations indexed by calendar months. By accounting for calendar time when producing such summary estimates, we avoid confounding by time-specific trends in enrollment and the outcome. To obtain a summary estimate of disparity between social groups with the same person-time experience of the health system, the summary must properly account for calendar time.

*3.3 Allowability*

In the IOM definition, disparity compares groups who are similarly situated (i.e., balanced) on "allowable" covariates. Allowable covariates are those whose differential distribution does not lead to inequitable outcomes. For a distributed good outcome $D$ (e.g., healthcare) they are factors that, on moral arguments, are appropriate for determining allocation. (Jackson 2021) For example, disparities in healthcare treat clinical need as allowable based on clinical guidelines (McGuire, Alegria, Cook et al. 2006, Cook, McGuire, Meara et al. 2009). For a state outcome $Y$ (e.g. health), the differential distribution of allowable covariates does not contribute to worse outcomes among the marginalized group. (Jackson 2021) For example, if the marginalized group is younger and increased age predicts worse hypertension control, the younger age of the Black population does not contribute to the disparate distribution of hypertension control at the population-level. Not treating age as allowable could mask disparity from barriers to hypertension control that the Black population disproportionately faces (e.g., neighborhood disadvantage and limited options for healthy diet, physical activity, and pharmacies (Mueller, Purnell, Mensah et al. 2015)).

*3.4 Is a Causal Framing of Disparity Necessary?*

A fundamental question in conceiving of disparity is how the groups come to be similarly situated (i.e., balanced) on the allowables. Many authors conceive of disparity as comparing populations that are similarly situated through an intervention where an external actor makes existing groups similar by

[1] Disparity compares across groups. It is broader than the legal notions of discrimination defined as disparate treatment (*Civil Rights Act of 1964*) and disparate impact (*Civil Rights Act of 1991*). It captures contemporary and intergenerational marginalization at personal, institutional, and societal levels that the restrict conditions and opportunities for a reasonably good life, whatever else one may desire in life (Diderichsen, Evans and Whitehead 2001, Braveman 2006, Powers and Faden 2019).

changing the value(s) of each person's allowable covariate(s).[2] (McGuire et al. 2006, Duan et al. 2008, Cook et al. 2009, Kaufman 2017) For example, disparate pulse oximeter performance is assessed in desaturation studies where hypoxia is induced among healthy volunteers. (Food and Drug Administration 2024) Disparate healthcare utilization is assessed in statistical analyses that hypothetically modify individuals' health and project utilization after this modification. This causal reading of the IOM definition is justified by its phrase "not due to", interpreted as "not caused by," where disparity compares social groups who are made similar (on the allowables) by intervention, to isolate the mediating role of inappropriate factors (e.g., SES) in producing differences in healthcare utilization (McGuire et al. 2006). But the phrase "not due to" also permits a non-causal framing where, by design, disparity compares social groups who are already alike on the allowables. Early work that applied the IOM definition of disparity was motivated by non-causal studies where patients of different social groups with the same underlying need for medical treatment are compared in their distribution of appropriate medical treatment received, and actually framed such studies as IOM concordant. (Cook et al. 2009:2-3) We argue that approaches that balance allowables by design (e.g., our model) align with the IOM definition. More broadly, arguments about the exact causes of disparity are not needed to view disparity with concern. (Braveman 2006) Moral concern may arise by the impact that disparity has on the human rights of marginalized groups. (Hutler 2022)

*3.5 Non-Random Sample Selection*

Some frameworks for health equity acknowledge that non-random sample selection may impact a disparity measure. (Kilbourne, Switzer, Hyman et al. 2006) Consider the causal graph of Figure 1A, where variables $\boldsymbol{W_k}$ establishing eligibility $Q_k$ (eligibility-related variables) are prior hypertension, established care in the health system, enrollment in the electronic portal program (EPPP), and current visit, with $k$ indexing the calendar time at which a variable is measured, and $J$ indexing the outcome's follow-up time.

Lack of generalizability occurs when a disparity measure is unbiased for the study sample (e.g., those enrolled in the EPPP) but biased for the broader population of interest (e.g., the entire health system). Lack of transportability occurs when the disparity measure is unbiased for the study sample but biased for a different population of interest (e.g., not enrolled in EPPP). (Smith 2020) Either scenario arises when (i) a risk factor $L_k$ (e.g., SES) has different associations with the outcome $Y_{k+J}$ (e.g., hypertension control) across social groups $R$ and (ii) the risk factor $L_k$'s distribution depends on eligibility-related variables $\boldsymbol{W_k}$ (e.g., it differs by EPPP enrollment), even with no difference in eligibility $Q_k$ across social groups $R$.[3] They also arise when the effect of eligibility-related variables on the outcome differs across social groups.

[2] An alternative causal model is to balance the allowables by hypothetically assigning at random which social group a person is perceived to belong to, to identify discrimination for a decision outcome $D$. (Greiner and Rubin 2011) However, this hypothetical intervention will also balance non-allowables implicated in disparity, leading to an over-adjusted measure of disparity.

[3] Similar arguments have been made about the underlying structure behind lack of generalizability for causal effects (Greenland 1977, Hernán 2017, Lu, Cole, Howe et al. 2022). We focus on a descriptive estimand where social group is not intervened on.

Collider stratification creates an association between social group $R$ and the outcome $Y_{k+J}$ among those who are eligible $Q_k = 1$. (VanderWeele and Robinson 2014) It can occur when eligibility-related variables $\boldsymbol{W_k}$ (e.g., EPPP enrollment) are affected by (i) social group $R$ and (ii) a risk factor $L_k$ (e.g. SES) for the outcome $Y_{k+J}$. (Hernán, Hernández-Díaz and Robins 2004, Elwert and Winship 2014) Recent work (Shahar and Shahar 2017, Nguyen, Dafoe and Ogburn 2019) implies that collider stratification occurs if the probability ratio of being eligible $Q_k = 1$ (comparing levels of social group $R$) varies across levels of a risk factor $L_k$ for the outcome $Y_{k+J}$, even if $L_k$ has homogeneous associations with $Y_{k+J}$ across social groups $R$.

Because collider stratification due to non-random sample selection can induce an association between social group and the outcome among the sample (e.g., those enrolled in the EPPP) that is not present among the broader population (e.g., irrespective of EPPP enrollment), it is often viewed as a bias. (VanderWeele and Robinson 2014, Knox et al. 2020, Rojas-Saunero, Glymour and Mayeda 2023) There are reasons to include contributions of collider stratification to disparity. First, if disparity is measured in a meaningfully defined population of interest,[4] the contributions are substantively grounded as they reflect that population of interest. (VanderWeele and Robinson 2014) Consider when eligibility is defined by a condition (e.g., hypertension) that gives meaning to the outcome (e.g., hypertension control). For example, persons without history of hypertension can have elevated blood pressure due to hypertension onset or due to exercise, but these reasons do not represent uncontrolled hypertension. The disparity is only defined among eligible persons. Second, for a meaningful population of interest, when collider stratification disadvantages the marginalized group on baseline covariates leading to a worse outcome distribution compared to the privileged group, this aligns with definitions of disparity (Section 3.1). Third, the contribution of collider stratification is amenable to intervention by changing how covariates affect eligibility or the outcome.[5] However, when collider stratification advantages the marginalized group, it may mask disparity from other sources and investigators may choose to exclude it from disparity.

## 4. A Target Study Conceptual Model for Measuring Disparity

### *4.1 Overview.*

We now describe the elements of our conceptual model for measuring disparity, the target study. In this heuristic, an eligible population (denoted as $Q_k$ [1: eligible, 0: otherwise]) from two or more social groups (e.g., Black $R = 1$ and White persons $R = 0$) are selected from an eligible source population (e.g., established care in system, prior hypertension, enrolled in EPPP, current visit) within a given moment or span of coarsened time $k$ representing the enrollment period (e.g., the month of January 2023). This selection occurs at the end of the enrollment window through a two-stage sampling strategy. The first stage

---

[4] We posit that a meaningful population is a set of persons who, within a short span of time, are bound together by a shared set of circumstances, conditions, experiences, or purpose making them distinct from others. See, e.g., Kindig and Stoddart (2003).

[5] Many interventions address disparity by altering how SES and comorbidities impact outcomes. (Mueller et al. 2015)

of sampling addresses non-random selection into the study. The second stage of sampling similarly situates (i.e., balances) the social groups on the allowable covariates (if any are chosen) so that for both social groups, the allowables follow a distribution from a within-sample standard population (chosen by the investigator). Thus, the disparity estimate in the final sample is not due to differences in the allowable distributions. After both stages of sampling, those enrolled are followed for a specified period of time. A specified statistical comparison of outcomes across social groups provides the measure of disparity $\psi_k$.[6] The additive and ratio disparity are contrasts of mean outcomes (prevalence or risk with binary outcomes):

$$\psi_k^{add} = \mu_k(1) - \mu_k(0) \text{ and } \psi_k^{rel} = \mu_k(1)/\mu_k(0) \tag{1}$$

where $\mu_k(r)$ denotes $E_{\boldsymbol{\Omega}}\left[Y_{k+J} \middle| Q_k = 1, R = r, k\right]$, the average outcome $Y_{k+J}$ in social group $R = r$ who are eligible $Q_k = 1$ (based on criteria for eligibility-related variables $\boldsymbol{W_k}$) and enrolled in the target study sample $\boldsymbol{\Omega}$ at calendar time $k$ with follow-up time $(0, \dots, J)$

Under this conceptual model, the target population (in which inference is made) operationally consists of the source population that, within the enrollment period, is eligible, sampled, and enrolled. That is, in real life, if one wanted to make inferences about disparity in a population that exists within a certain span of time, one would carry out the protocol of the target study. At any calendar time unit $k$, a person only enrolls once into a target study. (Each unit of calendar time is of equal length). Results of studies $\psi_k$ carried out at distinct calendar times $k$ can be aggregated into a summary measure of disparity $\Psi$. A weighted average of disparity measures $\psi_k$ indexed at each calendar time $k$ can be taken as:

$$\Psi = \frac{\sum_k \gamma_k \psi_k}{\sum_k \gamma_k} \tag{2}$$

where $\gamma_k$ is a weight specific to calendar time $k$

We discuss the choice of the weights $\gamma_k$ in Section 4.10 where we discuss the analysis of target studies.

We begin with our default model (Design 1) where we choose to enroll all eligible persons (or a simple random sample of eligible persons) during the first stage of sampling. Recall that eligibility criteria cause persons to be non-randomly selected from the source population, so our default model includes all contributions of this non-random selection of persons to disparity. Adaptations to deal with such non-random sample selection (i.e., Designs 2, 3, or 4) are discussed in Section 5. Design 1 has minimal structural constrains on the underlying causal relations between all relevant variables.[7]

*4.2 Enrollment Window(s)*

---

[6] If we interpret the definitions in Section 3.1 strictly, disparity is null when the marginalized group is not disadvantaged on the outcome. On the additive scale, disparity is $\psi_k^{+add} = (\psi_k^{add} - \Delta) \times I(\psi_k^{add} > \Delta)$ where $I(\cdot)$ is the indicator function (1: true, 0: false) and $\Delta = 0$. On the ratio scale, disparity is $\psi_k^{+rel} = \exp\left(\log\left(\psi_k^{rel}/\Delta\right) \times I(\psi_k^{rel} > \Delta)\right)$ where $\Delta = 1$. Then $\psi_k^{+}$ replaces $\psi_k$ everywhere. Disparity may be defined with a threshold $\Delta$, choosing $\Delta > 0$ on the additive scale or $\Delta > 1$ on the ratio scale.

[7] For a target study indexed at calendar time $k$, social group may affect eligibility-related variables or allowables (i.e., $R \rightarrow (\boldsymbol{W_k}, \boldsymbol{A_k})$ may exist) and allowables may affect eligibility-related variables or vice versa (i.e., $\boldsymbol{A_k} \rightarrow \boldsymbol{W_k}$ or $\boldsymbol{W_k} \rightarrow \boldsymbol{A_k}$ may exist). The outcome must not affect eligibility-related variables, allowables, or social group (i.e., $Y_{k+J} \rightarrow (\boldsymbol{W_k}, \boldsymbol{A_k}, \boldsymbol{R})$ may not exist).

To conduct a target study, we first choose a specific moment or narrow span in calendar time, denoted by $k$, to enroll persons. This requires choosing a level of granularity for calendar time (e.g., hours, days, months, years) and a specific moment $k$ as the enrollment period (e.g., the month of January 2023). For each person, all eligibility-related variables $\boldsymbol{W_k}$ (e.g., prior diagnosis of hypertension, established care in the health system, with a current visit in the window) and all allowable covariates $\boldsymbol{A_k}$ (e.g., age, sex) are defined and measured at or before the end of this period. Thus, relative to the end of the enrollment window, eligibility $Q_k$ is based on having acceptable current or prior values $\mathbb{w}_k$ of the eligibility-related variables $\boldsymbol{W_k}$. At any calendar time $k$, a person may only enroll in one study.

*4.3 Enrollment Groups*

The definitions of disparity in Section 3.1 compare groups with persistently different levels of social advantage, privilege, power, wealth, or prestige because of their position in society. (Braveman 2006) Within the United States, the NIMHD's concept of a disparity specifies social groups such as racial and ethnic minoritized groups (versus majoritized groups), underserved rural residents (versus urban residents), lower socioeconomic status (versus higher socioeconomic status), and sexual and gender minorities (versus sexual and gender majorities) (Duran and Pérez-Stable 2019) The National Institute of Mental Health (NIMH) further specifies groups with serious mental illness (versus those without) who have experienced long-standing stigmatization, discrimination, social exclusion, and loss of agency in society. Reflecting an intersectional perspective that mechanisms of social injustice combine to uniquely shape experience, (Collins and Bilge 2020) social groups may be defined by joint membership along multiple axes (e.g., Black women versus White men). (Jackson, Williams and VanderWeele 2016) This list is not exhaustive, and our model accommodates categorical[8] and time-varying definitions[9] of social groups.

*4.4 Eligibility Criteria*

The eligibility criteria can define the population of interest, reflecting issues of scope, societal level, and timing. In terms of scope, the criteria can restrict to places (e.g., the Mid-Atlantic region), institutions (e.g., a particular health system), or shared experiences or conditions (e.g., diagnosis of hypertension) that define a meaningful population. In terms of societal level, the criteria can focus on persons under the purview of a specific decision-maker (e.g., a clinical provider), facility (e.g., a clinic), or institution (e.g., a health system). In terms of timing, the criteria can focus on critical life stages, such as birth or a milestone event (e.g., myocardial infarction) where outcomes (e.g., appropriate medical treatment) are given meaning by that event. From here, we will use the following criteria: prior hypertension, established care in the health system, EPPP enrollment before calendar time $k$, and a recent primary care visit within calendar time $k$.

---

[8] If $R$ is categorical, including multiple marginalized groups, at time $k$ we may enroll all groups into the same target study or separate studies, each with one marginalized group and a random partition of the most privileged group. For the latter, the allowables (Section 4.5), standard population (Section 4.6), and sampling design (Sections 4.7 and 5) may differ across studies.

[9] For a social group $R_k$ that varies over time (e.g., mental illness), a person is classified to their social group as of time $k$.

*4.5 Allowable Covariates*

We choose the covariates that the social groups are to be similarly situated (i.e., balanced) on by the end of the enrollment process. These allowable covariates $\boldsymbol{A_k}$ are ones not implicated in generating disparate outcomes among the marginalized group, as described in Section 3.3. Because the allowables $\boldsymbol{A_k}$ are used to guide the enrollment process by defining the sampling fractions, they must be defined and measured by the end of the enrollment window when enrollment occurs. No allowable covariates may be chosen at all (i.e., $\boldsymbol{A_k} = \emptyset$), as discussed Section 4.7). Although the eligibility variables $\boldsymbol{W_k}$ (used to form the sampling frame) are separate from the allowables $\boldsymbol{A_k}$ (used to define the sampling fractions), $\boldsymbol{W_k}$ are conceptually deemed allowable as the enrollment process similarly situates social groups on them.

*4.6 Standard Distribution*

During enrollment we sample individuals so that the distribution of allowables $\boldsymbol{A_k}$ is the same for social groups, following a *within-sample* standard distribution chosen by the investigator. If the disparity varies across the allowables $\boldsymbol{A_k}$ (e.g., disparity is higher in mid-life) the choice of the standard population, denoted by $T = 1$, that defines the standard distribution may impact the magnitude and direction of the disparity measure $\psi_k$. The choice may be motivated by normative, theoretical, or practical concerns. If the marginalized group is the standard, its experience is emphasized. (Thurber, Thandrayen, Maddox et al. 2021) Then, the disparity measure $\psi_k$ compares the experience of the marginalized group (enrolled through simple random sampling) to the experience of a privileged group (enrolled through stratified sampling) that shares the marginalized group's distribution of allowable covariates $\boldsymbol{A_k}$ (e.g., its age structure).

To balance the allowables $\boldsymbol{A_k}$ across social groups $R$, the values of $\boldsymbol{A_k}$ found among the standard population must be within those of each social group $R = r$ at each time $k$, which is an overlap assumption:

$$P(\boldsymbol{A_k} = \boldsymbol{a_k} | Q_k = 1, R = r, k) > 0 \tag{3}$$

for all values $\boldsymbol{a_k}$ with $P(\boldsymbol{A_k} = \boldsymbol{a_k} | Q_k = 1, T = 1, k) > 0$ and all $k$

The overlap assumption (3) requires that at each time $k$ we look among the standard population (denoted by $T = 1$) and note the pattern of allowables $\boldsymbol{A_k}$ covariate values. Then for each of those strata, we need to find (among eligible persons) members of each social group $R = r$. Otherwise, the sampling strategies we now describe will not be able to balance the allowables according to the standard distribution.

*4.7 Enrollment Process (for Design 1)*

At any time $k$, each person enrolls (once) into one study through multiple stages of sampling.[10] Limiting participation to a single enrollment in a single study per unit of calendar time maps inference to well-defined populations at each unit of calendar time. There is a pre-stage where eligible individuals are selected, a first stage that addresses the contribution of selective mechanisms to disparity $\psi_k$, and a second stage that balances allowable covariates $\boldsymbol{A_k}$ in the final sample. For any target study design $\mathcal{D}$, at each stage $\ell$, a person is selected via a known sampling fraction $\alpha_{k,\ell}^{\mathcal{D}}$ defined as the ratio of sample $\mathbb{S}_{k,\ell}$'s size $\mathbb{N}_{k,\ell}$ to the sampling frame $\mathbb{S}_{k,\ell-1}$'s size $\mathbb{N}_{k,\ell-1}$. (Lohr 2022) The sampling fractions $\alpha_{k,\ell}^{\mathcal{D}}(\boldsymbol{v})$ may be stratified by and vary across covariate levels $\boldsymbol{V} = \boldsymbol{v}$.

We assume that the sampling is process is innocuous with respect to the outcome:

$$f_Y^{sampled}\left(Y_{k+J} \middle| Q_k = 1, \boldsymbol{V} = \boldsymbol{v}, k\right) = f_Y^{unsampled}\left(Y_{k+J} \middle| Q_k = 1, \boldsymbol{V} = \boldsymbol{v}, k\right) \tag{4}$$

where $f(\cdot)$ is the probability mass function, for all $k$

In words, the conditional distribution of the outcome given that persons are eligible and have covariate values $\boldsymbol{V} = \boldsymbol{v}$ is the same for those enrolled and those not enrolled. Sampling does not affect the outcome. A person is randomly selected (without replacement) using a probability equal to their rescaled sampling fraction $\alpha_{k,\ell}^{*,\mathcal{D}}(\boldsymbol{v})$ that is bounded between zero and one.[11] We draw a uniformly distributed random number bounded between zero and one (i.e., $U[0,1]$) and, if it is equal to or less than a person's rescaled sampling fraction, they are included. (Sunter 1977) The sampling process is separate for each social group $R = r$.

The pre-stage $\ell = 0$ is at the end of the enrollment window for time $k$. All eligibility criteria $\boldsymbol{W_k}$ and all covariates $\boldsymbol{V}$ used in the enrollment process are measured by this point. From the source population $\mathcal{P}_k(r)$, we select the eligible population $\mathbb{S}_{k,0}(r)$ of chosen size $\mathbb{N}_{k,0}(r)$. Full eligibility $Q_k$ is defined as:

$$Q_k = I(\boldsymbol{W_k} \in \mathbb{w}) \tag{5}$$

where $\mathbb{w}$ represents the eligible values $\boldsymbol{w}$ of the eligibility-related variables $\boldsymbol{W_k}$

In the first stage $\ell = 1$, a sample $\mathbb{S}_{k,1}^{\mathcal{D}1}(r)$ of chosen size $\mathbb{N}_{k,1}(r)$ is selected from the first-stage sampling frame $\mathbb{S}_{k,0}(r)$ of size $\mathbb{N}_{k,0}(r)$ (eligible persons) using the sampling fraction $\alpha_{k,1}^{\mathcal{D}1}(r)$:

$$\alpha_{k,1}^{\mathcal{D}1}(r) = \frac{\mathbb{N}_{k,1}(r)}{\mathbb{N}_{k,0}(r)} \tag{6}$$

Recall that eligible persons have been non-randomly chosen from the entire source population. They may all be selected at this stage (i.e., when $\mathbb{N}_{k,0}(r) = \mathbb{N}_{k,1}(r)$), otherwise $\alpha_{k,1}^{\mathcal{D}1}(r)$ leads to simple random sampling of eligible persons in each social group $R = r$. Under either choice, all selective mechanisms, including that of non-generalizability and collider-stratification, contribute to the disparity estimate $\psi_k$.

---

[10] If $k$ represents a wide span of calendar time (e.g., a month), individuals may be eligible at multiple instances during $k$ (e.g., many visits with a primary care provider in a month). The first instance during the unit of time $k$ would be used for enrollment.

[11] The sampling fraction $\alpha_{k,\ell}(\boldsymbol{v})$ in Sections 4 and 5 may be larger than one. To sample without replacement, we rescale it, i.e., $\alpha_{k,\ell}^{*,\mathcal{D}}(\boldsymbol{v}) = \mathbb{N}_{k,\ell}^{*}(\boldsymbol{v})/\mathbb{N}_{k,\ell-1}(\boldsymbol{v})$ where $\mathbb{N}_{k,\ell}^{*}(\boldsymbol{v}) = \mathbb{N}_{k,\ell}(\boldsymbol{v}) \times \min_v(\mathbb{N}_{k,\ell-1}(\boldsymbol{v})/\mathbb{N}_{k,\ell}(\boldsymbol{v}))$. Then $0 \leq \alpha_{k,\ell}^{*}(\boldsymbol{v}) \leq 1$ for all strata defined by $\boldsymbol{V} = \boldsymbol{v}$. $\alpha_{k,\ell}^{*,\mathcal{D}}(\boldsymbol{v})$ is a stage-specific probability of inclusion at time $k$. This constrains the sample size to $\mathbb{N}_{k,\ell}^{*}(\boldsymbol{v}) = \sum_v \mathbb{N}_{k,\ell}^{*}(\boldsymbol{v})$.

In the second stage $\ell = 2$, a sample $\mathbb{S}_{k,2}^{\mathcal{D}1}(r)$ of chosen size $\mathbb{N}_{k,2}(r)$ is selected from the second-stage sampling frame $\mathbb{S}_{k,1}^{\mathcal{D}1}$ (selected in stage one) of size $\mathbb{N}_{k,1}(r)$ using the sampling fraction $\alpha_{k,2}^{\mathcal{D}1}(r, \boldsymbol{a_k})$:

$$\alpha_{k,2}^{\mathcal{D}1}(r, \boldsymbol{a_k}) = \frac{\mathbb{N}_{k,1}(r)}{\mathbb{N}_{k,2}(r)} \times \frac{P(\boldsymbol{a_k}|Q_k=1, T=1, k)}{P(\boldsymbol{a_k}|Q_k=1, R=r, k)} \tag{7}$$

The final sample $\mathbb{S}_{k,2}^{\mathcal{D}1}$ [collapsed over $R$] is where disparity $\psi_k$ is measured. If no allowable covariates are specified, stage 2 enrolls all persons (i.e., when $\mathbb{N}_{k,1}(r) = \mathbb{N}_{k,2}(r)$) or selects by simple random sampling for each social group $R = r$. Otherwise, the sampling fractions $\alpha_{k,2}^{\mathcal{D}1}(r, \boldsymbol{a_k})$ similarly situates the social groups on the allowables $\boldsymbol{A_k}$ according to their distribution in the standard population, denoted $T = 1$.

*4.8 Time Zero*

Time zero indicates the temporal anchor during calendar time for the start of follow-up for the outcomes $Y_{k+j}$, where the time-scale of follow-up (i.e., time on study) is denoted as $j = 0,1,2, \dots, J$ (with $J$ the end of follow-up). Time-zero determines when outcomes are counted towards disparity. We propose to anchor time zero at calendar time $k$, for two reasons: 1) to avoid differential alignment of outcomes from the point of eligibility; 2) to avoid underestimating disparity from a relevant point in time. For example, disparities (e.g., in appropriate treatment) may occur early after enrollment (e.g., after hospital discharge for myocardial infarction). If early treatment is critical for preventing adverse outcomes (e.g., a second myocardial infarction), we want to characterize that disparity by setting time zero right after enrollment.

*4.9 Follow-up and Outcome Ascertainment*

We specify how outcomes are defined (e.g., incident or prevalent), what constructs are considered, how they are measured, and for how long they will be assessed. These details add precision that can aid future interventional work or policy actions to reduce disparity. For example, if our enrollment window is indexed around an incident diagnosis of hypertension, resolving disparities early on may require a focus on addressing patient knowledge, awareness, and structures that prevent adherence to a healthy diet and regular physical activity. Resolving disparities five years post-onset also involves supports to improve medication adherence, enable home-based blood-pressure monitoring, and resources and protocols to facilitate timely and appropriate treatment intensification by clinicians for patients with uncontrolled hypertension.

*4.10 Statistical Analysis*

Last, we need to specify how the data will be analyzed. We choose the scale (e.g., additive or ratio), choice of coding (shortfall [e.g., uncontrolled hypertension] or gain [controlled hypertension] for reporting disparity. For repeatedly measured outcomes or time-to-event outcomes, we also choose whether to present measures indexed at the end of follow-up (i.e., $k + J$) or to present a graphical summary of the disparity or of group-specific measures indexed at each time $k + j$ during follow-up (i.e., at time on study $j = 0, \ldots, J$) such as group-specific growth curves, cumulative incidence curves, or survival curves.

If there are multiple target studies across calendar times $k$, we can always present trends in disparity or group-specific outcomes across calendar time $k$. We may also provide a summary measure $\Psi$, which is a weighted average of calendar-specific disparity estimates $\psi_k$ for outcomes indexed at any point during follow-up $k + j$ (or the end of follow-up $k + J$), as in (2). To summarize the additive disparity $\psi_k^{add}$ using (2), we suggest the weight $\gamma_k = P(k|Q_k = 1, T = 1)$ which is the probability that an instance of enrollment among the standard population has calendar time $k$. To summarize the relative disparity $\psi_k^{rel}$, the weight $\gamma_k$ is multiplied by $\mu_k(0)$, the mean outcome among the enrolled privileged group at time $k$. (Miettinen 1972) This approach standardizes the additive disparity $\psi_k^{add}$ (or the relative disparity $\psi_k^{rel}$) to the distribution of enrollment timing among the standard population (Section 4.6), removing impacts of differential enrollment timing (Section 3.2).[12] The summary measure $\Psi$ is interpretable as a difference in standardized mean outcomes (for $\psi_k^{add}$) or a ratio of standardized mean outcomes (for $\psi_k^{rel}$). Thus, one may alternatively pool instances of enrollment and weight each instance in the statistical analysis by $\lambda_k(r)$:

$$\lambda_k(r) = \frac{P(k|Q_k=1,T=1)}{P(k|Q_k=1,R=r)} \tag{8}$$

Such a pooled analysis permits aggregation of trends in the outcome over the target study timescale $j$.

A person may be eligible many times (e.g., they may have prior hypertension at multiple visits). Of course, under certain eligibility criteria (e.g., recent onset of hypertension) a person may only be eligible at one point in calendar time. When persons enroll in multiple studies over calendar time, this leads to correlated outcomes which can be addressed by using a stratified cluster bootstrap (Davison and Hinkley 1997, Field and Welsh 2007, Ren, Lai, Tong et al. 2010, Huang 2018) to obtain confidence intervals.

## 5. Extension of the Target Study to Address Non-Random Sample Selection

### 5.1 *Overview*.

When investigators wish to include all contributions of non-random sample selection to disparity, the target study described in Section 4 is sufficient. To address non-random sample selection, we introduce sampling strategies that allow data from the eligible population described in Section 4, denoted by $Q_k = 1$, to infer about disparity that may exist in a broader population (Design 2), a different population (Design 3), or a counterfactual population (Design 4) in which collider stratification from selecting on full or partial eligibility does not occur. Designs 2 and 3 allow inference to studies where eligibility criteria are changed whereas Design 4 allows inference to target studies that, through intervention, change who is eligible. Design 4, if chosen, adds a causal element to the model in that the eligibility-related variables are intervened on, but Design 4 remains descriptive with respect to social group membership and allowables. We will see that, unlike Designs 2 and 3, Design 4 may be used when eligibility-related variables affect the outcome.

[12] The standardization of the disparity measure $\psi_k$ to the distribution of enrollment time among the standard population can be informally viewed as treating enrollment time as "allowable," though this is achieved through aggregation rather than sampling.

In addition to the innocuous sampling assumption (4) and variants of the overlap assumption (3), each modified sampling design relies on independence (or exchangeability) assumptions and positivity assumptions. In each design, these additional assumptions may partly depend on a set of non-allowable covariates $\boldsymbol{N_k}$ (e.g., socioeconomic status) that are measured by the time of enrollment $k$ and, when specified, are factored into the sampling design. Specifically, the first-stage and second-stage sampling fractions $\alpha_{k,1}^{\mathcal{D}}(\cdot)$ and $\alpha_{k,2}^{\mathcal{D}}(\cdot)$ may depend on the allowables $\boldsymbol{A_k}$ and non-allowables $\boldsymbol{N_k}$. When the target study design uses non-allowalbes $\boldsymbol{N_k}$, ultimately, it does not balance them across social groups $R = r$.[13]

Designs 2 and 3 operate under the same minimal structural constraints as Design 1.[14] The sampling strategy for Design 4, invoking counterfactuals, has more constraints which we discuss later. Aside from the sampling plan and aggregation over calendar time, the other elements are unchanged from Design 1.

*5.2 Design 2: Sampling as if from a Broader Population (Generalizability)*

As in Figure 1B, suppose that the indicator of full eligibility $Q_k$ (1: yes, 0: no) is based on partial eligibility $Q_k^{\ddagger}$ (1: yes if has prior hypertension, established care in the health system, with a visit within $k$, 0: no otherwise) and partial eligibility $Q_k^{\dagger} = 1$ (1: yes if enrolled in the EPPP, i.e., $Q_k = Q_k^{\dagger} \times Q_k^{\ddagger}$. Suppose we must study persons with prior hypertension, established care, a current visit who are enrolled in the EPPP (i.e., $Q_k = 1$ as in Figure 2A) but want to assess disparity regardless of EPPP enrollment (i.e., a broader version of eligibility $Q_k^{*}$=1 based on $Q_k^{\ddagger} = 1$ alone, as in Figure 2B). For this we use first-stage sampling fractions $\alpha_{k,1}^{\mathcal{D}2}(r, \boldsymbol{a_k}, \boldsymbol{n_k})$ that act as if we sample the broader population defined by $Q_k^{\ddagger} = 1$ alone:

$$\alpha_{k,1}^{\mathcal{D}2}(r, \boldsymbol{a_k}, \boldsymbol{n_k}) = \frac{\mathbb{N}_{k,1}(r)}{\mathbb{N}_{k,0}(r)} \times \frac{P(\boldsymbol{n_k}|Q_k^{\ddagger}=1,R=r,\boldsymbol{a_k},k)}{P(\boldsymbol{n_k}|Q_k=1,R=r,\boldsymbol{a_k},k)} \times \frac{P(\boldsymbol{a_k}|Q_k^{\ddagger}=1,R=r,k)}{P(\boldsymbol{a_k}|Q_k=1,R=r,k)} \quad (9)$$

$\mathbb{N}_{k,0}(r)$ is the size of the first-stage sampling frame $\mathbb{S}_{k,0}(r)$, the source population $\mathcal{P}_k$ with $Q_k = 1$. We use second-stage sampling fractions $\alpha_{k,2}^{\mathcal{D}2}(r, \boldsymbol{a_k})$ to create a final sample $\mathbb{S}_{k,2}^{\mathcal{D}2}(r)$ where the allowables $\boldsymbol{A_k}$ are balanced across groups to follow their distribution in the standard population, defined among $\mathbb{S}_{k,2}^{\mathcal{D}2}$:

$$\alpha_{k,2}^{\mathcal{D}2}(r, \boldsymbol{a_k}) = \frac{\mathbb{N}_{k,2}(r)}{\mathbb{N}_{k,1}(r)} \times \frac{P(\boldsymbol{a_k}|Q^{\ddagger}=1,T=1,k)}{P(\boldsymbol{a_k}|Q^{\ddagger}=1,R=r,k)} \quad (10)$$

When the marginalized group in the broader population is the standard population, it undergoes simple random sampling in stage 2, so that its expected outcome in the target study and in the broader population are the same (i.e., the inference for the marginalized group is purely descriptive).

---

[13] Each sampling strategy differs from the typical approach of adding "inappropriate" variables as allowable covariates, which can lead to an "overadjustment" by making social groups similar on the very factors that are implicated in disparity.

[14] It is okay if $R \rightarrow (\boldsymbol{W_k}, \boldsymbol{N_k}, \boldsymbol{A_k})$ exists and it is okay if $(\boldsymbol{A_k}, \boldsymbol{N_k}) \rightarrow \boldsymbol{W_k}$ or $\boldsymbol{W_k} \rightarrow (\boldsymbol{N_k}, \boldsymbol{A_k})$ exists, and any causal relations between $\boldsymbol{N_k}$ and $\boldsymbol{A_k}$ are permitted. However, it is not okay if $Y_{k+J} \rightarrow (\boldsymbol{W_k}, \boldsymbol{N_k}, \boldsymbol{A_k})$ exists.

The design permits inference to the broader population under an independence assumption:

$$Y_{k+J} \coprod Q_k^{\dagger} | Q_k^{\ddagger} = 1, R = r, \boldsymbol{N} = \boldsymbol{n}, \boldsymbol{A} = \boldsymbol{a}, k \text{ for all } k \tag{11}$$

In words, the outcome $Y_{k+J}$ (e.g., hypertension control) must be independent of partial eligibility $Q_{\boldsymbol{k}}^{\dagger}$ (e.g., based on EPPP enrollment) among the broader population (denoted by $Q_k^{\ddagger} = 1$) given social group $R = r$, the allowables $\boldsymbol{A_k}$ (e.g., age and sex), and non-allowables $\boldsymbol{N_k}$ (e.g., SES), . This assumption would hold in Figure 1B if $\boldsymbol{W_k^{\dagger}}$ (e.g., EPPP enrollment) did not affect the outcome $Y_{k+J}$ (i.e., the arrow $\boldsymbol{W_k^{\dagger}} \rightarrow Y_{k+J}$ is absent). A positivity assumption is also required:

$$P\left(Q_k^{\dagger} = 1 \middle| Q_k^{\ddagger} = 1, R = r, \boldsymbol{N_k} = \boldsymbol{n_k}, \boldsymbol{A_k} = \boldsymbol{a_k}, k\right) > 0 \tag{12}$$

$$\text{for all } (\boldsymbol{a_k}, \boldsymbol{n_k}) \text{ with } P\left(\boldsymbol{N_k} = \boldsymbol{n_k}, \boldsymbol{A_k} = \boldsymbol{a_k} | Q_k^{\ddagger} = 1, R = r, k\right) > 0 \text{ and all } k$$

For each social group $R = r$, we note at each time $k$ the pattern of allowable $\boldsymbol{A_k}$ and non-allowable $\boldsymbol{N_k}$ covariate values among the broader population (e.g., denoted only by $Q_k^{\ddagger} = 1$). For each pattern, we must observe persons who belong to the population that our target study enrolls (i.e., persons who meet our narrower version of full eligibility $Q_k = 1$ based on partial eligibility indicators $Q_k^{\dagger} = 1$ and $Q_k^{\ddagger} = 1$). The overlap assumption (3) needs to hold among the broader population defined by $Q_k^{\ddagger} = 1$.[15]

*5.3 Design 3: Sampling as if from a Different Population (Transportability)*

Suppose again that full eligibility $Q_k = 1$ is based on partial eligibility $Q_k^{\ddagger} = 1$ (prior hypertension, established care, current visit) and partial eligibility $Q_k^{\dagger} = 1$ (enrolled in the EPPP) as in Figure 1B. The target study again enrolls those who are fully eligible (i.e., $Q_k = 1$ as in Figure 2C) but we want to assess disparity for those not enrolled in the EPPP (i.e., a different version of full eligibility $Q_k^{**} = 1$ based on $Q_k^{\ddagger} = 1$ and $Q_k^{\dagger} = 0$, as in Figure 2D). For this we use modified first-stage sampling fractions $\alpha_{k,1}^{\mathcal{D}3}(r, \boldsymbol{a_k}, \boldsymbol{n_k})$ that act as if we sample the different population defined by $Q_k^{\ddagger} = 1$ and $Q_k^{\dagger} = 0$:

$$\alpha_{k,1}^{\mathcal{D}3}(r, \boldsymbol{a_k}, \boldsymbol{n_k}) = \frac{\mathbb{N}_{k,1}(r)}{\mathbb{N}_{k,0}(r)} \times \frac{P(\boldsymbol{n_k} | Q_k^{\dagger}=0, Q_k^{\ddagger}=1, R=r, \boldsymbol{a_k}, k)}{P(\boldsymbol{n_k} | Q_k=1, R=r, \boldsymbol{a_k}, k)} \times \frac{P(\boldsymbol{a_k} | Q_k^{\dagger}=0, Q_k^{\ddagger}=1, R=r, k)}{P(\boldsymbol{a_k} | Q_k=1, R=r, k)} \tag{13}$$

$\mathbb{N}_{k,0}(r)$ is the size of the first-stage sampling frame $\mathbb{S}_{k,0}(r)$, i.e., the source population $\mathcal{P}_k$ with $Q_k^{\ddagger} = 1$ and $Q_k^{\dagger} = 0$. We use second-stage sampling fractions $\alpha_{k,2}^{\mathcal{D}3}(r, \boldsymbol{a_k})$ to create a final sample $\mathbb{S}_{k,2}^{\mathcal{D}3}(r)$ where allowables are balanced across groups to their distribution in the standard population, defined among $\mathbb{S}_{k,2}^{\mathcal{D}3}$:

$$\alpha_{k,2}^{\mathcal{D}3}(r, \boldsymbol{a_k}) = \frac{\mathbb{N}_{k,2}(r)}{\mathbb{N}_{k,1}(r)} \times \frac{P(\boldsymbol{a_k} | Q_k^{\dagger}=0, Q_k^{\ddagger}=1, T=1, k)}{P(\boldsymbol{a_k} | Q_k^{\dagger}=0, Q_k^{\ddagger}=1, R=r, k)} \tag{14}$$

---

[15] (3) becomes $P\left(\boldsymbol{A_k} = \boldsymbol{a_k} | Q_k^{\ddagger} = 1, R = r, k\right) > 0$ for all $\boldsymbol{a_k}$ with $P\left(\boldsymbol{A_k} = \boldsymbol{a_k} | Q_k^{\ddagger} = 1, T = 1, k\right) > 0$ and all $k$

When the marginalized group in the different population is the standard population, it undergoes simple random sampling in stage 2, so that its expected outcome in the target study and in the different population is the same (i.e., the inference for the marginalized group is purely descriptive).

The design permits inference to the different population under the independence assumption (11) which, again, would hold in Figure 1B if $\boldsymbol{W}_{\boldsymbol{k}}^{\dagger}$ (e.g. EPPP enrollment) did not affect the outcome (e.g., hypertension control $Y_{k+J}$ (i.e., the arrow $\boldsymbol{W}_{\boldsymbol{k}}^{\dagger} \rightarrow Y_{k+J}$) is absent. A positivity assumption is also required:

$$P\left(Q_k^{\dagger} = 1 \middle| Q_k^{\ddagger} = 1, R = r, \boldsymbol{N}_{\boldsymbol{k}} = \boldsymbol{n}_{\boldsymbol{k}}, \boldsymbol{A}_{\boldsymbol{k}} = \boldsymbol{a}_{\boldsymbol{k}}, k\right) > 0 \tag{15}$$

$$\text{for all } (\boldsymbol{a}_{\boldsymbol{k}}, \boldsymbol{n}_{\boldsymbol{k}}) \text{ with } P\left(\boldsymbol{N}_{\boldsymbol{k}} = \boldsymbol{n}_{\boldsymbol{k}}, \boldsymbol{A}_{\boldsymbol{k}} = \boldsymbol{a}_{\boldsymbol{k}} | Q_k^{\dagger} = 0, Q_k^{\ddagger} = 1, R = r, k\right) > 0 \text{ and all } k$$

For each social group $R = r$, we note at each time $k$ the pattern of allowable $\boldsymbol{A}_{\boldsymbol{k}}$ and non-allowable $\boldsymbol{N}_{\boldsymbol{k}}$ covariate values among the different population (e.g., denoted by $Q_k^{\ddagger} = 1$ and $Q_k^{\dagger} = 0$). For each pattern, we must observe persons who belong to the population our target study enrolls (i.e., who meet our narrower version of full eligibility $Q_k = 1$ based on partial eligibility indicators $Q_k^{\dagger} = 1$ and $Q_k^{\ddagger} = 1$). The overlap assumption (3) needs to hold among the different population defined by $(Q_k^{\dagger} = 0, Q_k^{\ddagger} = 1)$.[16]

*5.4 Design 4: Sampling as if from a Counterfactual Population (Inference in a Selected Population)*

As in Figure 1C, now we express full eligibility $Q_k$ (1: yes, 0: no) with finer partial eligibility indicators $Q_k^{\ddagger}$ (prior hypertension, established care), $Q_k^{\dagger}$ (e.g., EPPP enrollment), and $Q_k^{\wr}$ (e.g., current visit), i.e., $Q_k = Q^{\wr} \times Q_k^{\dagger} \times Q_k^{\ddagger}$, where $\boldsymbol{W}_{\boldsymbol{k}}^{\ddagger}$ may affect $\boldsymbol{W}_{\boldsymbol{k}}^{\dagger}$ which may affect $\boldsymbol{W}_{\boldsymbol{k}}^{\wr}$.[17] Suppose we want to infer to those with full eligibility (Figure 2E) but worry that selecting persons enrolled in the EPPP (i.e., $Q_k^{\dagger} = 1$) induces collider-stratification that masks disparity. Suppose that we accept collider stratification through other partial eligibility indicators $Q_k^{\ddagger}$ (e.g., prior hypertension, established care) and $Q_k^{\wr}$ (e.g., current visit) as part of disparity. To avoid collider stratification through $Q_k^{\dagger}$, we infer to a counterfactual population where such collider-stratification is absent (Figure 2F). Denote $G_k$ as an intervention to allocate[18] the partial eligibility variables $\boldsymbol{W}_{\boldsymbol{k}}^{\dagger}$ (e.g., EPPP enrollment) according to a distribution $g_k(\cdot)$ that does not simultaneously depend on (i) social group $R$ and (ii) non-allowables $\boldsymbol{N}_{\boldsymbol{k}}$ (e.g., risk factors $L_k$) (Table 1). Let $V_k^{G_k}$ be the potential outcome of a variable $V_k$ under intervention $G_k$. Our use of a superscript to denote the intervention $G_k$ differs from our use of a superscript to denote a sampling design $\mathcal{D}$. We use sampling fractions $\alpha_{k,1}^{\mathcal{D}4}(r, \boldsymbol{a}_{\boldsymbol{k}}, \boldsymbol{n}_{\boldsymbol{k}})$ that act as if we sample the counterfactual population defined by $Q_k^{G_k} = 1$:

$$\alpha_{k,1}^{\mathcal{D}4}(r, \boldsymbol{a}_{\boldsymbol{k}}, \boldsymbol{n}_{\boldsymbol{k}}) = \frac{\mathbb{N}_{k,1}(r)}{\mathbb{N}_{k,0}(r)} \times \frac{P(\boldsymbol{n}_{\boldsymbol{k}} | Q_k^{G_k}=1, R=r, \boldsymbol{a}_{\boldsymbol{k}}, k)}{P(\boldsymbol{n}_{\boldsymbol{k}} | Q_k=1, R=r, \boldsymbol{a}_{\boldsymbol{k}}, k)} \times \frac{P(\boldsymbol{a}_{\boldsymbol{k}} | Q_k^{G_k}=1, R=r, k)}{P(\boldsymbol{a}_{\boldsymbol{k}} | Q_k=1, R=r, k)} \tag{16}$$

---

[16] (3) becomes $P(\boldsymbol{A}_{\boldsymbol{k}} = \boldsymbol{a}_{\boldsymbol{k}} | Q_k^{\dagger} = 0, Q_k^{\ddagger} = 1, R = r, k) > 0$ for all $\boldsymbol{a}_{\boldsymbol{k}}$ with $P(\boldsymbol{A}_{\boldsymbol{k}} = \boldsymbol{a}_{\boldsymbol{k}} | Q_{k'}^{\dagger} = 0, Q_k^{\ddagger} = 1, T = 1, k) > 0$ and all $k$

[17] Designs 2 and 3 do not distinguish eligibility-related variables that may be affected by $\boldsymbol{W}_{\boldsymbol{k}}^{\dagger}$, and thus subsume $\boldsymbol{W}_{\boldsymbol{k}}^{\wr}$ into $\boldsymbol{W}_{\boldsymbol{k}}^{\ddagger}$.

[18] Muñoz and van der Laan (2012) define causal effects of treatment allocation policies. We allocate eligibility-related variables.

$\mathbb{N}_{k,0}(r)$ is the size of the first-stage sampling frame $\mathbb{S}_{k,0}^{G_k}(r)$, i.e., the counterfactual source population $\mathcal{P}_k^{G_k}(r)$ with $Q_k^{G_k} = 1$. We use second-stage sampling fractions $\alpha_{k,2}^{\mathcal{D}4}(r, \boldsymbol{a_k})$ to create a final sample $\mathbb{S}_{k,2}^{\mathcal{D}4}(r)$ where the allowables $\boldsymbol{A_k}$ are balanced according to the standard distribution defined among $\mathbb{S}_{k,2}^{\mathcal{D}4}$:

$$\alpha_{k,2}^{\mathcal{D}4}(r, \boldsymbol{a_k}) = \frac{\mathbb{N}_{k,2}(r)}{\mathbb{N}_{k,1}(r)} \times \frac{P(\boldsymbol{a_k} | Q_k^{G_k} = 1, T = 1, k)}{P(\boldsymbol{a_k} | Q_k^{G_k} = 1, R = r, k)} \tag{17}$$

This design permits inference to the counterfactual population via an exchangeability assumption[19]:

$$\left(Y_{k+J}^{G_k}, Q_k^{\wr\, G_k}\right) \amalg Q_k^{\dagger} | Q_k^{\ddagger} = 1, R = r, \boldsymbol{N} = \boldsymbol{n}, \boldsymbol{A} = \boldsymbol{a}, k \quad \text{for all } k \tag{18}$$

In words, the potential outcome $Y_{k+J}^{G_k}$ (e.g., hypertension control) and the potential value of partial eligibility $Q_k^{\wr\, G_k}$ (e.g., current visit) must be jointly independent of observed partial eligibility $Q_k^{\dagger}$ (e.g., EPPP enrollment) given partial eligibility $Q_k^{\ddagger} = 1$, social group $R = r$, the allowables $\boldsymbol{A_k}$, and non-allowables $\boldsymbol{N_k}$, i.e., no unmeasured selection-bias. Positivity (12) is required as well as consistency: the intervention $G_k$ returns observed values for $Y_{k+J}$ and $Q_k^{\wr}$ when it assigns a person's observed values for $\boldsymbol{W_k^{\dagger}}$. The overlap assumption (3) needs to hold among the counterfactual population defined by $Q_k^{G_k} = 1$.[20] Now, if no partial eligibility variables occur after $\boldsymbol{W_k^{\dagger}}$ (i.e., if $\boldsymbol{W_k^{\wr}} = \emptyset$ the empty set) exchangeability (18) simplifies to:

$$Y_{k+J}^{G_k} \amalg Q_k^{\dagger} | Q_k^{\ddagger} = 1, R = r, \boldsymbol{N} = \boldsymbol{n}, \boldsymbol{A} = \boldsymbol{a}, k \quad \text{for all } k \tag{19}$$

The exchangeability assumptions (18) or (19) of Design 4 holds where the independence assumption (11) of Designs 2 and 3 fails: when $\boldsymbol{W_k^{\dagger}}$ affects the outcome $Y_{k+J}$, i.e., the arrow $\boldsymbol{W_k^{\dagger}} \rightarrow Y_{k+J}$ in Figure 1C (see Figure 4). Design 4 operates under additional structural constraints compared to Designs 1, 2, and 3.[21]

Table 1 specifies interventions for $G_k$ (Designs 4a, 4b, and 4c) that eliminate collider stratification through partial eligibility $Q_k^{\dagger}$. Design 4a makes partial eligibility $Q_k^{\dagger}$ random given partial eligibility $Q_k^{\ddagger} = 1$ and calendar time $k$. Design 4b makes partial eligibility $Q_k^{\dagger}$ random with respect to non-allowables $\boldsymbol{N_k}$ given partially eligibility $Q_k^{\ddagger} = 1$, social group $R = r$, the allowables $\boldsymbol{A_k}$ and calendar time $k$. Design 4c makes partial eligibility $Q_k^{\dagger}$ random with respect to social group $R$ given partial eligibility $Q_k^{\ddagger} = 1$, the allowables $\boldsymbol{A_k}$, a set of non-allowables $\boldsymbol{N_k}$ that satisfy exchangeability (18) or (19), and calendar time $k$. These designs target different counterfactual populations and may return different estimates of disparity. To choose, one may consider the design's feasibility (in how $\boldsymbol{W_k^{\dagger}}$ is allocated) or the design's inferential utility. When there are no downstream partial eligibility-related variables (i.e., $\boldsymbol{W}_{\boldsymbol{k}}^{\wr} = \emptyset$), Design 4a generalizes to the broader population defined by $Q_k^{\ddagger} = 1$ even when eligibility variables $\boldsymbol{W}_{\boldsymbol{k}}^{\dagger}$ affect the outcome $Y_{k+J}$ (unlike Design 2). Under the same conditions, when the marginalized group is the standard population, Design 4c reduces to simple random sampling among the marginalized group across both stages of sampling. Then, the model fully describes the fully eligible marginalized group defined by $Q_k = 1$. One may also consider meaningfulness. Designs 4b and 4c do not remove all social group differences in partial eligibility $Q_k^{\dagger}$, which may be a more "realistic" setting for characterizing disparity in outcomes. Finally, one may also consider effectiveness. When $\boldsymbol{N}_{\boldsymbol{k}}$ is multivariate, under certain causal structures Design 4c, unlike Designs 4a and 4b, may leave residual collider-stratification through conditioning on partial eligibility $Q_k^{\dagger}$.[22]

---

[19] Exchangeability (18), positivity (12), and consistency (above) help identify (in Section 6.6) the distributions $P(\boldsymbol{n_k} | Q_k^{G_k} = 1, R = r, \boldsymbol{a_k}, k)$ and $P(\boldsymbol{a_k} | Q_k^{G_k} = 1, T = 1, k)$ from the observed data. The sampling fractions $\alpha_{k,2}^{\mathcal{D}4}(\cdot)$ do not balance $\boldsymbol{N_k}$ across $R$.
[20] (3) becomes $P(\boldsymbol{A_k} = \boldsymbol{a_k} | Q_k^{G_k} = 1, R = r, k) > 0$ for $\boldsymbol{a_k}$ with $P(\boldsymbol{A_k} = \boldsymbol{a_k} | Q_k^{G_k} = 1, T = 1, k) > 0$ and all $k$
[21] It is okay if $R \rightarrow (\boldsymbol{W_k}, \boldsymbol{N_k}, \boldsymbol{A_k})$ exists and it is okay if $(\boldsymbol{A_k}, \boldsymbol{N_k}) \rightarrow \boldsymbol{W_k^{\ddagger}}$ or $\boldsymbol{W_k^{\ddagger}} \rightarrow (\boldsymbol{N_k}, \boldsymbol{A_k})$ exists, and any causal relations between $\boldsymbol{N_k}$ and $\boldsymbol{A_k}$ are permitted. It is not okay if $Y_{k+J} \rightarrow (\boldsymbol{W_k}, \boldsymbol{N_k}, \boldsymbol{A_k})$ exists or if $(\boldsymbol{W_k^{\dagger}}, \boldsymbol{W_k^{\wr}}) \rightarrow (\boldsymbol{N_k}, \boldsymbol{A_k})$ exists.

*5.5 Modified Statistical Analysis*

In Sections 4.1 and 4.10 we discussed procedures to aggregate results over calendar time use the distribution in the standard population implied by the design. This is the broader population under Design 2, the different population under Design 3, and the counterfactual population under Design 4.[23]

## 6. Emulation of the Target Study with Secondary Data

*6.1 Overview*

In theory, the target study protocol could be implemented in real life to measure disparity. Often, a target study will have to be emulated through the design and analysis of secondary data. We outline data structures and estimators to emulate the target study under our motivating example of assessing racial disparity in hypertension control in a healthcare system among those with prior hypertension, established care and a current visit who are (a) enrolled in the EPPP (Design 1, Section 4); (b) may or may not be enrolled in EPPP (Design 2, generalizability, Section 5.2); not enrolled in the EPPP (Design 3, transportability, Section 5.3); enrolled in the EPPP under a hypothetical allocation of EPPP (Design 4, inference in a selected population, Section 5.4). These applications are plausible when we only have outcomes $Y_{k+J}$ measured in the EPPP. The example target study protocols and emulation steps for each design are shown in Table 2. In this example, we assume multiple target studies across calendar time whose results are to be aggregated. Recall that with multiple target studies, a person may possibly be eligible for and enroll in multiple studies. We show how the emulation simplifies with one target study. For each design $\mathcal{D}$, we present estimators for each social group's mean outcome aggregated over calendar time $k$, $\tau^{\mathcal{D}}(r)$ (Sections 4.1, 4.10, 5.5). The aggregated additive disparity is $\Psi^{add} = \tau^{\mathcal{D}}(1) - \tau^{\mathcal{D}}(0)$ and the relative disparity is $\Psi^{rel} = \tau^{\mathcal{D}}(1)/\tau^{\mathcal{D}}(0)$.

---

[22] Under Design 4c, residual contributions from collider-stratification would occur on Figure 2C if, rather than the single node $L_k$, we had two nodes $L_{k(i)}$ and $L_{k(ii)}$ that each inherited all edges from $L_k$. The residual contribution would operate through the pathways $R \rightarrow L_{k(i)} \rightarrow W_k^{\dagger} \leftarrow L_{k(ii)} \rightarrow Y$ and $R \rightarrow L_{k(ii)} \rightarrow W_k^{\dagger} \leftarrow L_{k(i)} \rightarrow Y$. This would not occur under Designs 4a or 4b.

[23] To aggregate $\psi_k^{add}$, $\gamma_k = P(k|Q^{\ddagger} = 1, T = 1)$ for Design 2, $\gamma_k = P(k|Q_k^{\dagger} = 0, Q_k^{\ddagger} = 1, T = 1)$ for Design 3, and $\gamma_k = P(k|Q_k^{G_k} = 1, T = 1)$ for Design 4. To aggregate $\psi_k^{rel}$, $\gamma_k$ is multiplied by $\mu_k(0) = E_{\Omega}[Y_{k+J}|Q_k^{\ddagger} = 1, R = 0, k]$ for Design 2, by $\mu_k(0) = E_{\Omega}[Y_{k+J}|Q_k^{\dagger} = 0, Q_k^{\ddagger} = 1, R = 0, k]$ for Design 3, and by $\mu_k(0) = E_{\Omega}[Y_{k+J}|Q_k^{G_k} = 1, R = 0, k]$ for Design 4. For pooled analyses, $\lambda_k(r) = P(k|Q^{\ddagger} = 1, T = 1)/P(k|Q^{\ddagger} = 1, R = r)$ for Design 2, $\lambda_k(r) = P(k|Q_k^{\dagger} = 0, Q_k^{\ddagger} = 1, T = 1)/P(k|Q_k^{\dagger} = 0, Q_k^{\ddagger} = 1, R = r)$ for Design 3, and $\lambda_k = P(k|Q_k^{G_k} = 1, T = 1)/P(k|Q_k^{G_k} = 1, R = r)$ for Design 4.

We present two types of estimators that, given the appropriate data structure, are used to emulate the sampling-based enrollment and aggregation. G-computation, (Snowden, Rose and Mortimer 2011) akin to model-based standardization, sequentially regresses the outcome and predicted values. Weighting, (Hernán and Robins 2006) which takes a weighted average of the outcome, models membership in the social group $R = r$, the standard population $T = 1$ and, for Designs 2 through 4, indicators of partial eligibility. G-computation estimators are usually more efficient (Ren, Cislo, Cappelleri et al. 2023) but weighting is more objective as the weights are constructed without outcome data. To construct confidence intervals that account for clustering by individual, we use a cluster bootstrap that samples each individual with replacement. (Davison and Hinkley 1997, Field and Welsh 2007, Ren et al. 2010, Huang 2018). We abbreviate a weighted mean, $\sum_i Y_i \omega_i \,/\, \sum_i \omega_i$ where $i$ represents the unit of observation, as $E[Y \times \omega]$.

*6.2 Data Structure*

To emulate a target study, we specify a unit of calendar time for enrollment windows (Section 4.2) (e.g., months), enrollment groups $R$ (Section 4.3; e.g., Black persons [marginalized $R = 1$] and White persons [privileged $R = 0$]), and full eligibility $Q_k = 1$ (Section 4.4; e.g., prior hypertension, established care, and enrolled in EPPP before $k$, and current visit within $k$). For Design 1, we do not disaggregate full eligibility $Q_k$ into partial eligibility. For Designs 2 and 3, we distinguish partial eligibility $Q_k^{\dagger}$ that we generalize or transport over (e.g., EPPP enrollment) from partial eligibility $Q_k^{\ddagger}$ that we do not (e.g., prior hypertension, established care, current visit). For Design 4, we distinguish partial eligibility $Q_k^{\dagger}$ allocated by intervention (e.g., EPPP enrollment), partial eligibility $Q_k^{\wr}$ affected by the allocation (e.g., current visit), and partial eligibility $Q_k^{\ddagger}$ not affected by the allocation (e.g., prior hypertension, established care). We choose allowable covariates $\boldsymbol{A_k}$ to similarly situate social groups (Sections 3.3 and 4.5; e.g., age $X_k$) and the within-sample standard population, coded $T = 1$, determining their within sample distribution (Section 4.6; e.g., the Black group). We choose non-allowable covariates $\boldsymbol{N_k}$ to meet assumptions of independence (11) for Designs 2 and 3 or exchangeability (18) or (19) for Design 4 (Section 5.1; e.g., SES $L_k$).

We form a "long" dataset where every row is the vector $\mathcal{O}_{i,k} = (i, k, Q_k^{\ddagger}, Q_k^{\dagger}, Q_k^{\wr}, Q_k, R, T, X_k, L_k)$ for an individual $i$ at month $k$ (Figure 4). Each person $i$ contributes one record per calendar time $k$ (Sections 3.2 and 4.7). The data $\mathcal{O}$ only include calendar times $k$ where all social groups of interest are represented. The indicator $T$ of membership in the standard population is constructed (e.g., if the marginalized group $R = 1$ is the standard population, we set $T$ as equal to $R$). For Design 1, we subset the data $\mathcal{O}$ to those fully eligible at time $k$, i.e., by $Q_k = 1$. For Designs 2, 3, and 4, we subset the data $O$ to those partially eligible by $Q_k^{\ddagger} = 1$. We attach outcomes $Y_{k+J}$ at follow-up time $k + J$ to records indexed at time $k$.

*6.3 Identification and Estimation for Design 1 (Default Model)*

Under overlap (3) and innocuous sampling (4), we can identify the aggregated mean $\tau^{\mathcal{D}1}(r)$ as:

$$E_{\boldsymbol{a},k}(E[Y_{k+J}|Q_k = 1, R = r, \boldsymbol{a_k}, k]|Q_k = 1, T = 1) \tag{20}$$

where $E_{\boldsymbol{a},k}(\cdot)$ is over $P(\boldsymbol{a_k}, k|Q_k = 1, T = 1)$

To estimate (20) by G-computation, in step 1 we fit a model $\eta_1^{\mathcal{D}1}(\boldsymbol{a_k}, k)$ for the outcome $Y_{k+J}$ (e.g., hypertension control) given the allowables $\boldsymbol{A_k}$ (e.g. age) and calendar time $k$ (e.g., months) among those fully eligible $Q_k = 1$ (e.g., prior hypertension, established care, current visit, enrolled in EPPP) in the social group $R = r$. In step 2, we obtain predicted values $\wp_1^{\mathcal{D}1}$ from the model $\eta_1^{\mathcal{D}1}(\boldsymbol{a_k}, k)$ on those fully eligible $Q_k = 1$. In step 3, we average $\wp_1^{\mathcal{D}1}$ in the pooled, fully eligible $Q_k = 1$ standard population $T = 1$ to estimate the aggregated mean $\tau^{\mathcal{D}1}(r)$ (conditionally on calendar time $k$ for time-specific means, $\mu_k^{\mathcal{D}1}(r)$).

We may also estimate the aggregated mean outcome $\tau^{\mathcal{D}1}(r)$ by the weighting estimator:

$$E[Y_{k+J} \times \omega_{r,Q_k=1,k}^{\mathcal{D}1-pool}|Q_k = 1, R = r] \tag{21}$$

where $\omega_{r,Q_k=1,k}^{\mathcal{D}1-pool} = \frac{P(T=1|Q_k=1,\boldsymbol{a_k},k)}{P(R=r|Q_k=1,\boldsymbol{a_k},k)} \times \frac{P(R=r|Q_k=1)}{P(T=1|Q_k=1)}$

The first term of the weight is the ratio of (i) the probability of belonging to the standard population $T = 1$ among those fully eligible $Q_k = 1$, conditional on the allowables $\boldsymbol{A_k}$ and calendar time $k$ to (ii) the corresponding probability of belonging to the person's observed social group $R = r$. The second term of the weight is inverse of this ratio but with unconditional probabilities (with respect to $\boldsymbol{A_k}, k$). The probabilities in the first term may be estimated by predictions from models for $T = 1$ and $R = r$, and the probabilities in the second term are estimated directly. A weighted average of the outcome $Y_{k+J}$ (e.g., hypertension control) in the fully eligible $Q_k = 1$ social group $R = r$ estimates the aggregate mean $\tau^{\mathcal{D}1}(r)$. For time-specific means, $\mu_k^{\mathcal{D}1}(r)$, all terms condition on calendar time $k$.

*6.4 Identification and Estimation for Design 2 (Generalizability)*

Under a version of overlap (3) (see footnote 16), innocuous sampling (4), independence (11), and positivity (12), we can identify the aggregated mean $\tau^{\mathcal{D}2}(r)$ of the social group in the broader population $Q_k^{\ddagger} = 1$ (e.g., prior hypertension, established care in the health system, regardless of EPPP enrollment) as:

$$E_{\boldsymbol{a},k}[\![E_{\boldsymbol{n}}(E[Y_{k+J}|Q_k = 1, R = r, \boldsymbol{n_k}, \boldsymbol{a_k}, k]|Q_k^{\ddagger} = 1, R = r, \boldsymbol{a_k}, k)|Q_k^{\ddagger} = 1, T = 1]\!] \tag{22}$$

where $E_{\boldsymbol{n}}(\cdot)$ is over $P(\boldsymbol{n_k}|Q_k^{\ddagger} = 1, R = r, \boldsymbol{a_k}, k)$

and $E_{\boldsymbol{a},k}[\![\cdot]\!]$ is over $P(\boldsymbol{a_k}, k|Q_k^{\ddagger} = 1, T = 1)$

To estimate (22) by G-computation, in step 1 we fit a model $\eta_1^{\mathcal{D}2}(\boldsymbol{n_k}, \boldsymbol{a_k}, k)$ for the outcome $Y_{k+J}$ (e.g., hypertension control) given the allowables $\boldsymbol{A_k}$ (e.g. age), non-allowables $\boldsymbol{N_k}$ (e.g., SES), and calendar time $k$ (e.g., months) among the fully eligible $Q_k = 1$ (e.g., prior hypertension, established care, EPPP enrollment, current visit) social group $R = r$. In step 2, we obtain predicted values $\wp_1^{\mathcal{D}2}$ from the model $\eta_1^{\mathcal{D}2}(\boldsymbol{n_k}, \boldsymbol{a_k}, k)$ in broader population $Q_k^{\ddagger} = 1$ (e.g., regardless of EPPP enrollment). In step 3, we fit a model $\eta_2^{\mathcal{D}2}(\boldsymbol{a_k}, k)$ for the predicted values $\wp_1^{\mathcal{D}2}$ given the allowables $\boldsymbol{A_k}$ and calendar time $k$ on the broader $Q_k^{\ddagger} = 1$ social group $R = r$. In step 4, we obtain predicted values $\wp_2^{\mathcal{D}2}$ from the model $\eta_2^{\mathcal{D}2}(\boldsymbol{a_k}, k)$ on the broader population $Q_k^{\ddagger} = 1$. In step 5, we average $\wp_2^{\mathcal{D}2}$ in the pooled broader $Q_k^{\ddagger} = 1$ standard population $T = 1$ to estimate the aggregated mean $\tau^{\mathcal{D}2}(r)$ (conditionally on calendar time $k$ for time-specific means, $\mu_k^{\mathcal{D}2}(r)$).

We may also estimate the aggregated mean outcome $\tau^{\mathcal{D}2}(r)$ by the weighting estimator:

$$E[Y_{k+J} \times \omega_{r,Q_k=1,k}^{\mathcal{D}2-pool} | Q_k = 1, R = r] \tag{23}$$

$$\text{where } \omega_{r,Q_k=1,k}^{\mathcal{D}2-pool} = \frac{P(Q_k^{\dagger}=1|Q_k^{\ddagger}=1,R=r)}{P(Q_k^{\dagger}=1|Q_k^{\ddagger}=1,R=r,\boldsymbol{n_k},\boldsymbol{a_k},k)} \times \frac{P(T=1|Q_k^{\ddagger}=1,\boldsymbol{a_k},k)}{P(R=r|Q_k^{\ddagger}=1,\boldsymbol{a_k},k)} \times \frac{P(R=r|Q_k^{\ddagger}=1)}{P(T=1|Q_k^{\ddagger}=1)}$$

The second and third terms are similar to (21) but are defined by the broader $Q_k^{\ddagger} = 1$ population (e.g., prior hypertension, established care, current visit) rather than by those fully eligible $Q_k = 1$ (e.g., also enrolled in the EPPP). The first term is the ratio of (i) the unconditional (with respect to $\boldsymbol{A_k}, \boldsymbol{N_k}, k$) probability of partial eligibility $Q_k^{\dagger} = 1$ (e.g., enrolled in the EPPP) in the broader $Q_k^{\ddagger} = 1$ population (e.g., prior hypertension, established care, current visit) in the social group $R = r$, to (ii) the corresponding conditional probability given the non-allowables $\boldsymbol{N_k}$, allowables $\boldsymbol{A_k}$, and calendar time $k$. The denominator is estimated by predictions from a model for $Q_k^{\dagger} = 1$ and the numerator is estimated directly. A weighted average of the outcome $Y_{k+J}$ (e.g., hypertension control) in the fully eligible $Q_k = 1$ social group $R = r$ estimates the aggregate mean $\tau^{\mathcal{D}2}(r)$. For time-specific means, $\mu_k^{\mathcal{D}2}(r)$, all terms condition on calendar time $k$.

*6.5 Identification and Estimation for Design 3 (Transportability)*

Under a version of overlap (3) (see footnote 17), innocuous sampling (4), independence (11), and positivity (15), we identify the aggregated mean $\tau^{\mathcal{D}2}(r)$ of the social group in the different population ($Q_k^{\dagger} = 0, Q_k^{\ddagger} = 1$) (e.g., prior hypertension, established care, current visit, but *not* enrolled in the EPPP) as:

$$E_{\boldsymbol{a},k}[\![E_{\boldsymbol{n}}\left(E\left[Y_{k+J}\middle|Q_k = 1, R = r, \boldsymbol{n_k}, \boldsymbol{a_k}, k\right]\middle|Q_k^{\dagger} = 0, Q_k^{\ddagger} = 1, R = r, \boldsymbol{a_k}, k\right)|Q_k^{\dagger} = 0, Q_k^{\ddagger} = 1, T = 1]\!] \tag{24}$$

$$\text{where } E_{\boldsymbol{n}}(\cdot) \text{ is over } P\left(\boldsymbol{n_k}\middle|Q_k^{\dagger} = 0, Q_k^{\ddagger} = 1, R = r, \boldsymbol{a_k}, k\right)$$

$$\text{and } E_{\boldsymbol{a},k}[\![\cdot]\!] \text{ is over } P\left(\boldsymbol{a_k}, k\middle|Q_k^{\dagger} = 0, Q_k^{\ddagger} = 1, T = 1\right)$$

To estimate (24) by G-computation, in step 1 we fit a model $\eta_1^{\mathcal{D}3}(\boldsymbol{n_k}, \boldsymbol{a_k}, k)$ for the outcome $Y_{k+J}$ (e.g., hypertension control) given the allowables $\boldsymbol{A_k}$ (e.g. age), non-allowables $\boldsymbol{N_k}$ (e.g., SES), and calendar time $k$ (e.g., months) among the fully eligible $Q_k = 1$ (e.g., prior hypertension, established care, current visit, enrolled in EPPP) social group $R = r$. In step 2, we obtain predicted values $\mathscr{p}_1^{\mathcal{D}3}$ from the model $\eta_1^{\mathcal{D}3}(\boldsymbol{n_k}, \boldsymbol{a_k}, k)$ in the different population ($Q_k^{\dagger} = 0, Q_k^{\ddagger} = 1$) (e.g., not enrolled in EPPP). In step 3, we fit a model $\eta_2^{\mathcal{D}3}(\boldsymbol{a_k}, k)$ for the predicted values $\mathscr{p}_1^{\mathcal{D}3}$ given the allowables $\boldsymbol{A_k}$ and calendar time $k$ among the different ($Q_k^{\dagger} = 0, Q_k^{\ddagger} = 1$) social group $R = r$. In step 4, we obtain predicted values $\mathscr{p}_2^{\mathcal{D}3}$ from the model $\eta_2^{\mathcal{D}3}(\boldsymbol{a_k}, k)$ on the different population ($Q_k^{\dagger} = 0, Q_k^{\ddagger} = 1$). In step 5, we average $\mathscr{p}_2^{\mathcal{D}3}$ in the pooled different ($Q_k^{\dagger} = 0, Q_k^{\ddagger} = 1$) standard population $T = 1$ to estimate the aggregated mean $\tau^{\mathcal{D}3}(r)$ (conditionally on calendar time $k$ for time-specific means, $\mu_k^{\mathcal{D}3}(r)$).

We may also estimate the aggregated mean outcome $\tau^{\mathcal{D}3}(r)$ by the weighting estimator:

$$E[Y_{k+J} \times \omega_{r,Q_k=1,k}^{\mathcal{D}3-pool} | Q_k = 1, R = r] \tag{25}$$

where $\omega_{r,Q_k=1,k}^{\mathcal{D}3-pool} = \frac{P(Q_k^{\dagger}=0|Q_k^{\ddagger}=1,R=r,\boldsymbol{n_k},\boldsymbol{a_k},k)}{P(Q_k^{\dagger}=1|Q_k^{\ddagger}=1,R=r,\boldsymbol{n_k},\boldsymbol{a_k},k)} \times \frac{P(Q_k^{\dagger}=1|Q_k^{\ddagger}=1,R=r)}{P(Q_k^{\dagger}=0|Q_k^{\ddagger}=1,R=r)} \times \frac{P(T=1|Q_k^{\dagger}=0,Q_k^{\ddagger}=1,\boldsymbol{a_k},k)}{P(R=r|Q_k^{\dagger}=0,Q_k^{\ddagger}=1,\boldsymbol{a_k},k)} \times \frac{P(R=r|Q_k^{\dagger}=0,Q_k^{\ddagger}=1)}{P(T=1|Q_k^{\dagger}=0,Q_k^{\ddagger}=1)}$

The fourth and fifth terms of the weight are similar to those used in (21), except that they are among the different population ($Q_k^{\dagger} = 0, Q_k^{\ddagger} = 1$) (e.g., prior hypertension, established care, current visit, but not enrolled in EPPP) rather than those fully eligible $Q_k = 1$ (e.g., also enrolled in the EPPP). The first term of the weight is equivalent to the inverse odds of being fully eligible ($Q_k^{\dagger} = 1, Q_k^{\ddagger} = 1$) versus in the different population ($Q_k^{\dagger} = 0, Q_k^{\ddagger} = 1$), conditional on social group $R = r$, the non-allowables $\boldsymbol{N_k}$, allowables $\boldsymbol{A_k}$, and calendar time $k$. It is estimated by fitting a model for partial eligibility $Q_k^{\dagger} = 1$, making predictions, obtaining its complement, and taking the ratio of the compliment to the prediction. The second term is the odds but is unconditional (with respect to $\boldsymbol{A_k}, \boldsymbol{N_k}, k$) and is estimated directly. A weighted average of the outcome $Y_{k+J}$ (e.g., hypertension control) among the fully eligible $Q_k = 1$ social group $R = r$ who are estimates the aggregate mean $\tau^{\mathcal{D}3}(r)$. For calendar-time specific means, $\mu_k^{\mathcal{D}3}(r)$, the probabilities in the second term and fourth terms of the weight condition on calendar time $k$.

*6.5 Identification and Estimation for Design 4 (Inference in a Counterfactual Selected Population)*

Identification and estimation under Design 4 uses special weights $\phi_k$ and $\theta_k^{pool}$ that generally are:

$$\phi_k = \frac{\mathscr{q}_k(\cdot)P(Q_k^{\wr}=1|Q_k^{\dagger}=1,Q_k^{\ddagger}=1,R=r,\boldsymbol{n_k},\boldsymbol{a_k},k)}{E_{\boldsymbol{n}}[\mathscr{q}_k(\cdot)P(Q_k^{\wr}=1|Q_k^{\dagger}=1,Q_k^{\ddagger}=1,R=r,\boldsymbol{n_k},\boldsymbol{a_k},k)|Q_k^{\ddagger}=1,R=r,\boldsymbol{a_k},k]} \tag{26}$$

where $E_{\boldsymbol{n}}[\cdot]$ is over $P(\boldsymbol{n_k}|Q_k^{\ddagger} = 1, R = r, \boldsymbol{a_k}, k)$

$$\theta_k^{pool} = \frac{\boldsymbol{E_n}[E_r\{\mathscr{q}_k(\cdot)|Q^{\ddagger}=1,T=1,\boldsymbol{n_k},\boldsymbol{a_k},k\}P(Q^{\wr}=1|Q^{\dagger}=1,Q^{\ddagger}=1,R=r,\boldsymbol{n_k},\boldsymbol{a_k},k)|Q_k^{\ddagger}=1,T=1,\boldsymbol{a_k},k]}{E_{(\boldsymbol{n},\boldsymbol{a},\boldsymbol{k})}[E_r\{\mathscr{q}_k(\cdot)|Q^{\ddagger}=1,T=1,\boldsymbol{n_k},\boldsymbol{a_k},k\}P(Q^{\wr}=1|Q^{\dagger}=1,Q^{\ddagger}=1,T=1,\boldsymbol{n_k},\boldsymbol{a_k},k)|Q_k^{\ddagger}=1,T=1]} \tag{27}$$

where $E_r\{\cdot\}$ is over $P(r|Q_k^{\ddagger} = 1, T = 1, \boldsymbol{n_k}, \boldsymbol{a_k}, k)$,
$E_{\boldsymbol{n}}[\cdot]$ is over $P(\boldsymbol{n_k}|Q_k^{\ddagger} = 1, T = 1, \boldsymbol{a_k}, k)$,
and $E_{(\boldsymbol{n},\boldsymbol{a},\boldsymbol{k})}[\cdot]$ is over $P(\boldsymbol{n_k}, \boldsymbol{a_k}, k|Q_k^{\ddagger} = 1, T = 1)$

In (26) and (27) $\mathscr{q}_k(\cdot)$ is the probability of being partially eligible $Q_k^{\dagger} = 1$ (e.g., enrolled in EPPP) in the counterfactual source population given partial eligibility $Q_k^{\ddagger} = 1$ (e.g., prior hypertension, established care), social group $R = r$, allowables $\boldsymbol{A_k}$, non-allowables $\boldsymbol{N_k}$ and calendar time $k$ which, under the interventional distribution $\mathscr{g}_k(\cdot)$ for the Designs 4a, 4b, and 4c, is shown in the third column of Table 1. The term $\mathscr{q}_k(\cdot)$ can thus be estimated as the predicted value from an appropriate model for partial eligibility $Q_k^{\dagger} = 1$.[24]

The weights $\phi_k$ and $\theta_k^{pool}$ account for how non-random selection on $Q_k^{\wr\, G_k}$ affects the distribution of non-allowables $\boldsymbol{N_k}$ and allowables $\boldsymbol{A_k}$, as $\boldsymbol{W_k^{\wr}}$ (e.g., current visit) may be affected by both $\boldsymbol{W_k^{\dagger}}$ (e.g., EPPP enrollment) the allowables $\boldsymbol{A_k}$ and non-allowables $\boldsymbol{N_k}$ (e.g., as in Figure 1C and Figure 3). They incorporate a product $\rho_k$ of two parts (i) and (ii). The first part (i) is the conditional probability of partial eligibility $Q_k^{\wr} = 1$ (e.g., current visit) given other indicators of partial eligibility $Q_k^{\dagger}$ =1 (e.g., EPPP enrollment) and $Q_k^{\ddagger} = 1$ (e.g., prior hypertension, established care), non-allowables $\boldsymbol{N_k}$, allowables $\boldsymbol{A_k}$, calendar time $k$, and membership in the social group $R = r$ (in the case of $\phi_k$) or the standard population $T = 1$ (in the case of $\theta_k^{pool}$). The second part (ii) is either (ii-a) $\mathscr{q}_k(\cdot)$ as described above (in the case of $\phi_k$) or (ii-b) $\mathscr{q}_k(\cdot)$ standardized over the conditional distribution of social group $R = r$ within the standard population $T = 1$ (in the case of $\theta_k^{pool}$).[25] For $\phi_k$ the numerator is this product $\rho_k$ and the denominator standardizes $\rho_k$ over the conditional distribution of the non-allowables $\boldsymbol{N_k}$. For $\theta_k^{pool}$ the numerator standardizes $\rho_k$ over the conditional distribution of the non-allowables $\boldsymbol{N_k}$ while the denominator further standardizes $\rho_k$ over the conditional joint distribution of the allowables $\boldsymbol{A_k}$ and calendar time $k$. Both $\phi_k$ and $\theta_k^{pool}$ may be estimated by G-computation (see sample code in the Supplementary Material).

Under a version of overlap (3) (see footnote 21), innocuous sampling (4), exchangeability (18), positivity (12), and consistency, we identify the aggregated mean $\tau^{\mathcal{D}4}(r)$ of the social group in the fully eligible counterfactual population $Q_k^{G_k} = 1$ (e.g., prior hypertension, established care, enrolled in EPPP, current visit) after an intervention $G_k$ on partial eligibility-related variables $\boldsymbol{W_k^{\dagger}}$ (e.g., EPPP enrollment) as:

$$E_{\boldsymbol{a},k}[\![\theta_k^{pool} E_{\boldsymbol{n}}\big(\phi_k E\big[Y_{k+J} \big| Q_k = 1, R = r, \boldsymbol{n_k}, \boldsymbol{a_k}, k\big] \big| Q_k^{\ddagger} = 1, R = r, \boldsymbol{a_k}, k\big) | Q_k^{\ddagger} = 1, T = 1]\!] \qquad (28)$$

where $E_{\boldsymbol{n}}(\cdot)$ is over $P\big(\boldsymbol{n_k} \big| Q_k^{G_k} = 1, R = r, \boldsymbol{a_k}, k\big)$, identified by $\phi_k P\big(\boldsymbol{n_k} \big| Q_k^{\ddagger} = 1, R = r, \boldsymbol{a_k}, k\big)$, and $E_{\boldsymbol{a},k}[\![\cdot]\!]$ is over $P\big(\boldsymbol{a_k}, k \big| Q_k^{G_k} = 1, T = 1\big)$, identified by $\theta_k^{pool} P\big(\boldsymbol{a_k}, k \big| Q_k^{\ddagger} = 1, T = 1\big)$

To estimate (28) by G-computation, it suffices to follow the same procedure as for Design 2 (Section 6.3) with a slight change, to weight the model in step 3 by $\phi_k$ and use $\theta_k^{pool}$ as a weight for a weighted average for step 5. For calendar-time specific means, we condition the last expectation and all terms in $\theta_k^{pool}$ on $k$.

---

[24] For example, under design 4a, we would estimate $\mathscr{q}_k(\cdot)$ as $\hat{P}(Q_k^{\dagger} = 1 | Q_k^{\ddagger} = 1, k)$ from a model for $P(Q_k^{\dagger} = 1 | Q_k^{\ddagger} = 1, k)$.

[25] If, as in our example, a social group $R = r'$ represents the standard population $T = 1$, this calculation reduces to $\mathscr{q}_k(\cdot)$.

We may also estimate the aggregated mean outcome $\tau^{\mathcal{D}4}(r)$ by the weighting estimator:

$$E[Y_{k+J} \times \omega_{r,Q_k=1,k}^{\mathcal{D}4-pool} | Q_k = 1, R = r] \tag{29}$$

where

$$\omega_{r,Q_k=1,k}^{\mathcal{D}4} = \phi_k \times \theta_k^{pool} \times \frac{P(Q_k^{\wr}=1|Q_k^{\dagger}=1,Q_k^{\ddagger}=1,R=r)}{P(Q_k^{\wr}=1|Q_k^{\dagger}=1,Q_k^{\ddagger}=1,R=r,\boldsymbol{n_k},\boldsymbol{a_k},k)} \times \frac{P(Q_k^{\dagger}=1|Q_k^{\ddagger}=1,R=r)}{P(Q_k^{\dagger}=1|Q_k^{\ddagger}=1,R=r,\boldsymbol{n_k},\boldsymbol{a_k},k)} \times \frac{P(T=1|Q_k^{\ddagger}=1,\boldsymbol{a_k},k)}{P(R=r|Q_k^{\ddagger}=1,\boldsymbol{a_k},k)} \times \frac{P(R=r|Q_k^{\ddagger}=1)}{P(T=1|Q_k^{\ddagger}=1)}$$

The first two terms are $\phi_k$ and $\theta_k^{pool}$. The third term is the ratio of the unconditional probability (with respect to $\boldsymbol{A_k}, \boldsymbol{N_k}, k$) of being partially eligible $Q_k^{\wr} = 1$ (e.g., current visit) given other indicators of partial eligibility $Q_k^{\dagger}$ =1 (e.g., EPPP enrollment) and $Q_k^{\ddagger} = 1$ (e.g., prior hypertension, established care), and social group $R = r$ to the corresponding conditional probability given the non-allowables $\boldsymbol{N_k}$, allowables $\boldsymbol{A_k}$, and calendar time $k$. The remaining terms are identical to the expressions for the weight (23) used in Design 2. For calendar-time specific means, we condition all terms in the weight and all terms in $\theta_k^{pool}$ on $k$.

*6.6 Identification and Estimation of Design 4 Under Simplifying Conditions*

Emulation of Design 4a simplifies greatly with no indicators of partial eligibility $Q_k^{\wr}$ affected by an intervention on the partial eligibility-related variable $\boldsymbol{W_k^{\dagger}}$. All terms for $Q_k^{\wr}$ in $\phi_k$ (26), $\theta_k^{pool}$ (27), in the estimators (28) and (29) disappear. Under Design 4a (i.e., randomly assign $\boldsymbol{W_k^{\dagger}}$; Table 1), the term $\mathcal{q}_k(\cdot)$ cancels and the estimators (28) and (29) reduce to those of Design 2, i.e., (22) and (23). Then, Design 4a generalizes the results Ψ to those partially eligible by $Q_k^{\ddagger} = 1$ even when $\boldsymbol{W_k^{\dagger}}$ affects the outcome.

Under Design 4b, (i.e., randomly assign $\boldsymbol{W_k^{\dagger}}$ given $\boldsymbol{A_k}$ and $R = r$; Table 1)**,** the identifying expression for $\tau^{\mathcal{D}4b}(r)$ behind the G-computation estimator reduces to:

$$E_{\boldsymbol{a},k}[\![E_{\boldsymbol{n}}\big(E\big[Y_{k+J}\big|Q_k = 1, R = r, \boldsymbol{n_k}, \boldsymbol{a_k}, k\big]\big|Q_k^{\ddagger} = 1, R = r, \boldsymbol{a_k}, k\big)|Q_k = 1, T = 1]\!] \tag{30}$$

where $E_{\boldsymbol{n}}(\cdot)$ is over $P\big(\boldsymbol{n_k}\big|Q_k^{\ddagger} = 1, R = r, \boldsymbol{a_k}\ , k\big)$ and $E_{\boldsymbol{a},k}[\![\cdot]\!]$ is over $P(\boldsymbol{a_k}, k|Q_k = 1, T = 1)$

The weighting estimator reduces to:

$$E[Y_{k+J} \times \omega_{r,Q_k=1,k}^{\mathcal{D}4b-pool} | Q_k = 1, R = r] \tag{31}$$

where $\omega_{r,Q_k=1,k}^{\mathcal{D}4b-pool} = \frac{P(Q_k^{\dagger}=1|Q_k^{\ddagger}=1,R=r,\boldsymbol{a_k},k)}{P(Q_k^{\dagger}=1|Q_k^{\ddagger}=1,R=r,\boldsymbol{n_k},\boldsymbol{a_k},k)} \times \frac{P(T=1|Q_k=1,\boldsymbol{a_k},k)}{P(R=r|Q_k=1,\boldsymbol{a_k},k)} \times \frac{P(R=r|Q_k=1)}{P(T=1|Q_k=1)}$

Then, Design 4b addresses non-random selection while retaining social group differences in eligibility. Note that if the chosen allowables $\boldsymbol{A_k}$ are sufficient to satisfy independence (11) or exchangeability (18), so that no non-allowables $\boldsymbol{N_k}$ are needed, the estimators (30) and (31) reduce to (20) and (21) of Design 1.

Under Design 4c (i.e., randomly assign $\boldsymbol{W_k^{\dagger}}$ given $\boldsymbol{A_k}$ and $\boldsymbol{N_k}$ as in $T = 1$; Table 1), $\theta_k^{pool}$ reduces to one, and the identifying expression for $\tau^{\mathcal{D}4c}(r)$ behind the G-computation estimator reduces to:

$$E_{\boldsymbol{a},k}[\![E_{\boldsymbol{n}}\big(\phi_k^c E\big[Y_{k+J}\big|Q_k = 1, R = r, \boldsymbol{n_k}, \boldsymbol{a_k}, k\big]\big|Q_k^{\ddagger} = 1, R = r, \boldsymbol{a_k}, k\big)|Q_k = 1, T = 1]\!] \tag{32}$$

where $E_{\boldsymbol{n}}(\cdot)$ is over $P\big(\boldsymbol{n_k}\big|Q_k^{G_k} = 1, R = r, \boldsymbol{a_k}, k\big)$, identified by $\phi_k^c P\big(\boldsymbol{n_k}\big|Q_k^{\ddagger} = 1, R = r, \boldsymbol{a_k}, k\big)$
$E_{\boldsymbol{a},k}[\![\cdot]\!]$ is over $P(\boldsymbol{a_k}, k|Q_k = 1, T = 1)$,
and $\phi_k^c = \frac{P(Q_k^{\dagger}=1|Q_k^{\ddagger}=1,T=1,\boldsymbol{n_k},\boldsymbol{a_k},k)}{\boldsymbol{E_n}[P(Q_k^{\dagger}=1|Q_k^{\ddagger}=1,T=1,\boldsymbol{n_k},\boldsymbol{a_k},k)|Q_k^{\ddagger}=1,R=r,\boldsymbol{a_k},k]}$ with $E_{\boldsymbol{n}}[\cdot]$ over $P(\boldsymbol{n_k}|Q_k^{\ddagger} = 1, R = r, \boldsymbol{a_k}, k)$

The weighting estimator reduces to:

$$E[Y_{k+J} \times \omega_{r,Q_k=1,k}^{\mathcal{D}4c-pool} | Q_k = 1, R = r] \tag{33}$$

where $\omega_{r,Q_k=1,k}^{\mathcal{D}4c-pool} = \phi_k^c \times \frac{P(Q_k^\dagger=1|Q_k^\ddagger=1,R=r,\boldsymbol{a_k},k)}{P(Q_k^\dagger=1|Q_k^\ddagger=1,R=r,\boldsymbol{n_k},\boldsymbol{a_k},k)} \times \frac{P(T=1|Q_k=1,\boldsymbol{a_k},k)}{P(R=r|Q_k=1,\boldsymbol{a_k},k)} \times \frac{P(R=r|Q_k=1)}{P(T=1|Q_k=1)}$

Estimation under Design 4c is thus similar to that of Design 4b except with $\phi_k^c$ factored in. If the marginalized group is the standard population, its aggregated mean outcome (32) reduces to its pooled mean $E[Y_{k+J}|Q_k = 1, R = 1]$ and its weight (33) reduces to one. Then, the target study model is purely descriptive of the fully eligible $Q_k = 1$ marginalized group (i.e., its expected outcome in the target study and in the fully eligible population are the same), even when addressing non-random sample selection. It does so by making the selection process of the privileged group match that of the marginalized group.

## 7. Contributions and Comparison to Existing Literature

### *7.1 Target Trial Emulation*

Our target study model is inspired by the trial emulation framework (Hernán and Robins 2016) used to evaluate causal effects of treatment strategies. Our work differs in that (i) there are no treatment groups, only observed social groups; (ii) we balance allowable covariates by design (sampling) rather than balance confounders by intervention (randomization); (iii) under multiple studies across calendar time, we enforce one person per unit of time, all social groups compared must be represented at each unit of time, and our aggregated estimators adjust for calendar time enrollment by balancing it across groups. Our summary estimate is interpretable as a weighted average of disparity measures for populations indexed at different points in calendar time, and it avoids potential bias due to differential timing of enrollment by social groups. Our model can incorporate treatment strategies and within-group randomization post sampling to evaluate causal effects of treatment strategies on disparity. (Jackson, Hsu, Zalla et al. 2024)

### *7.2. Sampling as a Conceptual Model*

Tipton (2013), Westreich, Edwards, Lesko et al. (2017), and Dahabreh, Haneuse, Robins et al. (2021), use sampling designs of target populations to generalize or transport results from randomized trials. VanderWeele (2020) proposes simple random sampling to define causal effects of observing social groups to measure inequality. Lundberg (2022) uses simple random sampling to critique inference about causal effects on inequality in a superpopulation. Moreno-Betancur (2021), in a commentary on Jackson (2021), mentions enrollment in a target trial with balanced allowables to motivate causal inference with somewhat vague interventions. Our model uses stratified sampling of an eligible population to construct a sample with desirable properties (i.e., distributions of covariates that reflect populations of interest to address non-random sample selection, balance of allowable covariates across social groups) to measure disparity. Our formal presentation discusses the sampling designs and emulation procedures in considerable detail.

*7.3 Alternative Conceptual Models of Disparity*

Other conceptual models that employ allowability are widely used to define disparity in healthcare or used to audit or improve algorithmic fairness. We outline these models and then compare them to our model. We ignore issues of non-random sample selection to focus on core differences between the models. Each of these models, including our own, presumes that the constructs measured by allowables have the same meaning for each social group (i.e., that there is no differential interpretation or measurement error).

Cook et al. (2009) frame disparity as the difference in healthcare utilization outcomes unexplained by differences in the allowable covariates, a generalization of the Oaxaca-Blinder Decomposition (Blinder 1973, Oaxaca 1973) originally used to measure labor discrimination.[26] Under this interpretation, they define an IOM concordant disparity that compares outcomes between an observed marginalized group $R = 1$ and a counterfactual privileged group $R = 0$ with certain distributional properties. Its joint distribution of the allowables $\boldsymbol{A}$ and non-allowables $\boldsymbol{N}$, $f^*_{\boldsymbol{A},\boldsymbol{N}}(\boldsymbol{A}, \boldsymbol{N}|R = 0)$, must: (a) return $f_{\boldsymbol{A}}(\boldsymbol{A}|R = 1)$ the factual marginal distribution of $\boldsymbol{A}$ among the marginalized group $R = 1$ when integrated over $\boldsymbol{N}$; (b) return $f_{\boldsymbol{N}}(\boldsymbol{N}|R = 0)$ the factual marginal distribution of $\boldsymbol{N}$ among the privileged group $R = 0$ when integrated over $\boldsymbol{A}$. Criterion (a) ensures balance of the allowables $\boldsymbol{A}$. Criterion (b) picks up the mediating role of $\boldsymbol{N}$ leading to differences in healthcare utilization in the overall population. They avoid strict causal assumptions by not placing further constraints on $f^*_{\boldsymbol{A},\boldsymbol{N}}(\boldsymbol{A}, \boldsymbol{N}|R = 0)$. They adapt a "Rank and Replace" procedure (McGuire et al. 2006) to implement this model by producing a counterfactual privileged population satisfying criteria (a) and (b).

Duan et al. (2008) agree with the decomposition perspective but argue that Cook et al. (2009)'s criteria and "Rank and Replace" procedure lead to implausible populations not relevant for policymaking. The deficiency in their view is that a decomposition involves an intervention to assign the privileged group the allowable $\boldsymbol{A}$ distribution of the marginalized group, and such intervention will impact the non-allowable $\boldsymbol{N}$ distribution when $\boldsymbol{A}$ causes $\boldsymbol{N}$. They propose a causal decomposition, involving an intervention on the allowables among the privileged group, that compares the observed marginalized group $R = 1$ to a counterfactual privileged group $R = 0$ where the joint distribution of $\boldsymbol{A}$ and $\boldsymbol{N}$ depend on the causal relationship between them. When $\boldsymbol{N}$ causes $\boldsymbol{A}$ (Figure 5A), the joint distribution is $f_{\boldsymbol{N}}(\boldsymbol{N}|R = 0) \times f_{\boldsymbol{A}|\boldsymbol{N}}(\boldsymbol{A}|R = 1, \boldsymbol{N})$, i.e., their "conditional framework", because they assign $\boldsymbol{A}$ within levels of $\boldsymbol{N}$. Whereas when $\boldsymbol{A}$ causes $\boldsymbol{N}$ (Figure 5B), the joint distribution is $f_{\boldsymbol{N}|\boldsymbol{A}}(\boldsymbol{N}|R = 0, \boldsymbol{A}) \times f_{\boldsymbol{A}}(\boldsymbol{A}|R = 1)$, their "marginal framework", because they assign $\boldsymbol{A}$ irrespective of $\boldsymbol{N}$. For implementation, they propose a density ratio weighting procedure. The model relies on "nature preserving assumptions" that allow the counterfactuals to be mapped to data, wherein the intervention on $\boldsymbol{A}$ does not change how the outcome is conditionally distributed. The model does not extend to settings where $\boldsymbol{N}$ and $\boldsymbol{A}$ cause each other over time (Figure 5C).

[26] The decompositions of Cook et al. (2009) and Duan et al. (2008) use a counterfactual privileged population $E[Y|R = 0]_{cf}$ to separate the observed difference $E[Y|R = 1] - E[Y|R = 0]$ into a portion $E[Y|R = 0]_{cf} - E[Y|R = 0]$ due to, and a portion $E[Y|R = 1] - E[Y|R = 0]_{cf}$ not due to, differences in the distribution of allowables across social groups $R = 1$ and $R = 0$.

Many authors, (Pearl 2001, Zhang, Wu and Wu 2016, Kilbertus, Rojas-Carulla, Parascandolo et al. 2017, Nabi and Shpitser 2018, Zhang and Bareinboim 2018, Chiappa 2019, Weinberger 2022) define discrimination as a direct effect of assigning social group membership $R$ (or its perception) on an outcome $D$ made by a decider.[27] The direct effect does not occur through allowable covariates $\boldsymbol{A}$ appropriate for decision-making, capturing inappropriate causal paths.[28] In Figures 5A and 5B, discrimination is reflected by the direct path set: $R \rightarrow D$ and $R \rightarrow \boldsymbol{N} \rightarrow D$ (again, we say "direct" as the path set avoids $\boldsymbol{A}$). On Figure 5C, discrimination is reflected by the path set: $R \rightarrow D$, $R \rightarrow \boldsymbol{N_1} \rightarrow D$, $R \rightarrow \boldsymbol{N_2} \rightarrow D$, and $R \rightarrow \boldsymbol{N_1} \rightarrow \boldsymbol{N_2} \rightarrow D$. A direct effect can be defined by potential outcomes $D^{r,\boldsymbol{A}^r}$ under interventions that jointly assign social group $R$ and allowables $\boldsymbol{A}$. For exposition, consider the direct effect defined among the marginalized group, where persons social group $R$ is set to marginalized $r = 1$ versus privileged $r = 0$ but the allowables $\boldsymbol{A}$ are held to what they would be under marginalized status, i.e., $\boldsymbol{A}^{r=1}$ (to measure discrimination):

$$E[D^{r=1,\boldsymbol{A}^{r=1}}|R = 1] - E[D^{r=0,\boldsymbol{A}^{r=1}}|R = 1]$$

Under consistency and composition assumptions, (VanderWeele and Vansteelandt 2009) the expression $E[D^{r=1,\boldsymbol{A}^{r=1}}|R = 1]$ is identified as the marginalized group's factual mean, $E[Y|R = 1]$. Under Figure 5B, if the graph includes all confounders of $R$'s effects on $\boldsymbol{A}$ and $D$ (e.g., $H$), all confounders of $\boldsymbol{A}$'s effect on $D$, and no confounder is affected by $R$ (i.e., the recanting witness criterion holds (Chen, Shpitser and Pearl 2005, Shpitser 2013)), the "cross-world" expression $E[D^{r=0,\boldsymbol{A}^{r=1}}|R = 1]$ is identified as:

$$\sum_{h,\boldsymbol{a},\boldsymbol{n}} E[D|R = 0, \boldsymbol{n}, \boldsymbol{a}, h]P(\boldsymbol{n}|R = 0, \boldsymbol{a}, h)P(\boldsymbol{a}, h|R = 1)$$

The expression shows a joint distribution $f_{\boldsymbol{N}|\boldsymbol{A},H}(\boldsymbol{N}|R = 0, \boldsymbol{A}, H) \times f_{\boldsymbol{A},H}(\boldsymbol{A}, H|R = 1)$ where the confounder $H$ is treated as allowable.[29] Under Figures 5A and 5C, the direct effect is not identified because the non-allowables $\boldsymbol{N}$ confound the effect of the allowables $\boldsymbol{A}$ on the outcome $D$ but are affected by social group $R$.

Our model defines disparity by comparing groups who are (distributionally) similarly situated (i.e., balanced) on the allowables $\boldsymbol{A}$ by design. We argued in Section 3.4 that this design is IOM concordant.[30] Our model's estimand is also interpretable as a standardized measure, where a disparity metric is calculated for each level of the allowables $\boldsymbol{A} = \boldsymbol{a}$ and these metrics are standardized to a common distribution. In contrast, the Cook et al. (2009), Duan et al. (2008), and direct effect models decompose either a crude difference or a total effect into a portion that captures unjust differences (e.g., disparity, discrimination) and another that does not.[31] Direct effects are defined by nuanced interventions that either act in a "cross-world" sense (Andrews and Didelez 2021), act on distinct mechanisms of assigning social group $R$ (Robins, Richardson and Shpitser 2021), or change how information flows from assigning $R$ (Díaz 2023).

---

[27] In the direct effect model considered here for exposition, the decomposition uses a counterfactual $E[D^{r=0,\boldsymbol{A}^{r=1}}|R = 1]$ to separate the total effect of assigning social group (or its perception) among the marginalized group $E\left[D^{r=1,\boldsymbol{A}^{r=1}}\middle|R = 1\right] - E[D^{r=0,\boldsymbol{A}^{r=0}}|R = 1]$ into an indirect effect $E\left[D^{r=0,\boldsymbol{A}^{r=1}}\middle|R = 1\right] - E[D^{r=0,\boldsymbol{A}^{r=0}}|R = 1]$ due to social group's effect on the allowables and a direct effect $E\left[D^{r=1,\boldsymbol{A}^{r=1}}\middle|R = 1\right] - E[D^{r=0,\boldsymbol{A}^{r=1}}|R = 1]$ not due to social group's effect on the allowables.

[28] A causal path is a directed sequence of edges out a parent node and into a child node in a directed acyclic graph.

[29] If the path $R \leftarrow H \rightarrow \boldsymbol{N} \rightarrow Y$ in Figure 5 disadvantages the marginalized group, making $H$ allowable underestimates disparity.

[30] Cook et al. (2009) and Li and Li (2023) describe propensity score weighting based on allowables alone as not IOM concordant. We argued that our model and approaches that emulate it (e.g., propensity score weighting (21)) are indeed IOM concordant.

Our default model, Design 1, assumes overlap (3) of allowables $\boldsymbol{A}$[32] and does not specify $H$ (i.e., determinants of $R$ as in Figure 5) or non-allowables $\boldsymbol{N}$. That is, unlike the Cook et al. (2009), Duan et al. (2008), and direct effect models, our model does not require investigators to understand all the non-allowable factors that lead to the outcome. If we do express $H$ and $\boldsymbol{N}$ and choose the marginalized group as the standard population, our model would compare the observed outcomes of the marginalized group to a privileged group with a joint distribution $f_{\boldsymbol{N},H|\boldsymbol{A}}(\boldsymbol{N},H|R=0,\boldsymbol{A}) \times f_{\boldsymbol{A}}(\boldsymbol{A}|R=1)$ for $\boldsymbol{A}$, $\boldsymbol{N}$, and $H$ (i.e., $H$ is treated as if it were non-allowable). Our default model is agnostic about causal structure. Our model is identified under Figures 5A, 5B, and 5C, whereas the Duan et al. (2008) model does not cover Figure 5C, and the direct effect model is not identified in Figures 5A or 5C. Our model also assumes that the sampling process does not impact the data-generating process (4). The Duan et al. (2008) and direct effect models assume that their interventions leave aspects (e.g., the conditional outcome distribution) of the data generation process intact. This assumption is very demanding given how society and social groups are currently structured. (Kohler-Hausmann 2019, Jackson and Arah 2020).

*7.4 Selection Bias (Including Generalizability and Transportability)*

Non-random sample selection can bias the causal effect of assigning perceived group membership (e.g., a measure of discrimination) (Greiner and Rubin 2011, Malinsky, Shpitser and Richardson 2019, Knox et al. 2020, Gaebler et al. 2022, Strensrud, Robins, Sarvet et al. 2022) It also impacts a descriptive measure of inequality. (VanderWeele and Robinson 2014) We underscored that non-randomly selected populations can be inherently meaningful and that collider stratification may affect baseline covariates to further disadvantage a marginalized group on outcomes, aligning with the Healthy People 2020 and NIMHD definitions of disparity (Sections 3.1 and 3.5). We provided Design 1 for use in these settings.

We argued that non-random sample selection may be addressed whenever (a) it limits the ability to infer to a population of interest or (b) masks disparity through collider-stratification. We provided Designs 2 and 3 for use in setting (a), and Design 4 for use in setting (b). Design 4 may be contrasted with Design 1 to understand how much non-random sample selection impacts disparity. Designs 2 and 3 use similar assumptions used to generalize or transport descriptive measures and the results of randomized trials. (Degtiar and Rose 2023) When there are no allowables specified and the combined sample [collapsed over social group $R$] is the standard population, and there is only one study (i.e., indexed at a single moment of calendar time), then the identifying expressions of Designs 2 and 3 reduce to those of Bareinboim, Tian and Pearl (2014), the weighting estimators for Design 2 reduce to inverse selection weights of Cole and Stuart (2010), the weighting estimators for Design 3 reduce to the inverse odds sampling weights of Westreich et al. (2017), and the G-computation estimators for Design 2 reduce to those of Lesko, Buchanan, Westreich et al. (2017) and also Dahabreh, Robertson, Steingrimsson et al. (2020).

[31] Standardization has been proposed for decomposition (Kitagawa 1955) but we do not adopt that perspective here.

[32] The Cook et al. (2009), Duan et al. (2008), and the direct effect models also require a form of overlap in $\boldsymbol{A}$.

Design 4 is a novel approach to address non-random sample selection. It envisions an intervention to allocate *certain* (i.e., not all) eligibility-related variables, even when other eligibility-related variables (not intervened upon) are affected by the intervention. Dahabreh, Robins, Haneuse et al. (2019) envision an intervention to scale up trial participation, the last step in study enrollment. The exchangeability assumption of Design 4, like Dahabreh et al. (2019), holds when eligibility-related variables or study participation affect the outcome. The independence assumption of Desings 2 and 3 used to generalize or transport does not. Estimation under Design 4 simplifies greatly when the intervention point is the last step in enrollment (e.g., an intervention to allocate study participation).

We focused on non-random sample selection. Missing data, loss to follow-up, and competing risks during follow-up may bias a disparity estimate (Howe and Robinson 2018). These may be addressed in the statistical analysis of the target study and its emulation. Events (e.g., death, hospital discharge for study in inpatient disparity in prognosis) before enrollment may affect who (non-randomly) selects into the study sample (Rojas-Saunero et al. 2023) If survivors at enrollment (i.e., those without the event) are not considered to be a meaningful population, an extension of Design 4 with a sustained intervention may help if the event is manipulable, but such an extension is not developed here and is saved for future work.

## 8. Discussion

We have proposed a conceptual model for measuring disparity and a framework to emulate it with secondary data. Through a sampling plan, the model similarly situates social groups on allowable covariates at baseline to map to meaningful definitions of disparity (Design 1). The model extends to address non-random sample selection in various ways to permit generalizability (Design 2), transportability (Design 3), or inference in a selected population without inducing undesirable forms of collider-stratification (Design 4). We motivated Design 4 for when collider stratification due to non-random sample selection attenuates disparity, but it is difficult to know when this will occur (Nguyen et al. 2019). Investigators may emulate both Designs 1 and 4 to report how selective mechanisms may contribute to or attenuate disparity. Unlike existing models used to measure disparity, our model involves no intervention to assign social group membership and no intervention to manipulate the allowable covariates. Only under Design 4 does the model intervene on variables used to establish eligibility. Under Designs 1, 2, and 3, the model can recover the crude expected outcomes of a social group (e.g., the marginalized group) by using it to determine the standard distribution of the sampling plan. This is also true of Designs 4a and 4c under simplifying conditions (Section 6.6). We have described data structures and provided weighting and G-computation estimators to emulate the model in complex data. Our model and emulation procedures avoid bias due to differential enrollment over calendar time. The summary estimates are weighted averages of estimates for populations at points in calendar time. In the Supplementary Material we provide sample code, a data application to electronic medical records, and proof of all results.

Our model has translational value for advancing public health and clinical medicine. First, it relies on minimal assumptions and is therefore a practical measure for advancing social justice. Second, its features accommodate specific populations during critical life stages: eligibility, time zero, follow-up, and outcome definition, aspects which actual interventions must consider in practice. Third, it maps to definitions of disparity that have strong moral foundations and have long guided public health action. Fourth, it can be extended to evaluate causal effects of (i) hypothetical interventions to inform future interventions (Jackson 2021) and (ii) actual interventions in (non)-randomized trials (Jackson et al. 2024).

Our model also has conceptual value. Our Designs 1, 2, and 3 are grounded in the observed world. They pick up the realized effects of unjust mechanisms as they operate in this world. Mechanisms of injustice are exquisitely complex, inter-dependent, mutually constituted, and dynamically reinforcing. (Reskin 2012) Causal approaches that leverage observational data assume that the way outcomes are conditionally distributed in the factual world will be unchanged in the counterfactual world created by hypothetical interventions, ignoring this complexity. (Jackson and Arah 2020) Our Designs 1, 2, and 3 capture the impact of this complexity as observed without specifying how this complexity works or assuming it away. Our Design 4 invokes a consistency assumption, though, and is subject to this limitation.

**Author's Note**

This manuscript used data from electronic patient medical records of patients seen within a large healthcare system. To protect patient privacy and to comply with HIPAA, we are unable to share or post the data with third parties for re-analysis. The data application code, and sample code to implement all estimators, is available at: https://osf.io/ta7vw/ (Open Science Framework) and https://github.com/jwjackson/targetstudy (GitHub).

## Figures

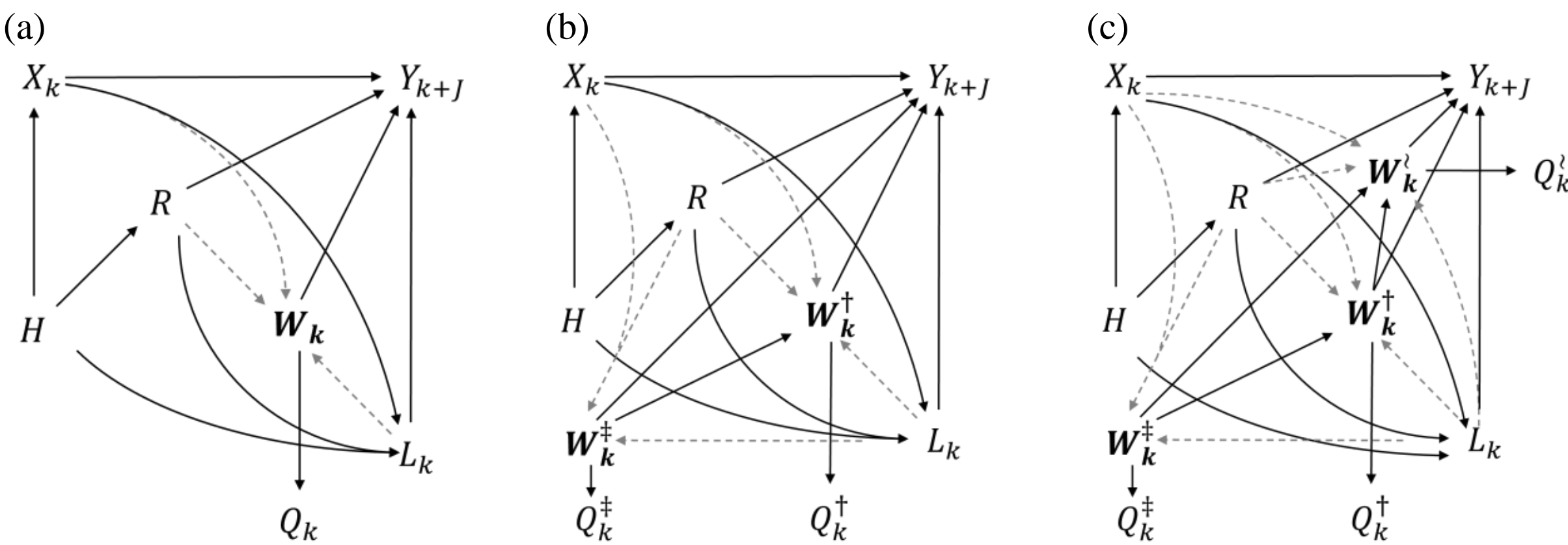

**Figure 1**. Causal directed acyclic graphs depicting causal relationships between historical processes $H$, race $R$, demographics age and sex $X_k$, and comorbidities and adult socioeconomic status $L_k$, hypertension control $Y_{k+J}$, and eligibility-related variables: $\boldsymbol{W_k}$ (prior hypertension, established care, electronic patient portal program [EPPP] enrollment, current visit [in (a)]), $\boldsymbol{W_k^\ddagger}$ (prior hypertension and established care [in (c)], and additionally current visit [in (b)]), $\boldsymbol{W_k^\dagger}$ (EPPP enrollment [in (b) and (c)], and $\boldsymbol{W_k^\wr}$ (current visit [in (c)]), all measured at calendar time $k$. Full eligibility $Q_k$ is based on all eligibility-related variables $\boldsymbol{W_k}$. Similarly, indicators of partial eligibility $Q_k^\ddagger$, $Q_k^\dagger$, $Q_k^\wr$ are based on their corresponding subsets of eligibility-related variables $\boldsymbol{W_k^\ddagger}$, $\boldsymbol{W_k^\dagger}$, $\boldsymbol{W_k^\wr}$. The subscript $k$ marks calendar time and $J$ the interval of time until $Y_{k+J}$ is measured. Dashed lines emphasize selective paths that become unblocked when conditioning on the indicators of eligibility $Q_k$ or all indicators of partial eligibility ($Q_k^\ddagger$, $Q_k^\dagger$, and $Q_k^\wr$). Note that the subscript $k$ refers to the calendar time at which a variable (e.g., a persons level of SES in adulthood) is measured (e.g., the calendar time of the visit) rather than the time at which its value was realized (e.g., some time before the visit).

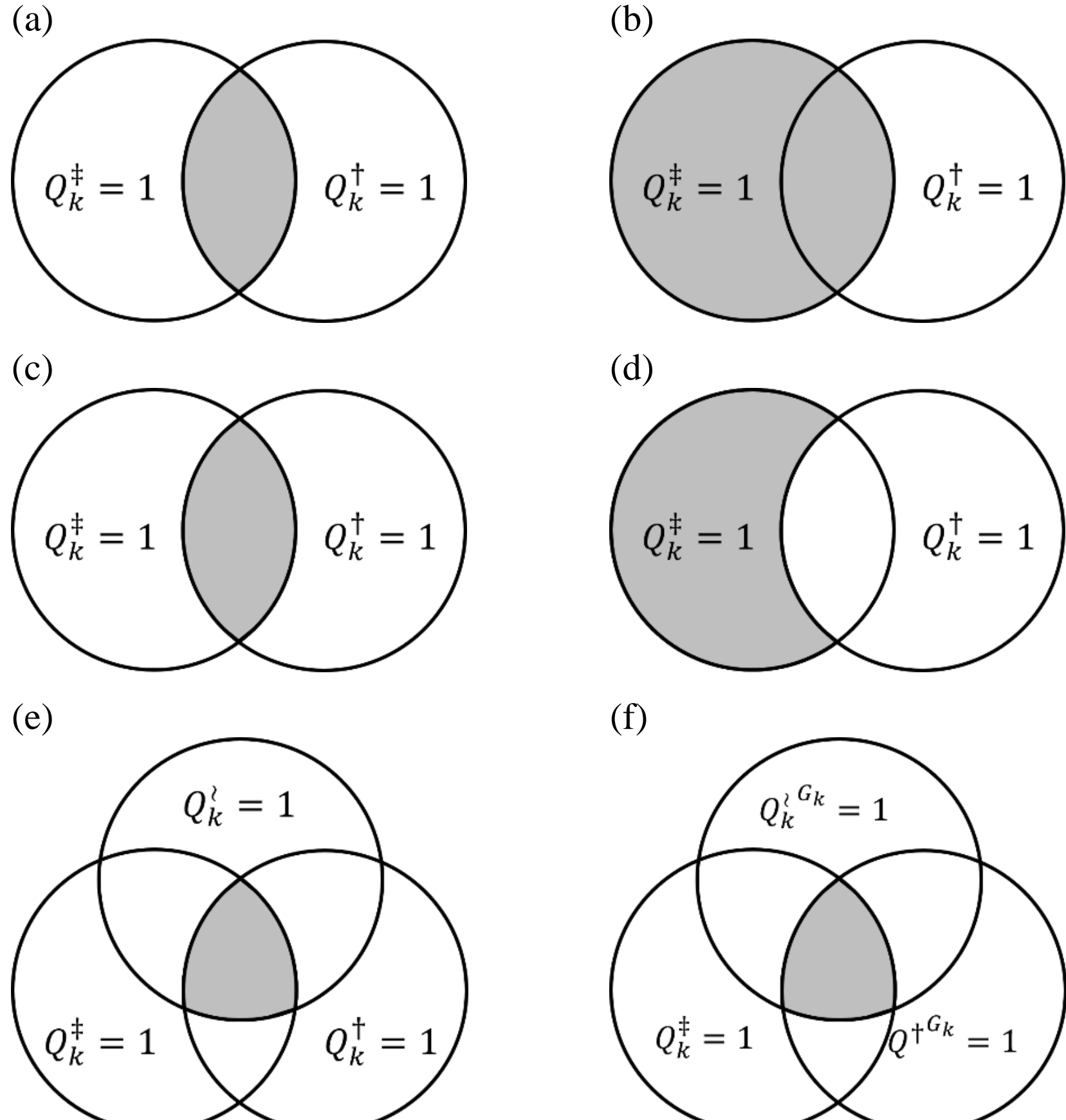

**Figure 2.** Venn Diagrams depicting populations eligible and inferred to under Designs 2, 3, and 4 using partial eligibility indicators (labeled $Q_k^\ddagger$, $Q_k^\dagger$, and $Q_k^\wr$ (1:yes, 0: no)) based on eligibility-related variables $\boldsymbol{W}_{\boldsymbol{k}}^\ddagger$, $\boldsymbol{W}_{\boldsymbol{k}}^\dagger$, and $\boldsymbol{W}_{\boldsymbol{k}}^\wr$ (where $\boldsymbol{W}_{\boldsymbol{k}}^\dagger$ affects $\boldsymbol{W}_{\boldsymbol{k}}^\wr$). The first row pertains to a target study (Design 2, Section 5.2) that (a) enrolls a fully eligible population defined by $Q_k = 1$, i.e., $Q_k^\ddagger = 1$ and $Q_k^\dagger = 1$ but (b) infers to a broader population defined by $Q_k^\ddagger = 1$. The second row pertains to a target study (Design 3, Section 5.3) that (c) enrolls a fully eligible population defined by $Q_k = 1$, i.e., $Q_k^\ddagger = 1$ and $Q_k^\dagger = 1$ but (d) infers to a different population defined by $Q_k^\ddagger = 1$ and $Q_k^\dagger = 0$. The third row pertains a target study (Design 4, Section 5.4) that (e) enrolls a fully eligible population defined by $Q_k = 1$, i.e., $Q_k^\ddagger = 1$, $Q_k^\dagger = 1$, and $Q_k^\wr = 1$ but (d) infers to a fully eligible counterfactual population defined by $Q_k^{G_k} = 1$, i.e., $Q_k^\ddagger = 1$, ${Q_k^\dagger}^{G_k} = 1$, and ${Q_k^\wr}^{G_k} = 1$ after an intervention $G_k$ that eliminates collider stratification through $\boldsymbol{W}_{\boldsymbol{k}}^\dagger$. A target study (Design 1, Section 4) may enroll and infer to the same fully eligible population defined by $Q_k = 1$.

(a)

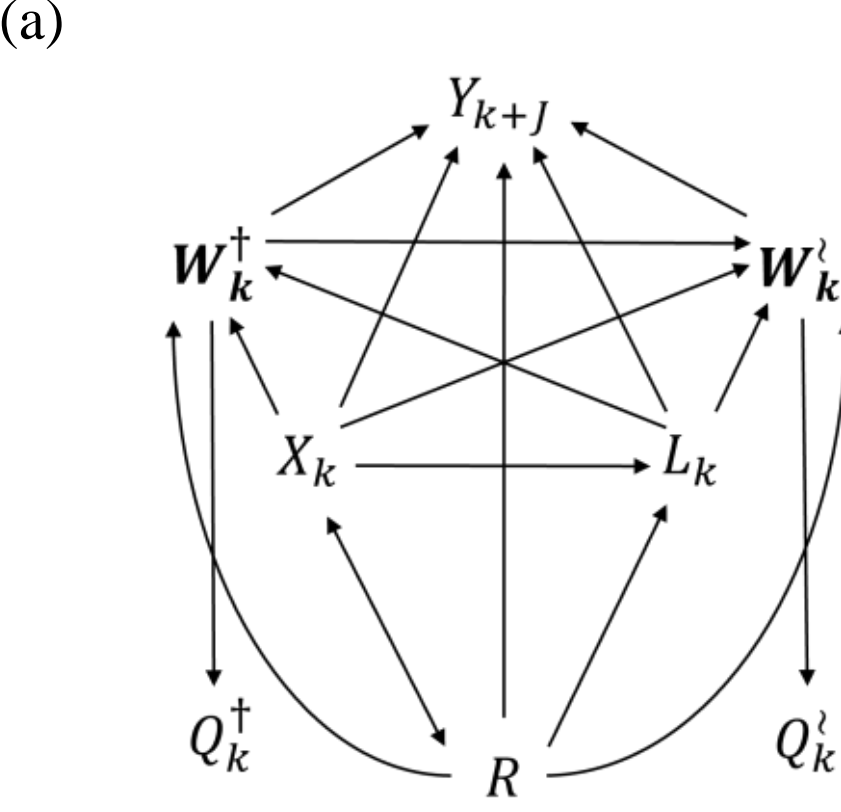

(b)

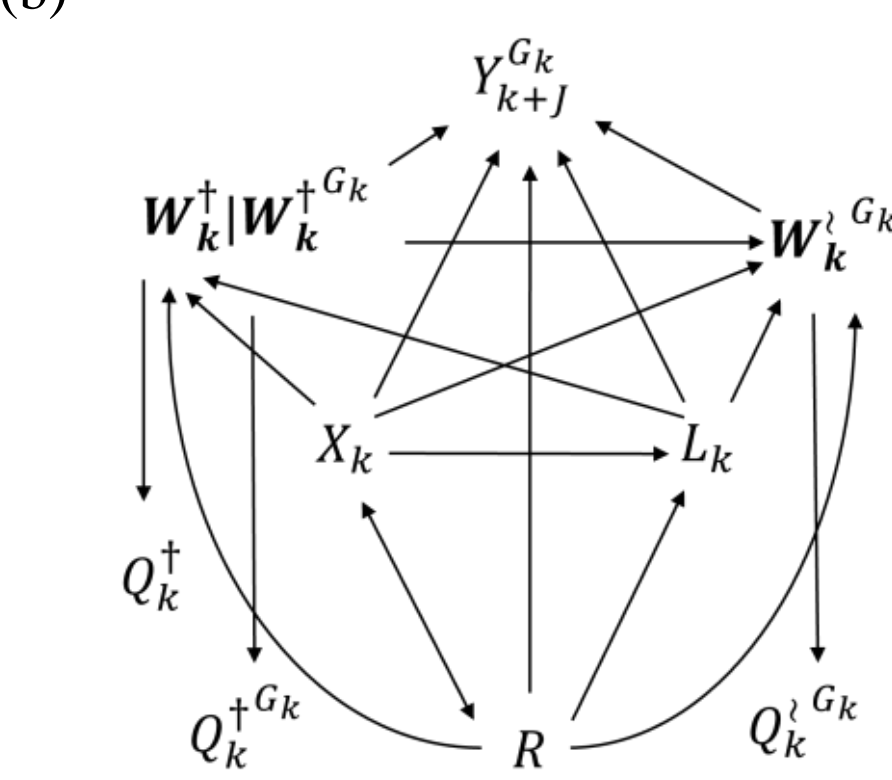

**Figure 3**. Single World Intervention Graph (Richardson and Robins 2013) depicting Design 4a (a) without intervention (b) with intervention on $G_k$ to set the partial eligibility variable $\boldsymbol{W}_{\boldsymbol{k}}^{\dagger}$ (e.g., enrollment in an electronic patient portal program (EPPP)) according to a random draw. $Y_{k+J}$ is the outcome (e.g., hypertension control) at time $k+J$, $R$ represents social group membership (e.g., race), $X_k$ (e.g., age) and $L_k$ (e.g., SES) are covariates that may be deemed allowable $\boldsymbol{A}_{\boldsymbol{k}}$ or non-allowable $\boldsymbol{N}_{\boldsymbol{k}}$. $Q_k^{\dagger}$ is the partial eligibility indicator (1: yes, 0: no) for the partial eligibility variable $\boldsymbol{W}_{\boldsymbol{k}}^{\dagger}$ (e.g., EPPP enrollment)). $\boldsymbol{W}_{\boldsymbol{k}}^{\wr}$ represents another partial eligibility variable (e.g., current visit) affected by $\boldsymbol{W}_{\boldsymbol{k}}^{\dagger}$, with its partial eligibility indicator $Q_k^{\wr}$ (1: yes, 0: no). For simplicity, the historical process variable $H$ and the partial eligibility variables $\boldsymbol{W}_{\boldsymbol{k}}^{\ddagger}$ (e.g., prior hypertension, established care in health system) and their indicator $Q_k^{\ddagger}$ that appear on Figure 1C are omitted but if included would inherit their causal relationships from Figure 1C. Note that on (a) independence (11) $Y_{k+J} \coprod Q_k^{\dagger} | X_k, L_k, R$ does not hold because $\boldsymbol{W}_{\boldsymbol{k}}^{\dagger}$ affects $Y_{k+J}$. However, on (b) exchangeability (18) $\left(Y_{k+J}^{G_k}, Q_k^{\wr\, G_k}\right) \coprod Q_k^{\dagger} | X_k, L_k, R$ does hold. Note also that (i) though $\boldsymbol{W}_{\boldsymbol{k}}^{\dagger G_k}$ is randomly assigned, $(X_k, L_k)$ are not independent of $Q_k^{\dagger G_k}$ given $Q_k^{\wr\, G_k} = 1$ (ii) there is no collider stratification between $R$ and $Y_{k+J}^{G_k}$ from conditioning on $Q_k^{\dagger G_k}$.

| $i$ | $k$ | $Q_k^{\ddagger}$ | $Q_k^{\dagger}$ | $Q_k^{\wr}$ | $Q_k$ | $R_i$ | $T_i$ | $X_k$ | $L_k$ | $Y_{k+J}$ |
|---|---|---|---|---|---|---|---|---|---|---|
| 1 | 1 | 1 | 0 | 0 | 0 | 1 | 1 | 23 | 0 | 0 |
| 1 | 3 | 1 | 0 | 0 | 0 | 1 | 1 | 23 | 0 | 0 |
| 1 | 4 | 1 | 1 | 1 | 1 | 1 | 1 | 23 | 0 | 1 |
| 1 | 5 | 1 | 1 | 0 | 0 | 1 | 1 | 24 | 1 | 0 |
| 1 | 10 | 1 | 1 | 1 | 1 | 1 | 1 | 24 | 1 | 0 |
| 1 | 12 | 1 | 0 | 0 | 0 | 1 | 1 | 24 | 1 | 0 |
| 1 | 15 | 1 | 0 | 1 | 0 | 1 | 1 | 25 | 1 | 1 |
| 2 | 1 | 1 | 1 | 0 | 0 | 0 | 0 | 40 | 1 | 0 |
| 2 | 4 | 1 | 0 | 0 | 0 | 0 | 0 | 40 | 1 | 1 |
| 2 | 7 | 1 | 1 | 1 | 1 | 0 | 0 | 41 | 1 | 0 |

**Figure 4** Sample data structure to emulate the target study $(i, k, Q_k^{\ddagger}, Q_k^{\dagger}, Q_k^{\wr}, R, T, X_k, L_k, Y_{k+J})$ where $i$ is a person identifier, $k$ representes the coarsened moment of calendar time, $R$ represents social group membership, $T$ is an indicator of membership in the standard population, and the covariates $X_k, L_k$ are chosen from to select allowable covariates $\boldsymbol{A_k}$ (for all designs) and, if needed, non-allowable covariates $\boldsymbol{N_k}$ (for Designs 2, 3, and 4 that address non-random sample selection), $Q_k$ represents full eligibility and $Q_k^{\ddagger}$ and $Q_k^{\dagger}$ are partial eligibility indicators for Designs 2 and 3, and $Q_k^{\wr}$ is an additional partial eligibility indicator for Design 4 (see Section 5.4 for details), and $Y_{k+J}$ is the outcome. Assuming Design 4, Person 1 is evaluated for seven studies and eligible (i.e., $Q_k = 1$) for two (at $k = 4{,}10$). Person 2 is evaluated for three studies and eligible for one (at $k = 7$). For Design 1, we subset the data to those fully eligible $Q_k = 1$. For Designs 2, 3, and 4 we subset the data to those partially eligible by $Q_k^{\ddagger} = 1$.

(a)

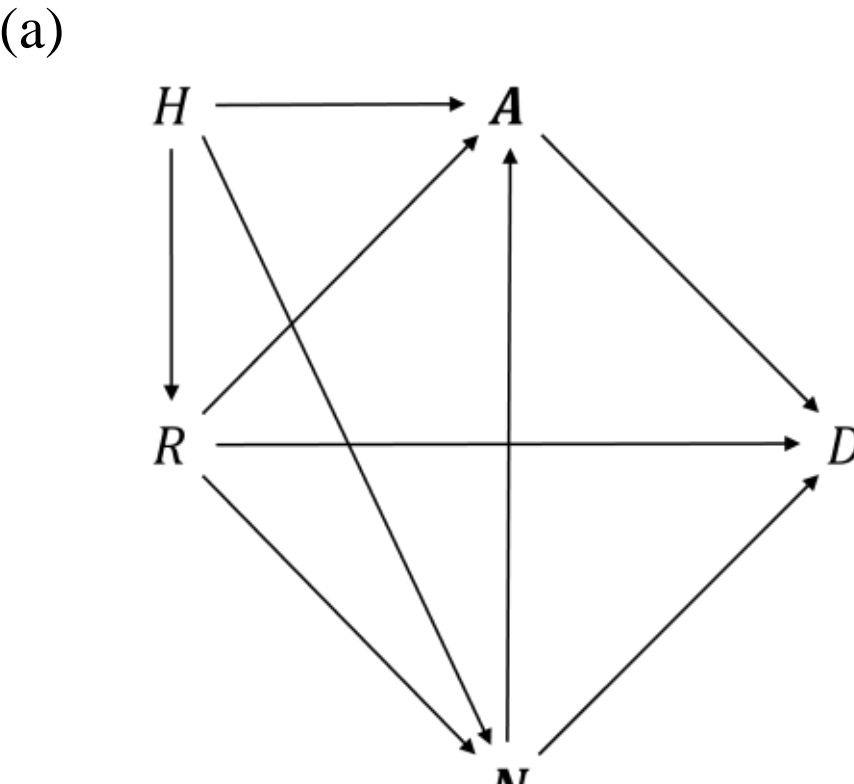

(b)

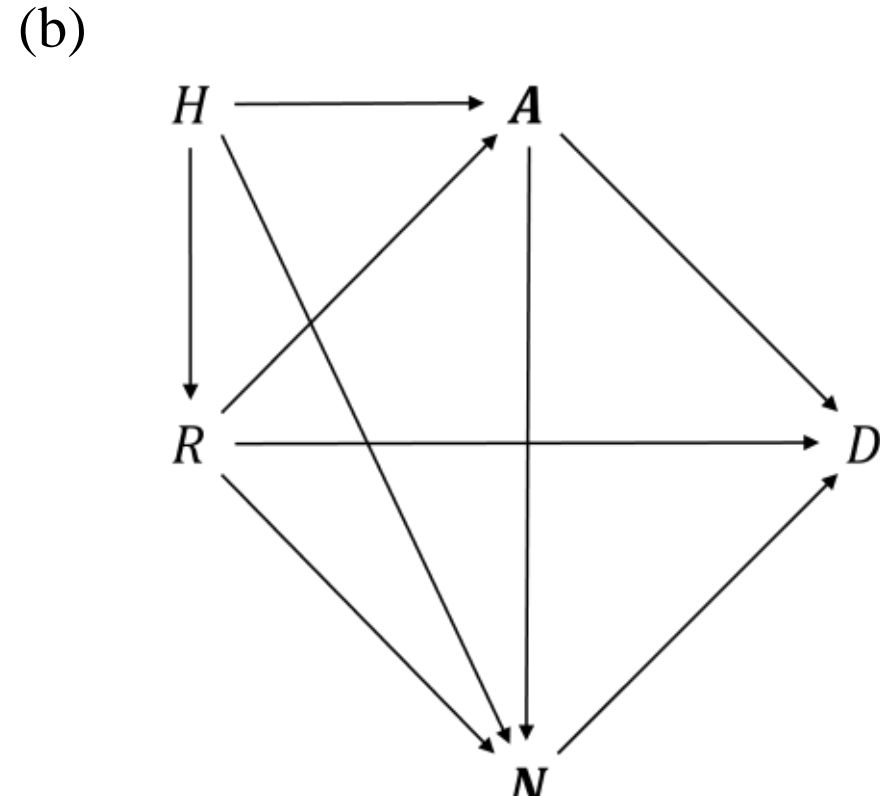

(c)

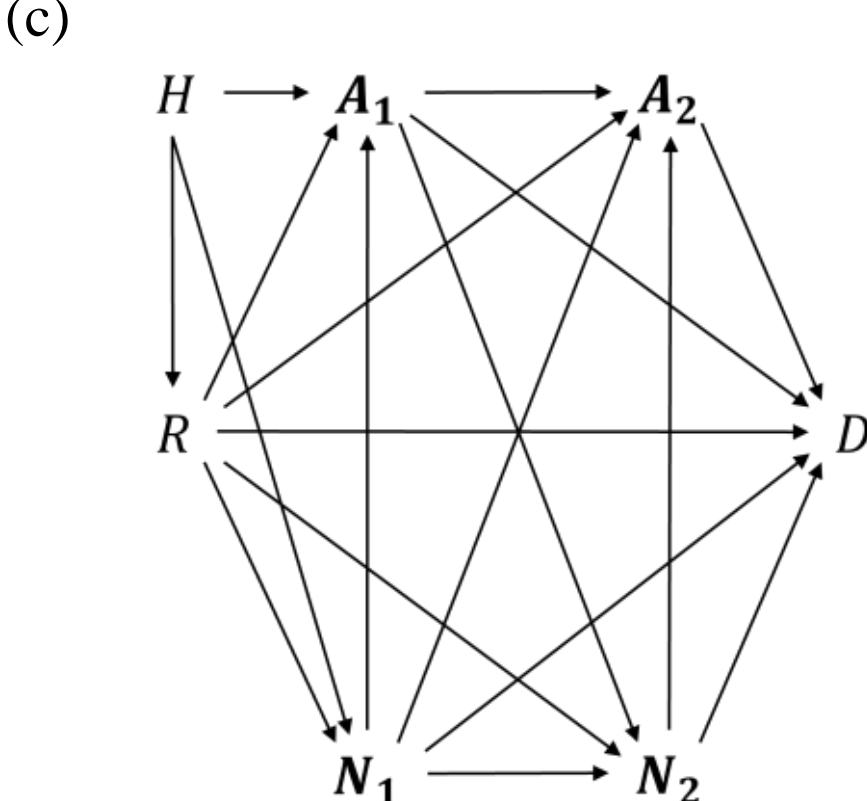

**Figure 5**. Directed Acyclic Graphs depicting causal relations between historical processes $H$, social group $R$, a set of allowable covariates $\boldsymbol{A}$, a set of non-allowable covariates $\boldsymbol{N}$, and a decision-based outcome $D$. In (a), the non-allowables affect the allowables. In (b) the allowables affect the non-allowables. In (c) there is causal feedback between allowables $\boldsymbol{A}$ and non-allowables $\boldsymbol{N}$ over time.

**Table 1. Example Interventions to Eliminate Forms of Collider-Stratification Under Design 4**

| Sub-design | Definition of the intervention $G_k^\dagger$ to allocate $\boldsymbol{W_k^\dagger}$ according to the distribution $\mathcal{g}_k(\cdot)$ | The allocation strategy $\mathcal{g}_k(\cdot)$ for $\boldsymbol{W_k^\dagger}$ under $G_k$ | Distribution $\mathcal{q}_k(\cdot)$ of partial eligibility $Q_k^\dagger = 1$ under $G_k$ |
|---|---|---|---|
| 4a | $\boldsymbol{W_k^\dagger} \sim \mathcal{g}_k(\cdot) = P(\boldsymbol{w_k^\dagger} \mid Q_k^\ddagger = 1, k)$ | Randomly assign $\boldsymbol{W_k^\dagger}$ (e.g., EPPP enrollment) by the observed probability of $\boldsymbol{W_k^\dagger} = \boldsymbol{w}^\dagger$ among those with $Q_k^\ddagger = 1$ (e.g., prior hypertension) at calendar time $k$. This removes associations between $\boldsymbol{W_k^\dagger}$ and each of the allowables $\boldsymbol{A_k}$ (e.g, age $X_k$)**,** non-allowables $\boldsymbol{N_k}$ (e.g., SES $L_k$), and social group $R$. | $P(Q_k^\dagger = 1 \mid Q_k^\ddagger = 1, k)$ |
| 4b | $\boldsymbol{W_k^\dagger} \sim \mathcal{g}_k(\cdot) = P(\boldsymbol{w_k^\dagger} \mid Q_k^\ddagger = 1, R = r, \boldsymbol{a_k}, k)$ | Randomly assign $\boldsymbol{W_k^\dagger}$ (e.g., EPPP enrollment) by the observed probability of $\boldsymbol{W_k^\dagger} = \boldsymbol{w}^\dagger$ among those with $Q_k^\ddagger = 1$ (e.g., prior hypertension) given social group $R = r$, the allowables $\boldsymbol{A_k}$ (e.g., age $X_k$) at calendar time $k$. This removes direct associations (with respect to social group $R$ and allowables $\boldsymbol{A_k}$) between $\boldsymbol{W_k^\dagger}$ and non-allowables $\boldsymbol{N_k}$ (e.g., SES $L_k$). | $P(Q_k^\dagger = 1 \mid Q_k^\ddagger = 1, R = r, \boldsymbol{a_k}, k)$ |
| 4c | $\boldsymbol{W_k^\dagger} \sim \mathcal{g}_k(\cdot) = P(\boldsymbol{w_k^\dagger} \mid Q_k^\ddagger = 1, T = 1, \boldsymbol{n_k}, \boldsymbol{a_k}, k)$ | Randomly assign $\boldsymbol{W_k^\dagger}$ (e.g., EPPP enrollment) by the observed probability of $\boldsymbol{W_k^\dagger} = \boldsymbol{w}^\dagger$ among those with $Q_k^\ddagger = 1$ (e.g., prior hypertension) in the standard population $T = 1$ given the allowables $\boldsymbol{A_k}$ (e.g., age $X_k$) and a set of non-allowables $\boldsymbol{N_k}$ (e.g., SES $L_k$) that satisfy exchangeability (18) or (19)) at calendar time $k$. This removes direct associations (with respect to allowables $\boldsymbol{A_k}$ and non-allowables $\boldsymbol{N_k}$) between $\boldsymbol{W_k^\dagger}$ and social group $R$.[a] | $P(Q_k^\dagger = 1 \mid Q_k^\ddagger = 1, T = 1, \boldsymbol{n_k}, \boldsymbol{a_k}, k)$ |

[a]As explained at the end of Section 5.4 and in Footnote 22, when the non-allowables $\boldsymbol{N_k}$ are multivariate and follow certain causal structures, Design 4c may leave residual contributions from collider stratification that would be eliminated under Designs 4a and 4b.

**Table 2. Example Target Study Protocol Specification and its Emulation with Secondary Data**

| | Target Study | Emulation with EMR data |
|---|---|---|
| Enrollment windows | Weekly over 2015 | Same |
| Enrollment groups | Self-reported Non-Hispanic Black persons and Non-Hispanic White Persons | Same, implemented as self-reported race/ethnicity as recorded in EMR |
| Eligibility criteria | Established care in health system, prior diagnosis of hypertension, not pregnant, not diagnosed with ESKD, enrolled in EPPP, and current visit for primary care | Same, implemented as 2+ primary care visit in past 2 years, diagnosis of hypertension, in past 2 years, not pregnant, not diagnosed with ESKD, enrolled in EPPP before current visit |
| Allowable covariates | Age and sex assigned at birth | Same, implemented using age and sex[a] in EMR at current visit |
| Standard population | Black population | Same |
| Enrollment Process | Stratified sampling… | With pooled data (Figure 4), apply… |
| *…Design 1 for inference in the fully eligible population* | …to balance age and sex (allowable), sampling from the eligible population | …G-Computation (20) or weighting (21) using age and sex (allowable) |
| *…Design 2: for inference in the broader population (e.g., regardless of EPPP enrollment)* | … to balance age and age (allowable), and account for comorbidity and SES (non-allowable), sampling as if from the population regardless of EPPP enrollment | …G-Computation (22) or weighting (23) using age, sex[a] (allowable), and comorbidity, and SES[b] (non-allowable) |
| *…Design 3: for inference in a different population (e.g., not enrolled in EPPP)* | …to balance age and sex (allowable), and account for comorbidity and SES (non-allowable), sampling as if from the population not enrolled in EPPP | …G-Computation (24) or weighting (25) using age, sex[a] (allowable), and comorbidity, and SES[b] (non-allowable) |
| *…Design 4: for inference in a counterfactual population after intervention to allocate EPPP enrollment (e.g., to remove the impact of collider-stratification)* | …to balance age and sex (allowable), and account for comorbidity and SES (non-allowable), sampling as if from a counterfactual eligible population after intervening to allocate EPPP enrollment | …G-Computation (28) or weighting (29) using age, sex[a] (allowable), and comorbidity, and SES[b] (non-allowable); or simplified versions of G-computation and weighting, i.e., (22) and (23) under Design 4a, (30) and (31) under Design 4b, or (32) and (33) under Design 4c |
| Time zero | Time of current visit | Same |
| Outcome assessment | Uncontrolled hypertension at current visit (i.e., systolic blood pressure ≥140 mm Hg or diastolic blood pressure ≥90 mm Hg) | Same |
| Statistical analysis | Mean difference in uncontrolled hypertension | Same |
| *…Aggregation of results* | Weighted average of results according to the distribution of calendar time enrollment in the Black population | Same |

Abbreviations: Electronic Medical Records (EMR), Electronic Patient Portal Program (EPPP), End Stage Kidney Disease (ESKD), Socioeconomic Status (SES), Mercury (Hg)
[a]Sex as recorded in the EMR
[b]In the EMR, SES is approximated by health insurance type and categorized CDC Social Vulnerability Index

Supplemental Material for "The Target Study: A Conceptual Model and Framework for Measuring Disparity" by John W. Jackson, Yea-Jen Hsu, Raquel C. Greer, Tony Boonyasai, Chanelle J. Howe

DATA APPLICATION

*Methods*

We sought to measure racial disparities in the prevalence of uncontrolled hypertension at primary care visits in a large Mid-Atlantic healthcare system (n=29 clinics) during 2015. The prevalence of uncontrolled hypertension is a commonly used metric to promote healthcare quality and equity. (Merai, Siegel, Rakotz et al. 2016) Our data consisted of electronic medical records of healthcare encounters with data on the encounter type, encounter date, clinic, vital signs including systolic and diastolic blood pressure, race/ethnicity (1=Black, 0=White), demographics (age in years, sex assigned at birth), measures of socioeconomic status (primary type of health insurance, CDC social vulnerability index linked by zip code), state of residence, presence of comorbidity and enrollment into an electronic patient portal program (EPPP) for care management. The data were arranged in "long" format where the person and encounter visit ids uniquely identify a record. Our study was reviewed by the Johns Hopkins Medicine Institutional Review Board.

Consider the target study specification in Table 2 of the main text. We sought to emulate Design 1 to assess the average weekly racial disparity (comparing Black to White persons) in uncontrolled hypertension in 2015 among adult Maryland residents aged 18 years or older meeting the eligibility criteria: (a) no diagnosis of end-stage kidney disease (ESKD) documented or recorded as starting before the index visit; (b) no pregnancy documented or recorded as starting within the twelve months before the index visit; (c) prior self-reported diagnosis of hypertension, documented at least six months before the index visit or recorded as starting at least six months before the index visit; (d) engaged in care within the health system, defined as two or more primary care visits within the twelve months before the index visit as well as a primary care visit during 2014 with the index visit during 2015; (e) enrolled in the health system's EPPP. Aside from criterion (e), these criteria align (with operational exceptions) with the denominator of the National Quality Forum metric #0018 for controlling high blood pressure in the Merit-Based Incentive Payment System.

We chose age and assigned sex at birth (sex) as allowable because the eligible Black population were not disadvantaged on these covariates (they were younger on average [59 vs. 61 years, respectively] with a higher proportion of females [64% vs. 57%, respectively], both of which predict better hypertension control). (Jackson 2021) Failing to adjust for age and sex would potentially mask disparity from other mechanisms. Measures of socioeconomic status were not treated as allowable because their effects are, broadly speaking, amenable to intervention and Black persons are disadvantaged on these. (Jackson 2021) Compared to White persons, Black persons have a higher percentage of Medicaid insurance [15% vs. 12%] and worse social vulnerability scores on average [53% vs. 19% with a higher score indicating greater disadvantage], both of which predict worse hypertension control. We thus view worse socioeconomic status among Black persons as leading to worse (and disparate) hypertension control outcomes compared to eligible White persons. We chose the Black population as the standard population to obtain their observed prevalence of uncontrolled hypertension in disparity estimate. Figure S1 shows a sample size flowchart for obtaining the analytic dataset. We applied G-computation (20) and weighting (21) estimators of the main text to estimate the average weekly disparity in the eligible population. Recall that these estimators aggregate the weekly studies in such a way as to balance calendar time across race, which eliminates confounding by any racial differences in calendar time enrollment under seasonal trends in blood pressure and hypertension control.

Selective mechanisms involving enrollment into the EPPP could affect the disparity estimated among this population. Fewer Black persons were enrolled in the EPPP than White persons (31% vs. 43%). Likewise, fewer persons with Medicaid insurance (25%) were enrolled compared to those with Medicare (29%) or private insurance (49%), and type of health insurance coverage is associated with hypertension control. The disparity estimated by our emulation of Design 1 captures the contribution of this selective mechanism. Due to the contribution of the selective mechanism, we may not be able to use our estimate of disparity to (i) infer about disparity to the entire health system regardless of EPPP enrollment (i.e., it may not be generalizable) and (ii) infer about disparity to those not enrolled in the EPPP (i.e., it may not be transportable). Furthermore, even if we restrict our attention to those enrolled in the EPPP, it is possible that the selective mechanism may have attenuated disparity in this subpopulation.

To address these issues using the approaches outlined in the main text, for the sake of demonstration, we proceed as if outcomes were only measured among those in the EPPP. (In more realistic settings, certain outcomes (e.g., measurement of home-based blood pressures) may only be available among those enrolled in the EPPP). Thus, for our demonstration, we emulated Design 2 by combining outcome and covariate data from the EPPP with covariate data among those regardless of EPPP enrollment, to infer about disparity regardless of EPPP enrollment. We emulated Design 3 by combining outcome and covariate data from the EPPP with covariate data among those not enrolled in the EPPP, to infer about disparity in those not enrolled in the EPPP. We emulated Design 4 to assess what disparity would be if selection into the EPPP was not jointly dependent on race and non-allowable covariates, so that selecting on EPPP enrollment would not induce collider stratification. This gives us a sense of how much the selective mechanism contributes to disparity among those enrolled in the EPPP. To accomplish this we combined outcome and covariate data from those enrolled in the EPPP with covariate data among those regardless of EPPP enrollment. In these emulations, measures of socioeconomic status (i.e., type of medical insurance and social vulnerability index) and medical comorbidity were treated as non-allowable, and age and sex were treated as allowable. Figure S2 shows a sample size flowchart for obtaining the analytic dataset. We applied the G-computation ((22), (24), (29)) and weighting ((23), (25), (30)) estimators for these respective designs to estimate the average weekly disparity in these populations. Recall that these estimators aggregate the weekly studies in such a way as to balance calendar time across race, which eliminates confounding by any racial differences in calendar time enrollment under seasonal trends in blood pressure or hypertension control.

As described before, for the sake of demonstration, we emulated Designs 2 and 3 by proceeding as if outcomes were only measured among those in the EPPP. In more realistic settings, certain outcomes (e.g., measurement of home-based blood pressures) may only be available among those enrolled in the EPPP. Nonetheless, we had visit-based blood pressures which were available irrespective of EPPP enrollment. Therefore, to check the quality of the emulations for our Designs 2 and 3 demonstrations, we emulated Design 1 for those irrespective of EPPP enrollment, and again for those not enrolled in the EPPP, which provide ground truth estimates for the emulations of Designs 2 and 3, respectively.

*Results*

Across all weighting estimators, the mean of the weights ranged from 0.993 to 1.004 and ranged from 0.093 to 7.21, which is consistent with the properties of correctly estimated weights. (Cole and Hernán 2008). Table S1 shows the results for emulations of all target study designs. For Design 1, under the G-computation estimator, the proportion of uncontrolled hypertension was 56% (95%CI 53%, 59%) among the Black population and 48% (95%CI 45%, 52%) among the White population, leading to a disparity of 8% on the additive scale (95%CI 3%, 12%) and a disparity of 1.16 on the relative scale (95%CI 1.06, 1.26). Similar results were obtained for the weighting estimator of Design 1. The agreement between different estimators for Design 1 reassures us that our results are not biased by model misspecification. Furthermore, we found similar results across all designs, regardless of the estimation approach. This suggests that selective mechanisms through EPPP do not materially impact the estimate of disparity among those in the EPPP and furthermore, that the disparity within the EPPP may generalize to those beyond the EPPP and also be transported to those not enrolled in the EPPP.

Our results for Design 1 assume racial overlap in the allowable covariates (age and sex) distributions and innocuous sampling. Our results for Designs 2 and 3 further assume no model misspecification, positivity, and conditional independence of EPPP enrollment given racial group membership, the allowable covariates (age and sex) and non-allowable covariates (type of insurance, social vulnerability index, and comorbidity). Designs 4a, 4b, and 4c share a similar set of assumptions but replace the conditional independence assumption with a conditional exchangeability assumption given racial group membership, the allowable covariates (age and sex) and non-allowable covariates (type of insurance, social vulnerability index, and comorbidity). Most of these assumptions are unverifiable in practice. The additional requirement of Designs 4a, 4b, and 4c that certain eligibility criteria used (e.g., prior diagnosis of hypertension, engagement in care) precede the eligibility criterion of EPPP enrollment, and other criteria (e.g., index visit in 2015) occur after EPPP enrollment, are reasonably satisfied in our data application. There is likely to be some residual confounding as area-level social vulnerability as measured by the CDC index and type of insurance do not capture individual-level socioeconomic status or social needs. Due to lack of measurement, we were unable to control for patient-activation which may likely be a common cause of EPPP enrollment and hypertension control and thus confound the effect of EPPP on uncontrolled hypertension.

Table S2 shows the results of our empirical checks for the emulations of Designs 2 and 3 by applying Design 1 to the population irrespective of EPPP enrollment (to check the emulation of Design 2) and to the population not enrolled in the EPPP (to check the emulation of Design 3). The generalized results from emulating Design 2 were similar to results from emulating Design 1 among those irrespective of EPPP enrollment. The transported results from emulating Design 3 were similar to results from emulating Design 1 among those not enrolled in the EPPP, although the prevalence estimates were slightly underestimated. These results add confidence that the assumptions invoked by Design 2 and Design 3 held approximately.

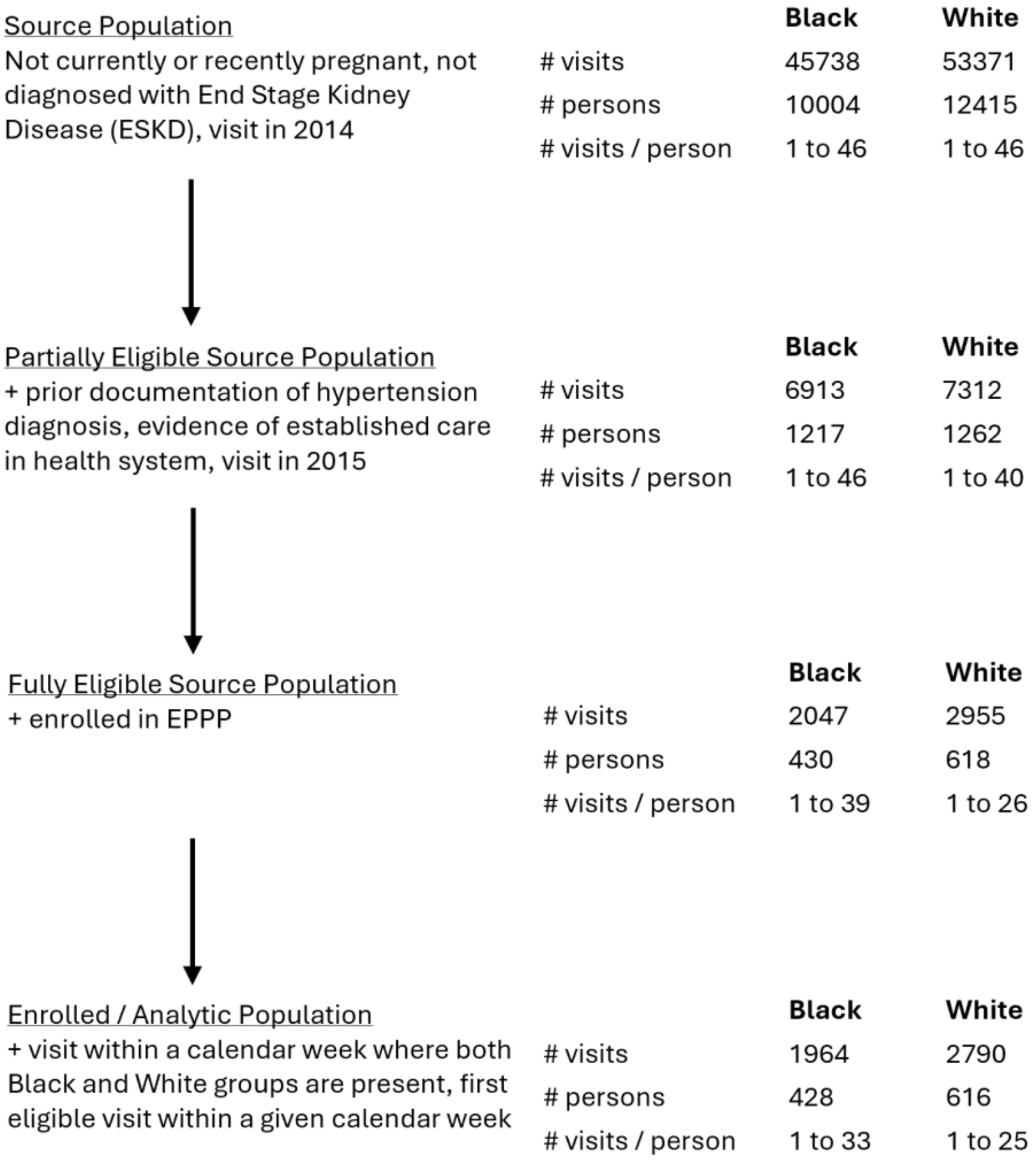

Figure S1. Flowchart describing creation of the analytic study population for emulating Designs 1, 2 and 3

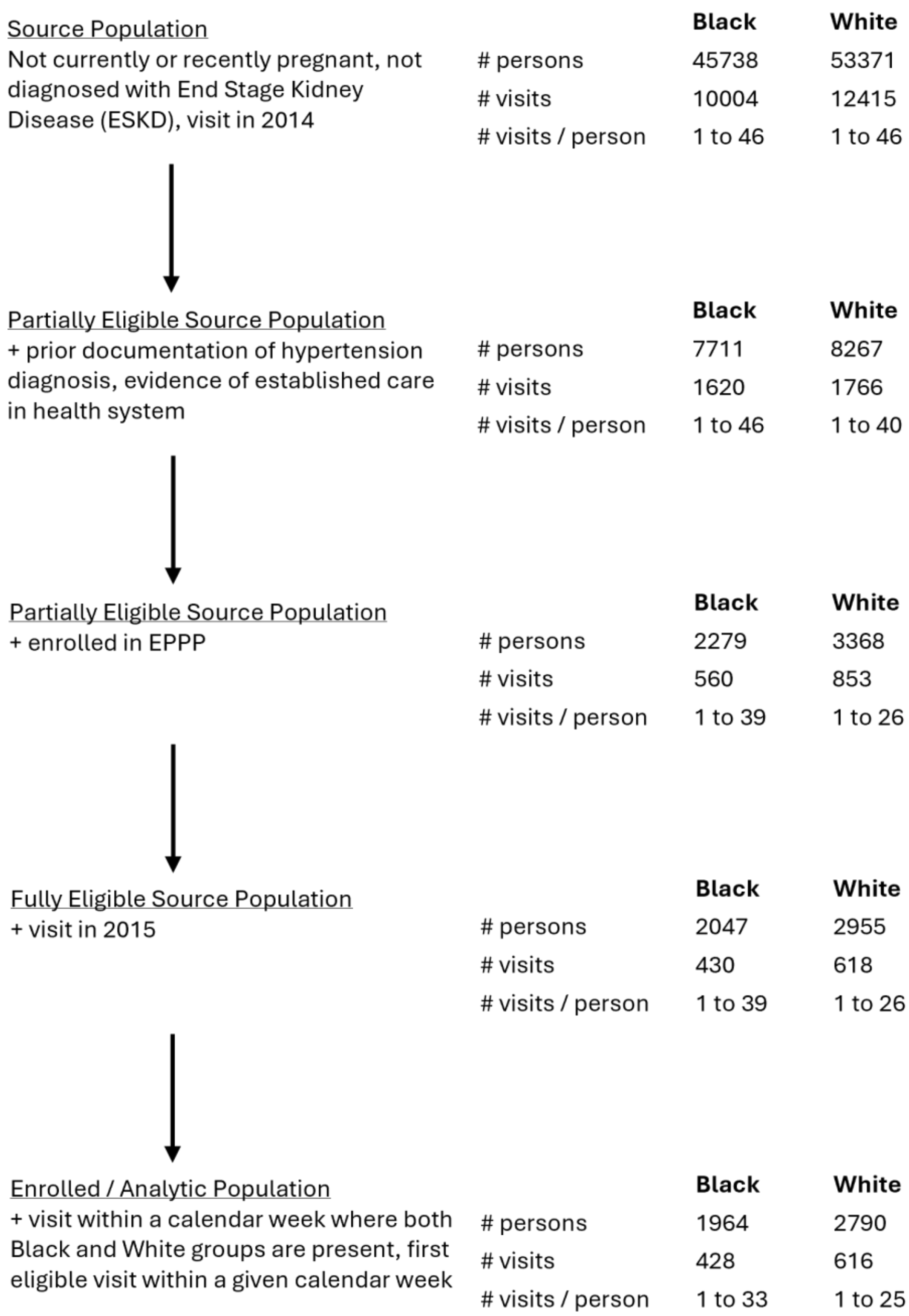

Figure S2. Flowchart describing creation of the analytic study population for emulating Design 4

| **Table S1. Prevalence and Disparity in Uncontrolled Hypertension in a Mid-Atlantic Healthcare System in 2015** | | | | |
|---|---|---|---|---|
| | Prevalence of Uncontrolled Hypertension | | Estimate of Disparity | |
| | *Black Population* | *White Population* | *Prevalence Difference* | *Prevalence Ratio* |
| *G-computation* | | | | |
| Design 1 | 0.56 (0.53, 0.59) | 0.48 (0.45, 0.52) | 0.08 (0.03, 0.12) | 1.16 (1.06, 1.26) |
| Design 2 | 0.57 (0.53, 0.60) | 0.50 (0.46, 0.53) | 0.07 (0.02, 0.12) | 1.14 (1.04, 1.25) |
| Design 3 | 0.57 (0.53, 0.61) | 0.50 (0.47, 0.54) | 0.07 (0.01, 0.12) | 1.13 (1.02, 1.25) |
| Design 4a | 0.56 (0.53, 0.60) | 0.49 (0.46, 0.53) | 0.07 (0.02, 0.12) | 1.14 (1.03, 1.25) |
| Design 4b | 0.56 (0.53, 0.60) | 0.48 (0.45, 0.52) | 0.08 (0.04, 0.12) | 1.16 (1.07, 1.27) |
| Design 4c | 0.56 (0.53, 0.59) | 0.48 (0.45, 0.51) | 0.08 (0.04, 0.13) | 1.17 (1.08, 1.28) |
| *Weighting* | | | | |
| Design 1 | 0.56 (0.53, 0.59) | 0.48 (0.45, 0.51) | 0.08 (0.04, 0.12) | 1.16 (1.07, 1.27) |
| Design 2 | 0.57 (0.53, 0.60) | 0.50 (0.47, 0.53) | 0.07 (0.01, 0.11) | 1.13 (1.03, 1.24) |
| Design 3 | 0.57 (0.52, 0.61) | 0.51 (0.47, 0.54) | 0.06 (0.00, 0.11) | 1.12 (1.00, 1.23) |
| Design 4a | 0.56 (0.53, 0.60) | 0.50 (0.47, 0.53) | 0.06 (0.02, 0.11) | 1.13 (1.03, 1.24) |
| Design 4b | 0.56 (0.53, 0.60) | 0.49 (0.45, 0.52) | 0.07 (0.02, 0.12) | 1.15 (1.05, 1.26) |
| Design 4c | 0.56 (0.53, 0.59) | 0.48 (0.45, 0.51) | 0.08 (0.04, 0.13) | 1.17 (1.08, 1.28) |

Design 1 (inference to those enrolled in an electronic patient portal program (EPPP)); Design 2 (inference to those regardless of EPPP enrollment (generalizability)); Design 3 (inference to those not enrolled in EPPP (transportability)); Design 4 (inference to those enrolled in EPPP under an intervention to assign EPPP randomly [4a], given age, sex, and race only [4b], given age, sex, socioeconomic status and comorbidity only [4c]).

| **Table S2. Prevalence and Disparity in Uncontrolled Hypertension in a Mid-Atlantic Healthcare System in 2015 Estimated Directly (using Design 1) Among Those Irrespective of EPPP Enrollment and Among Those Not Enrolled in the EPPP** | | | | |
|---|---|---|---|---|
| | Prevalence of Uncontrolled Hypertension | | Estimate of Disparity | |
| | *Black Population* | *White Population* | *Prevalence Difference* | *Prevalence Ratio* |
| *G-computation* | | | | |
| Irrespective of EPPP | 0.56 (0.53, 0.59) | 0.48 (0.45, 0.52) | 0.08 (0.03, 0.12) | 1.16 (1.06, 1.26) |
| Not in EPPP | 0.61 (0.59, 0.63) | 0.54 (0.51, 0.56) | 0.08 (0.04, 0.11) | 1.14 (1.08, 1.21) |
| *Weighting* | | | | |
| Irrespective of EPPP | 0.56 (0.53, 0.59) | 0.48 (0.45, 0.51) | 0.08 (0.04, 0.12) | 1.16 (1.07, 1.27) |
| Not in EPPP | 0.61 (0.59, 0.63) | 0.53 (0.51, 0.56) | 0.08 (0.04, 0.11) | 1.14 (1.08, 1.21) |

Design 1 (inference to those enrolled in an electronic patient portal program (EPPP)).

PROOFS

Recall the variables used to establish eligiblity $\boldsymbol{W_k} = (\boldsymbol{W_k^{\ddagger}}, \boldsymbol{W_k^{\dagger}}, \boldsymbol{W_k^{\wr}})$ and their corresponding range of values that establish eligibility $\mathbb{w} = (\mathbb{w}^{\ddagger}, \mathbb{w}^{\dagger}, \mathbb{w}^{\wr})$ where $\boldsymbol{W_k^{\ddagger}} = \varnothing$, $\boldsymbol{W_k^{\dagger}} = \varnothing$, or $\boldsymbol{W_k^{\wr}} = \varnothing$ is permitted. Indicators of eligibility are $Q_k = Q_k^{\ddagger} \times Q_k^{\dagger} \times Q_k^{\wr}$ where $Q_k = I(\boldsymbol{W_k} \in \mathbb{w})$, $Q_k^{\ddagger} = I(\boldsymbol{W_k^{\ddagger}} \in \mathbb{w}^{\ddagger})$, $Q_k^{\dagger} = I(\boldsymbol{W_k^{\dagger}} \in \mathbb{w}^{\dagger})$, $Q_k^{\wr} = I(\boldsymbol{W_k^{\wr}} \in \mathbb{w}^{\wr})$.

For any design $\mu_k^{\mathcal{D}}(r)$ is the mean outcome $Y_{k+J}$ and $f_{k,\ell}^{\mathcal{D}}$ the joint distribution of the outcome, allowables $\boldsymbol{A_k}$ (and if relevant, the non-allowables $\boldsymbol{N_k}$) among the social group $R = r$ enrolled in the target study at calendar time $k$. Note that $f_{k,\ell}^{\mathcal{D}}$ is specific to the stage $\ell$ of sampling during enrollment $\ell$.

Throughout the proofs and in the main text, we abbreviate a weighted mean, $\sum_i Y_i \omega_i \,/\, \sum_i \omega_i$ where $i$ represents the unit of observation, as $E[Y_i \times \omega_i]$ (in the text and in the proofs we drop the subscript $i$ to simplify notation).

*AGGREGATION OF $\psi_k$ OVER CALENDAR TIME $k$*

Let $\gamma_k = P(k|Q_k = 1, T = 1)$

$$
\begin{aligned}
&\Psi^{add} \\
&= \frac{\sum_k \gamma_k \psi_k^{add}}{\sum_k \gamma_k} \\
&= \frac{\sum_k \gamma_k \left( \mathbb{E}_\Omega\left[Y_{k+J}\middle|Q_k = 1, R = 1, k\right] - \mathbb{E}_\Omega\left[Y_{k+J}\middle|Q_k = 1, R = 0, k\right] \right)}{\sum_k \gamma_k} \\
&= \frac{\sum_k P(k|Q_k = 1, T = 1) \left( \mathbb{E}_\Omega\left[Y_{k+J}\middle|Q_k = 1, R = 1, k\right] - \mathbb{E}_\Omega\left[Y_{k+J}\middle|Q_k = 1, R = 0, k\right] \right)}{\sum_k P(k|Q_k = 1, T = 1)} \\
&= \sum_k \begin{pmatrix} \mathbb{E}_\Omega\left[Y_{k+J}\middle|Q_k = 1, R = 1, k\right] P(k|Q_k = 1, T = 1) \\ -\mathbb{E}_\Omega\left[Y_{k+J}\middle|Q_k = 1, R = 0, k\right] P(k|Q_k = 1, T = 1) \end{pmatrix} \\
&= \sum_k \begin{pmatrix} \mathbb{E}_\Omega\left[Y_{k+J}\middle|Q_k = 1, R = 1, k\right] P(k|Q_k = 1, R = 1) \times \frac{P(k|Q_k = 1, T = 1)}{P(k|Q_k = 1, R = 1)} \\ -\mathbb{E}_\Omega\left[Y_{k+J}\middle|Q_k = 1, R = 0, k\right] P(k|Q_k = 1, R = 0) \times \frac{P(k|Q_k = 1, T = 1)}{P(k|Q_k = 1, R = 0)} \end{pmatrix} \\
&= \mathbb{E}_\Omega\left[Y_{k+J} \times \lambda_k(r)\middle|Q_k = 1, R = 1\right] - \mathbb{E}_\Omega\left[Y_{k+J} \times \lambda_k(r)\middle|Q_k = 1, R = 1\right]
\end{aligned}
$$

where $\lambda_k(r) = P(k|Q_k = 1, T = 1)/P(k|Q_k = 1, R = r)$

The fourth equality establishes that $\Psi^{add}$, which is a standardized difference of means, is equivalent to the difference of standardized means with $P(k|Q_k = 1, T = 1)$ as the weight. The last equality establishes that $\Psi^{add}$ is equivalent to a weighted analysis taking a difference of means, each pooled over calendar time $k$, with $\lambda_k(r)$ as the weight for each instance of enrollment.

$$
\begin{aligned}
&\Psi^{rel} \\
&= \frac{\sum_k \gamma_k \mathbb{E}_\Omega\left[Y_{k+J}\middle|Q_k = 1, R = 0, k\right] \psi_k^{rel}}{\sum_k \gamma_k \mathbb{E}_\Omega\left[Y_{k+J}\middle|Q_k = 1, R = 0, k\right]} \\
&= \frac{\sum_k \gamma_k \, \mathbb{E}_\Omega\left[Y_{k+J}\middle|Q_k = 1, R = 0, k\right] \left( \frac{\mathbb{E}_\Omega\left[Y_{k+J}\middle|Q_k = 1, R = 1, k\right]}{\mathbb{E}_\Omega\left[Y_{k+J}\middle|Q_k = 1, R = 0, k\right]} \right)}{\sum_k \gamma_k \mathbb{E}_\Omega\left[Y_{k+J}\middle|Q_k = 1, R = 0, k\right]} \\
&= \frac{\sum_k P(k|Q_k = 1, T = 1) \, \mathbb{E}_\Omega\left[Y_{k+J}\middle|Q_k = 1, R = 0, k\right] \left( \frac{\mathbb{E}_\Omega\left[Y_{k+J}\middle|Q_k = 1, R = 1, k\right]}{\mathbb{E}_\Omega\left[Y_{k+J}\middle|Q_k = 1, R = 0, k\right]} \right)}{\sum_k P(k|Q_k = 1, T = 1) \mathbb{E}_\Omega\left[Y_{k+J}\middle|Q_k = 1, R = 0, k\right]} \\
&= \frac{\sum_k \mathbb{E}_\Omega\left[Y_{k+J}\middle|Q_k = 1, R = 1, k\right] P(k|Q_k = 1, T = 1)}{\sum_k \mathbb{E}_\Omega\left[Y_{k+J}\middle|Q_k = 1, R = 0, k\right] P(k|Q_k = 1, T = 1)} \\
&= \frac{\sum_k \mathbb{E}_\Omega\left[Y_{k+J}\middle|Q_k = 1, R = 1, k\right] P(k|Q_k = 1, T = 1) \times \frac{P(k|Q_k = 1, T = 1)}{P(k|Q_k = 1, R = 1)}}{\sum_k \mathbb{E}_\Omega\left[Y_{k+J}\middle|Q_k = 1, R = 0, k\right] P(k|Q_k = 1, T = 1) \times \frac{P(k|Q_k = 1, T = 1)}{P(k|Q_k = 1, R = 0)}} \\
&= \mathbb{E}_\Omega\left[Y_{k+J} \times \lambda_k(r)\middle|Q_k = 1, R = 1\right] / \mathbb{E}_\Omega\left[Y_{k+J} \times \lambda_k(r)\middle|Q_k = 1, R = 1\right]
\end{aligned}
$$

where $\lambda_k(r) = P(k|Q_k = 1, T = 1)/P(k|Q_k = 1, R = r)$

The fourth equality establishes that $\Psi^{rel}$, which is a standardized ratio of means, is equivalent to the ratio of standardized means with $P(k|Q_k = 1, T = 1)\mathbb{E}_\Omega\left[Y_{k+J}\middle|Q_k = 1, R = 0, k\right]$ as the weight. The last equality establishes that $\Psi^{rel}$ is equivalent to a weighted analysis taking a ratio of means, each pooled over calendar time $k$, with $\lambda_k(r)$ as the weight for each instance of enrollment.

*DESIGN 1*

Joint distribution of $(Y_{k+J}, \boldsymbol{A_k})$ given $Q_k = 1$ and $R = r$ in the Eligible Source Population

$$f_{k,0}^{\mathcal{D}1} = P\big(Y_{k+J} = y \big| Q_k = 1, R = r, \boldsymbol{a_k}, k\big) P(\boldsymbol{a_k} | Q_k = 1, R = r, k)$$

Joint distribution of $(Y_{k+J}, \boldsymbol{A_k})$ given $Q_k = 1$ and $R = r$ after Stage One of Sampling

$$f_{k,1}^{\mathcal{D}1} = P\big(Y_{k+J} = y \big| Q_k = 1, R = r, \boldsymbol{a_k}, k\big) P(\boldsymbol{a_k} | Q_k = 1, R = r, k)$$

Joint distribution of $(Y_{k+J}, \boldsymbol{A_k})$ given $Q_k = 1$ and $R = r$ after Stage Two of Sampling

$$f_{k,2}^{\mathcal{D}1} = P\big(Y_{k+J} = y \big| Q_k = 1, R = r, \boldsymbol{a_k}, k\big) P(\boldsymbol{a_k} | Q_k = 1, T = 1, k)$$

These joint distributions hold by design of the target study and under overlap (3) and innocuous sampling (4) in the main text.

Mean of $Y_{k+J}$ given $R = r$ in the target study

$$\mu_k^{\mathcal{D}1}(r)$$
$$= \textstyle\sum_{\boldsymbol{a}} E\big(Y_{k+J} \big| Q_k = 1, R = r, \boldsymbol{a_k}, k\big) P(\boldsymbol{a_k} | Q_k = 1, T = 1, k)$$

By definition of a mean

Sampling Fractions

$$\alpha_{k,1}^{\mathcal{D}1}(r)$$
$$= \frac{\mathbb{N}_{k,1}(r)}{\mathbb{N}_{k,0}(r)}$$

$$\alpha_{k,2}^{\mathcal{D}1}(\boldsymbol{a_k}, r)$$
$$= \frac{\mathbb{N}_{k,2}(r, \boldsymbol{a_k})}{\mathbb{N}_{k,1}(r, \boldsymbol{a_k})}$$
$$= \frac{\mathbb{N}_{k,2}(r)}{\mathbb{N}_{k,1}(r)} \times \frac{f_{k,2}^{\mathcal{D}1}}{f_{k,1}^{\mathcal{D}1}}$$
$$= \frac{\mathbb{N}_{k,2}(r)}{\mathbb{N}_{k,1}(r)} \times \frac{P(\boldsymbol{a_k} | Q_k = 1, T = 1, k)}{P(\boldsymbol{a_k} | Q_k = 1, R = r, k)}$$

G-Computation Estimator

$$\mu_k^{\mathcal{D}1}(r)$$
$$= \textstyle\sum_{\boldsymbol{a}} E\big(Y_{k+J} \big| Q_k = 1, R = r, \boldsymbol{a_k}, k\big) P(\boldsymbol{a_k} | Q_k = 1, T = 1, k)$$
$$= E_{\boldsymbol{a}}(E[Y_{k+J} | Q_k = 1, R = r, \boldsymbol{a_k}, k] | Q_k = 1, T = 1, k)$$

where $E_{\boldsymbol{a}}(\cdot)$ is over $P(\boldsymbol{a_k} | Q_k = 1, T = 1, k)$

By definition of an iterated expectation

Weighting Estimator

$$\mu_k^{\mathcal{D}1}(r)$$
$$= \textstyle\sum_{\boldsymbol{a}} E\big(Y_{k+J} \big| Q_k = 1, R = r, \boldsymbol{a_k}, k\big) P(\boldsymbol{a_k} | Q_k = 1, T = 1, k)$$
$$= \textstyle\sum_{\boldsymbol{a}} E\big(Y_{k+J} \big| Q_k = 1, R = r, \boldsymbol{a_k}, k\big) P(\boldsymbol{a_k} | Q_k = 1, R = r, k) \times \frac{P(\boldsymbol{a_k} | Q_k = 1, T = 1, k)}{P(\boldsymbol{a_k} | Q_k = 1, R = r, k)}$$
$$= \textstyle\sum_{\boldsymbol{a}} E\big(Y_{k+J} \big| Q_k = 1, R = r, \boldsymbol{a_k}, k\big) P(\boldsymbol{a_k} | Q_k = 1, R = r, k) \times \frac{P(T = 1 | Q_k = 1, \boldsymbol{a_k}, k)}{P(R = r | Q_k = 1, \boldsymbol{a_k}, k)} \times \frac{P(R = r | Q_k = 1, k)}{P(T = 1 | Q_k = 1, k)}$$
$$= E[Y_{k+J} \times \omega_{r,Q_k=1,k}^{\mathcal{D}1} | Q_k = 1, R = r, k]$$

where

$$\omega_{r,Q_k=1,k}^{\mathcal{D}1} = \frac{P(T = 1 | Q_k = 1, \boldsymbol{a_k}, k)}{P(R = r | Q_k = 1, \boldsymbol{a_k}, k)} \times \frac{P(R = r | Q_k = 1, k)}{P(T = 1 | Q_k = 1, k)}$$

Aggregated G-Computation Estimator (over calendar time $k$)

$\tau^{\mathcal{D}1}(r) = \sum_k \mu_k^{\mathcal{D}1}(r)\, P(k|Q_k = 1, T = 1)$
$= \sum_{\boldsymbol{a},k} E\left(Y_{k+J}\middle|Q_k = 1, R = r, \boldsymbol{a_k}, k\right)P(\boldsymbol{a_k}|Q_k = 1, T = 1, k)P(k|Q_k = 1, T = 1)$
$= \sum_{\boldsymbol{a},k} E\left(Y_{k+J}\middle|Q_k = 1, R = r, \boldsymbol{a_k}, k\right)P(\boldsymbol{a_k}, k|Q_k = 1, T = 1)$
$= E_{\boldsymbol{a},k}(E[Y_{k+J}|Q_k = 1, R = r, \boldsymbol{a_k}, k]|Q_k = 1, T = 1)$
where $E_{\boldsymbol{a},k}(\cdot)$ is over $P(\boldsymbol{a_k}, k|Q_k = 1, T = 1)$
By definition of an iterated expectation

Aggregated Weighting Estimator (over calendar time $k$)

$\tau^{\mathcal{D}1}(r) = \sum_k \mu_k^{\mathcal{D}1}(r)\, P(k|Q_k = 1, T = 1)$
$= \sum_{\boldsymbol{a},k} E\left(Y_{k+J}\middle|Q_k = 1, R = r, \boldsymbol{a_k}, k\right)P(\boldsymbol{a_k}|Q_k = 1, T = 1, k)P(k|Q_k = 1, T = 1)$
$= \sum_{\boldsymbol{a},k} E\left(Y_{k+J}\middle|Q_k = 1, R = r, \boldsymbol{a_k}, k\right)P(\boldsymbol{a_k}, k|Q_k = 1, R = r)$
$\times \frac{P(\boldsymbol{a_k}, k|Q_k=1, T=1)}{P(\boldsymbol{a_k}, k|Q_k=1, R=r)}$
$= \sum_{\boldsymbol{a},k} E\left(Y_{k+J}\middle|Q_k = 1, R = r, \boldsymbol{a_k}, k\right)P(\boldsymbol{a_k}, k|Q_k = 1, R = r)$
$\times \frac{P(T=1|Q_k=1, \boldsymbol{a_k}, k)}{P(R=r|Q_k=1, \boldsymbol{a_k}, k)} \times \frac{P(R=r|Q_k=1)}{P(T=1|Q_k=1)}$
$= E[Y_{k+J} \times \omega_{r,Q_k=1,k}^{\mathcal{D}1-pool}|Q_k = 1, R = r]$
where
$\omega_{r,Q_k=1,k}^{\mathcal{D}1-pool} = \frac{P(T=1|Q_k=1, \boldsymbol{a_k}, k)}{P(R=r|Q_k=1, \boldsymbol{a_k}, k)} \times \frac{P(R=r|Q_k=1)}{P(T=1|Q_k=1)}$

*DESIGN 2*

Joint distribution of $(Y_{k+J}, \boldsymbol{A_k}, \boldsymbol{N_k})$ given $Q_k^{\ddagger} = 1$ and $R = r$ in the Broader Eligible Source Population

$$f_{k,*}^{\mathcal{D}2} = P\left(Y_{k+J} = y \middle| Q_k^{\ddagger} = 1, R = r, \boldsymbol{n_k}, \boldsymbol{a_k}, k\right) P(\boldsymbol{n_k} | Q_k^{\ddagger} = 1, R = r, \boldsymbol{a_k}, k) P\left(\boldsymbol{a_k} \middle| Q_k^{\ddagger} = 1, R = r, k\right)$$

Joint distribution of $(Y_{k+J}, \boldsymbol{A_k}, \boldsymbol{N_k})$ given $Q_k = 1$ and $R = r$ in the Available Eligible Source Population

$$f_{k,0}^{\mathcal{D}2} = P\left(Y_{k+J} = y \middle| Q_k^{\ddagger} = 1, R = r, \boldsymbol{n_k}, \boldsymbol{a_k}, k\right) P(\boldsymbol{n_k} | Q_k = 1, R = r, \boldsymbol{a_k}, k) P(\boldsymbol{a_k} | Q_k = 1, R = r, k)$$

Joint distribution of $(Y_{k+J}, \boldsymbol{A_k}, \boldsymbol{N_k})$ given $Q_k = 1$ and $R = r$ after Stage One of Sampling

$$f_{k,1}^{\mathcal{D}2} = P\left(Y_{k+J} = y \middle| Q_k^{\ddagger} = 1, R = r, \boldsymbol{n_k}, \boldsymbol{a_k}, k\right) P(\boldsymbol{n_k} | Q_k^{\ddagger} = 1, R = r, \boldsymbol{a_k}, k) P\left(\boldsymbol{a_k} \middle| Q_k^{\ddagger} = 1, R = r, k\right)$$

Joint distribution of $(Y_{k+J}, \boldsymbol{A_k}, \boldsymbol{N_k})$ given $Q_k = 1$ and $R = r$ after Stage Two of Sampling

$$f_{k,2}^{\mathcal{D}2} = P\left(Y_{k+J} = y \middle| Q_k^{\ddagger} = 1, R = r, \boldsymbol{n_k}, \boldsymbol{a_k}, k\right) P(\boldsymbol{n_k} | Q_k^{\ddagger} = 1, R = r, \boldsymbol{a_k}, k) P\left(\boldsymbol{a_k} \middle| Q_k^{\ddagger} = 1, T = 1, k\right)$$

These joint distributions hold by design of the target study and under a version of overlap (3) (see footnote 16), innocuous sampling (4), independence (11) and positivity (12) in the main text.

Mean of $Y_{k+J}$ given $R = r$ in the target study

$$\mu_k^{\mathcal{D}2}(r)$$
$$= \textstyle\sum_{\boldsymbol{a},\boldsymbol{n}} E\left(Y_{k+J} \middle| Q_k^{\ddagger} = 1, R = r, \boldsymbol{n_k}, \boldsymbol{a_k}, k\right) P\left(\boldsymbol{n_k} \middle| Q_k^{\ddagger} = 1, R = r, \boldsymbol{a_k}, k\right) P\left(\boldsymbol{a_k} \middle| Q_k^{\ddagger} = 1, T = 1, k\right)$$
$$= \textstyle\sum_{\boldsymbol{a},\boldsymbol{n}} E\left(Y_{k+J} \middle| Q_k = 1, R = r, \boldsymbol{n_k}, \boldsymbol{a_k}, k\right) P\left(\boldsymbol{n_k} \middle| Q_k^{\ddagger} = 1, R = r, \boldsymbol{a_k}, k\right) P\left(\boldsymbol{a_k} \middle| Q_k^{\ddagger} = 1, T = 1, k\right)$$

By definition of a mean, overlap (3) (footnote 17), innocuous sampling (4), independence (11) and positivity (12).

Sampling Fractions

$$\alpha_{k,1}^{\mathcal{D}2}(r, \boldsymbol{a_k}, \boldsymbol{n_k})$$
$$= \frac{\mathbb{N}_{k,1}^{*}(r, \boldsymbol{a_k}, \boldsymbol{n_k})}{\mathbb{N}_{k,0}(r, \boldsymbol{a_k}, \boldsymbol{n_k})}$$
$$= \frac{\mathbb{N}_{k,1}(r)}{\mathbb{N}_{k,0}(r)} \times \frac{f_{k,1}^{\mathcal{D}2}}{f_{k,0}^{\mathcal{D}2}}$$
$$= \frac{\mathbb{N}_{k,1}(r)}{\mathbb{N}_{k,0}(r)} \times \frac{P(\boldsymbol{n_k} | Q_k^{\ddagger}=1, R=r, \boldsymbol{a_k}, k)}{P(\boldsymbol{n_k} | Q_k=1, R=r, \boldsymbol{a_k}, k)} \times \frac{P(\boldsymbol{a_k} | Q_k^{\ddagger}=1, R=r, k)}{P(\boldsymbol{a_k} | Q_k=1, R=r, k)}$$

*Remark 1.*

For intuition about the first equality, note that if we could directly sample the broader source population $\mathbb{S}_{k,0}$ defined by $Q_k^{\ddagger} = 1$ alone with simple random sampling, the resulting sample $\mathbb{S}_{k,1}^{\mathcal{D}2}$ would have the joint distribution $f_{k,*}^{\mathcal{D}2}$. The sampling fraction would equal $\mathbb{N}_{k,1}^{*}/\mathbb{N}_{k,0}^{*} = \mathbb{N}_{k,1}^{*}(\boldsymbol{v})/\mathbb{N}_{k,0}^{*}(\boldsymbol{v})$, where $\mathbb{N}_{k,\ell}^{*}$ and $\mathbb{N}_{k,\ell}^{*}(\boldsymbol{v})$ are, respectively, the marginal and conditional sample sizes during stage $\ell$ at time $k$ when sampling the broader population.

We cannot sample all of $\mathbb{S}_{k,0}$ defined by $Q_k^{\ddagger} = 1$ alone. We can only sample $\mathbb{S}_{k,0}$ defined by $Q_k = 1$. To construct a sample that recovers a joint distribution $f_{k,1}^{\mathcal{D}2} = f_{k,*}^{\mathcal{D}2}$, we modify the sampling fraction to incorporate a correction factor $(\mathbb{N}_{k,0}^{*}(\boldsymbol{v})/\mathbb{N}_{k,0}(\boldsymbol{v}))$ to account for the fact that $\mathbb{N}_{k,0}^{*}(\boldsymbol{v}) \neq \mathbb{N}_{k,0}(\boldsymbol{v})$, where $\mathbb{N}_{k,0}(\boldsymbol{v})$ is the conditional sample size when sampling the population defined by $Q_k = 1$. The sampling fraction then reduces to $\mathbb{N}_{k,1}^{*}(\boldsymbol{v})/\mathbb{N}_{k,0}(\boldsymbol{v})$.

For intuition about the second equality, note that $\mathbb{N}_{k,1}^{*}(\boldsymbol{v})$ can be re-expressed as $\mathbb{N}_{k,1}^{*} \times f_{k,*}^{\mathcal{D}2}$, which is equivalent to $\mathbb{N}_{k,1} \times f_{k,1}^{\mathcal{D}2}$ because $\mathbb{N}_{k,1}^{*}$ is chosen by the investigator (so we can drop the superscript) and because $f_{k,*}^{\mathcal{D}2} = f_{k,1}^{\mathcal{D}2}$.

$$\alpha_{k,2}^{\mathcal{D}2}(r, \boldsymbol{a_k})$$
$$= \frac{\mathbb{N}_{k,2}(r, \boldsymbol{a_k})}{\mathbb{N}_{k,1}(r, \boldsymbol{a_k})}$$
$$= \frac{\mathbb{N}_{k,2}(r)}{\mathbb{N}_{k,1}(r)} \times \frac{f_{k,2}^{\mathcal{D}2}}{f_{k,1}^{\mathcal{D}2}}$$
$$= \frac{\mathbb{N}_{k,2}(r)}{\mathbb{N}_{k,1}(r)} \times \frac{P(\boldsymbol{a_k} | Q_k^{\ddagger}=1, T=1, k)}{P(\boldsymbol{a_k} | Q_k^{\ddagger}=1, R=r, k)}$$

G-Computation Estimator

$$
\begin{aligned}
&\mu_k^{\mathcal{D}2}(r)\\
&= \textstyle\sum_{\boldsymbol{a},\boldsymbol{n}} E\left(Y_{k+J} \middle| Q_k = 1, R = r, \boldsymbol{n_k}, \boldsymbol{a_k}, k\right) P\left(\boldsymbol{n_k} \middle| Q_k^{\ddagger} = 1, R = r, \boldsymbol{a_k}, k\right) P\left(\boldsymbol{a_k} \middle| Q_k^{\ddagger} = 1, T = 1, k\right)\\
&= E_{\boldsymbol{a}}\left\llbracket E_{\boldsymbol{n}}\left(E\left[Y_{k+J} \middle| Q_k = 1, R = r, \boldsymbol{n_k}, \boldsymbol{a_k}, k\right] \middle| Q_k^{\ddagger} = 1, R = r, \boldsymbol{a_k}, k\right) | Q_k^{\ddagger} = 1, T = 1, k \right\rrbracket
\end{aligned}
$$

where $E_{\boldsymbol{n}}(\cdot)$ is over $P\left(\boldsymbol{n_k} \middle| Q_k^{\ddagger} = 1, R = r, \boldsymbol{a_k}, k\right)$ and $E_{\boldsymbol{a}}\llbracket\cdot\rrbracket$ is over $P\left(\boldsymbol{a_k} \middle| Q_k^{\ddagger} = 1, T = 1, k\right)$

By definition of an iterated expectation

Weighting Estimator

$$
\begin{aligned}
&\mu_k^{\mathcal{D}2}(r)\\
&= \textstyle\sum_{\boldsymbol{a},\boldsymbol{n}} E\left(Y_{k+J} \middle| Q_k = 1, R = r, \boldsymbol{n_k}, \boldsymbol{a_k}, k\right) P\left(\boldsymbol{n_k} \middle| Q_k^{\ddagger} = 1, R = r, \boldsymbol{a_k}, k\right) P\left(\boldsymbol{a_k} \middle| Q_k^{\ddagger} = 1, T = 1, k\right)\\
&= \textstyle\sum_{\boldsymbol{a},\boldsymbol{n}} E\left(Y_{k+J} \middle| Q_k = 1, R = r, \boldsymbol{n_k}, \boldsymbol{a_k}, k\right) P\left(\boldsymbol{n_k} | Q_k = 1, R = r, \boldsymbol{a_k}, k\right) P\left(\boldsymbol{a_k} | Q_k = 1, R = r, k\right)\\
&\quad \times \frac{P(\boldsymbol{n_k} | Q_k^{\ddagger}=1, R=r, \boldsymbol{a_k}, k)}{P(\boldsymbol{n_k} | Q_k=1, R=r, \boldsymbol{a_k}, k)} \times \frac{P(\boldsymbol{a_k} | Q_k^{\ddagger}=1, T=1, k)}{P(\boldsymbol{a_k} | Q_k=1, R=r, k)}\\
&= \textstyle\sum_{\boldsymbol{a},\boldsymbol{n}} E\left(Y_{k+J} \middle| Q_k = 1, R = r, \boldsymbol{n_k}, \boldsymbol{a_k}, k\right) P\left(\boldsymbol{n_k} | Q_k = 1, R = r, \boldsymbol{a_k}, k\right) P\left(\boldsymbol{a_k} | Q_k = 1, R = r, k\right)\\
&\quad \times \frac{P(Q_k^{\dagger}=1 | Q_k^{\ddagger}=1, R=r, \boldsymbol{a_k}, k)}{P(Q_k^{\dagger}=1 | Q_k^{\ddagger}=1, R=r, \boldsymbol{n_k}, \boldsymbol{a_k}, k)} \times \frac{P(Q_k^{\dagger}=1 | Q_k^{\ddagger}=1, R=r, k)}{P(Q_k^{\dagger}=1 | Q_k^{\ddagger}=1, R=r, \boldsymbol{a_k}, k)} \times \frac{P(T=1 | Q_k^{\ddagger}=1, \boldsymbol{a_k}, k)}{P(R=r | Q_k^{\ddagger}=1, \boldsymbol{a_k}, k)} \times \frac{P(R=r | Q_k^{\ddagger}=1, k)}{P(T=1 | Q_k^{\ddagger}=1, k)}\\
&= \textstyle\sum_{\boldsymbol{a},\boldsymbol{n}} E\left(Y_{k+J} \middle| Q_k = 1, R = r, \boldsymbol{n_k}, \boldsymbol{a_k}, k\right) P\left(\boldsymbol{n_k} | Q_k = 1, R = r, \boldsymbol{a_k}, k\right) P\left(\boldsymbol{a_k} | Q_k = 1, R = r, k\right)\\
&\quad \times \frac{P(Q_k^{\dagger}=1 | Q_k^{\ddagger}=1, R=r, k)}{P(Q_k^{\dagger}=1 | Q_k^{\ddagger}=1, R=r, \boldsymbol{n_k}, \boldsymbol{a_k}, k)} \times \frac{P(T=1 | Q_k^{\ddagger}=1, \boldsymbol{a_k}, k)}{P(R=r | Q_k^{\ddagger}=1, \boldsymbol{a_k}, k)} \times \frac{P(R=r | Q_k^{\ddagger}=1, k)}{P(T=1 | Q_k^{\ddagger}=1, k)}\\
&= E[Y_{k+J} \times \omega_{r,Q_k=1,k,}^{\mathcal{D}2} | Q_k = 1, R = r, k]
\end{aligned}
$$

where

$$
\omega_{r,Q_k=1,k}^{\mathcal{D}2} = \frac{P(Q_k^{\dagger}=1 | Q_k^{\ddagger}=1, R=r, k)}{P(Q_k^{\dagger}=1 | Q_k^{\ddagger}=1, R=r, \boldsymbol{n_k}, \boldsymbol{a_k}, k)} \times \frac{P(T=1 | Q_k^{\ddagger}=1, \boldsymbol{a_k}, k)}{P(R=r | Q_k^{\ddagger}=1, \boldsymbol{a_k}, k)} \times \frac{P(R=r | Q_k^{\ddagger}=1, k)}{P(T=1 | Q_k^{\ddagger}=1, k)}
$$

Aggregated G-Computation Estimator (over calendar time $k$)

$$
\begin{aligned}
&\tau^{\mathcal{D}2}(r) = \textstyle\sum_k \mu_k^{\mathcal{D}2}(r)\, P(k | Q_k^{\ddagger} = 1, T = 1)\\
&= \textstyle\sum_{\boldsymbol{a},\boldsymbol{n},k} E\left(Y_{k+J} \middle| Q_k = 1, R = r, \boldsymbol{n_k}, \boldsymbol{a_k}, k\right) P\left(\boldsymbol{n_k} \middle| Q_k^{\ddagger} = 1, R = r, \boldsymbol{a_k}, k\right) P\left(\boldsymbol{a_k} \middle| Q_k^{\ddagger} = 1, T = 1, k\right) P(k | Q_k^{\ddagger} = 1, T = 1)\\
&= \textstyle\sum_{\boldsymbol{a},\boldsymbol{n},k} E\left(Y_{k+J} \middle| Q_k = 1, R = r, \boldsymbol{n_k}, \boldsymbol{a_k}, k\right) P\left(\boldsymbol{n_k} \middle| Q_k^{\ddagger} = 1, R = r, \boldsymbol{a_k}, k\right) P\left(\boldsymbol{a_k}, k \middle| Q_k^{\ddagger} = 1, T = 1\right)\\
&= E_{\boldsymbol{a},k}\left\llbracket E_{\boldsymbol{n}}\left(E\left[Y_{k+J} \middle| Q_k = 1, R = r, \boldsymbol{n_k}, \boldsymbol{a_k}, k\right] \middle| Q_k^{\ddagger} = 1, R = r, \boldsymbol{a_k}, k\right) | Q_k^{\ddagger} = 1, T = 1 \right\rrbracket
\end{aligned}
$$

where $E_{\boldsymbol{n}}(\cdot)$ is over $P\left(\boldsymbol{n_k} \middle| Q_k^{\ddagger} = 1, R = r, \boldsymbol{a_k}, k\right)$ and $E_{\boldsymbol{a},k}\llbracket\cdot\rrbracket$ is over $P\left(\boldsymbol{a_k}, k \middle| Q_k^{\ddagger} = 1, T = 1\right)$

By definition of an iterated expectation

Aggregated Weighting Estimator (over calendar time $k$)

$$
\begin{aligned}
&\tau^{\mathcal{D}2}(r) = \textstyle\sum_k \mu_k^{\mathcal{D}2}(r)\, P(k | Q_k^{\ddagger} = 1, T = 1)\\
&= \textstyle\sum_{\boldsymbol{a},\boldsymbol{n},k} E\left(Y_{k+J} \middle| Q_k = 1, R = r, \boldsymbol{n_k}, \boldsymbol{a_k}, k\right) P\left(\boldsymbol{n_k} \middle| Q_k^{\ddagger} = 1, R = r, \boldsymbol{a_k}, k\right) P\left(\boldsymbol{a_k} \middle| Q_k^{\ddagger} = 1, T = 1, k\right) P(k | Q_k^{\ddagger} = 1, T = 1)\\
&= \textstyle\sum_{\boldsymbol{a},\boldsymbol{n},k} E\left(Y_{k+J} \middle| Q_k = 1, R = r, \boldsymbol{n_k}, \boldsymbol{a_k}, k\right) P\left(\boldsymbol{n_k} | Q_k = 1, R = r, \boldsymbol{a_k}, k\right) P\left(\boldsymbol{a_k}, k | Q_k = 1, R = r\right)\\
&\quad \times \frac{P(\boldsymbol{n_k} | Q_k^{\ddagger}=1, R=r, \boldsymbol{a_k}, k)}{P(\boldsymbol{n_k} | Q_k=1, R=r, \boldsymbol{a_k}, k)} \times \frac{P(\boldsymbol{a_k}, k | Q_k^{\ddagger}=1, T=1)}{P(\boldsymbol{a_k}, k | Q_k=1, R=r)}\\
&= \textstyle\sum_{\boldsymbol{a},\boldsymbol{n}} E\left(Y_{k+J} \middle| Q_k = 1, R = r, \boldsymbol{n_k}, \boldsymbol{a_k}, k\right) P\left(\boldsymbol{n_k} | Q_k = 1, R = r, \boldsymbol{a_k}, k\right) P\left(\boldsymbol{a_k}, k | Q_k = 1, R = r\right)\\
&\quad \times \frac{P(Q_k^{\dagger}=1 | Q_k^{\ddagger}=1, R=r)}{P(Q_k^{\dagger}=1 | Q_k^{\ddagger}=1, R=r, \boldsymbol{n_k}, \boldsymbol{a_k}, k)} \times \frac{P(T=1 | Q_k^{\ddagger}=1, \boldsymbol{a_k}, k)}{P(R=r | Q_k^{\ddagger}=1, \boldsymbol{a_k}, k)} \times \frac{P(R=r | Q_k^{\ddagger}=1)}{P(T=1 | Q_k^{\ddagger}=1)}\\
&= E[Y_{k+J} \times \omega_{r,Q_k=1,k,}^{\mathcal{D}2-pool} | Q_k = 1, R = r]
\end{aligned}
$$

where

$$
\omega_{r,Q_k=1,k}^{\mathcal{D}2-pool} = \frac{P(Q_k^{\dagger}=1 | Q_k^{\ddagger}=1, R=r)}{P(Q_k^{\dagger}=1 | Q_k^{\ddagger}=1, R=r, \boldsymbol{n_k}, \boldsymbol{a_k}, k)} \times \frac{P(T=1 | Q_k^{\ddagger}=1, \boldsymbol{a_k}, k)}{P(R=r | Q_k^{\ddagger}=1, \boldsymbol{a_k}, k)} \times \frac{P(R=r | Q_k^{\ddagger}=1)}{P(T=1 | Q_k^{\ddagger}=1)}
$$

*DESIGN 3*

<u>Joint distribution of $(Y_{k+J}, \boldsymbol{A_k}, \boldsymbol{N_k})$ given $Q_k^{\dagger} = 0, Q_k^{\ddagger} = 1$, and $R = r$ in the Different Eligible Source Population</u>

$$f_{k,**}^{\mathcal{D}3} = P\left(Y_{k+J} = y \middle| Q_k^{\dagger} = 0, Q_k^{\ddagger} = 1, R = r, \boldsymbol{n_k}, \boldsymbol{a_k}, k\right) P(\boldsymbol{n_k} | Q_k^{\dagger} = 0, Q_k^{\ddagger} = 1, R = r, \boldsymbol{a_k}, k) P\left(\boldsymbol{a_k} \middle| Q_k^{\dagger} = 0, Q_k^{\ddagger} = 1, R = r, k\right)$$

<u>Joint distribution of $(Y_{k+J}, \boldsymbol{A_k}, \boldsymbol{N_k})$ given $Q_k = 1$, and $R = r$ in the Available Eligible Source Population</u>

$$f_{k,0}^{\mathcal{D}3} = P\left(Y_{k+J} = y \middle| Q_k^{\dagger} = 0, Q_k^{\ddagger} = 1, R = r, \boldsymbol{n_k}, \boldsymbol{a_k}, k\right) P(\boldsymbol{n_k} | Q_k = 1, R = r, \boldsymbol{a_k}, k) P(\boldsymbol{a_k} | Q_k = 1, R = r, k)$$

<u>Joint distribution of $(Y_{k+J}, \boldsymbol{A_k}, \boldsymbol{N_k})$ given $Q_k = 1$, and $R = r$ after Stage One of Sampling</u>

$$f_{k,1}^{\mathcal{D}3} = P\left(Y_{k+J} = y \middle| Q_k^{\dagger} = 0, Q_k^{\ddagger} = 1, R = r, \boldsymbol{n_k}, \boldsymbol{a_k}, k\right) P(\boldsymbol{n_k} | Q_k^{\dagger} = 0, Q_k^{\ddagger} = 1, R = r, \boldsymbol{a_k}, k) P\left(\boldsymbol{a_k} \middle| Q_k^{\dagger} = 0, Q_k^{\ddagger} = 1, R = r, k\right)$$

<u>Joint distribution of $(Y_{k+J}, \boldsymbol{A_k}, \boldsymbol{N_k})$ given $Q_k = 1$, and $R = r$ after Stage Two of Sampling</u>

$$f_{k,2}^{\mathcal{D}3} = P\left(Y_{k+J} = y \middle| Q_k^{\dagger} = 0, Q_k^{\ddagger} = 1, R = r, \boldsymbol{n_k}, \boldsymbol{a_k}, k\right) P(\boldsymbol{n_k} | Q_k^{\dagger} = 0, Q_k^{\ddagger} = 1, R = r, \boldsymbol{a_k}, k) P\left(\boldsymbol{a_k} \middle| Q_k^{\dagger} = 0, Q_k^{\ddagger} = 1, T = 1, k\right)$$

These joint distributions hold by design of the target study and under a version of overlap (3) (see footnote 17), innocuous sampling (4), independence (11) and positivity (15) in the main text.

<u>Mean of $Y_{k+J}$ given $R = r$ in the target study</u>

$$\mu_k^{\mathcal{D}3}(r)$$
$$= \sum_{\boldsymbol{a},\boldsymbol{n}} E\left(Y_{k+J} \middle| Q_k^{\dagger} = 0, Q_k^{\ddagger} = 1, R = r, \boldsymbol{n_k}, \boldsymbol{a_k}, k\right) P\left(\boldsymbol{n_k} \middle| Q_k^{\dagger} = 0, Q_k^{\ddagger} = 1, R = r, \boldsymbol{a_k}, k\right) P\left(\boldsymbol{a_k} \middle| Q_k^{\dagger} = 0, Q_k^{\ddagger} = 1, T = 1, k\right)$$
$$= \sum_{\boldsymbol{a},\boldsymbol{n}} E\left(Y_{k+J} \middle| Q_k = 1, R = r, \boldsymbol{n_k}, \boldsymbol{a_k}, k\right) P\left(\boldsymbol{n_k} \middle| Q_k^{\dagger} = 0, Q_k^{\ddagger} = 1, R = r, \boldsymbol{a_k}, k\right) P\left(\boldsymbol{a_k} \middle| Q_k^{\dagger} = 0, Q_k^{\ddagger} = 1, T = 1, k\right)$$

By definition of a mean, overlap (3) (footnote 17), innocuous sampling (4), independence (11) and positivity (15).

<u>Sampling Fractions</u>

$$\alpha_{k,1}^{\mathcal{D}3}(r, \boldsymbol{a_k}, \boldsymbol{n_k})$$
$$= \frac{\mathbb{N}_{k,1}^{*}(r, \boldsymbol{a_k}, \boldsymbol{n_k})}{\mathbb{N}_{k,0}(r, \boldsymbol{a_k}, \boldsymbol{n_k})}$$
$$= \frac{\mathbb{N}_{k,1}(r)}{\mathbb{N}_{k,0}(r)} \times \frac{f_{k,1}^{\mathcal{D}3}}{f_{k,0}^{\mathcal{D}3}}$$
$$= \frac{\mathbb{N}_{k,1}(r)}{\mathbb{N}_{k,0}(r)} \times \frac{P(\boldsymbol{n_k} | Q_k^{\dagger}=0, Q_k^{\ddagger}=1, R=r, \boldsymbol{a_k}, k)}{P(\boldsymbol{n_k} | Q_k=1, R=r, \boldsymbol{a_k}, k)} \times \frac{P(\boldsymbol{a_k} | Q_k^{\dagger}=0, Q_k^{\ddagger}=1, R=r, k)}{P(\boldsymbol{a_k} | Q_k=1, R=r, k)}$$

*Remark 2.*

For intuition about the first equality, note that if we could directly sample the different source population $\mathbb{S}_{k,0}$ defined by $Q_k^{\dagger} = 0$ and $Q_k^{\ddagger} = 1$ with simple random sampling, the resulting sample $\mathbb{S}_{k,1}^{\mathcal{D}3}$ would have the joint distribution $f_{k,**}^{\mathcal{D}3}$. The sampling fraction would equal $\mathbb{N}_{k,1}^{**}/\mathbb{N}_{k,0}^{**} = \mathbb{N}_{k,1}^{**}(\boldsymbol{v})/\mathbb{N}_{k,0}^{**}(\boldsymbol{v})$, where $\mathbb{N}_{k,\ell}^{**}$ and $\mathbb{N}_{k,\ell}^{**}(\boldsymbol{v})$ are, respectively, the marginal and conditional sample sizes in stage $\ell$ at time $k$ when sampling the different population.

We cannot sample $\mathbb{S}_{k,0}$ defined by $Q_k^{\dagger} = 0$ and $Q_k^{\ddagger} = 1$. We can only sample $\mathbb{S}_{k,0}$ defined by $Q_k = 1$. To construct a sample that recovers a joint distribution $f_{k,1}^{\mathcal{D}3} = f_{k,**}^{\mathcal{D}3}$, we modify the sampling fraction to incorporate a correction factor $(\mathbb{N}_{k,0}^{**}(\boldsymbol{v})/\mathbb{N}_{k,0}(\boldsymbol{v}))$ to account for the fact that $\mathbb{N}_{k,0}^{**}(\boldsymbol{v}) \neq \mathbb{N}_{k,0}(\boldsymbol{v})$, where $\mathbb{N}_{k,0}(\boldsymbol{v})$ is the conditional sample size when sampling the population defined by $Q_k = 1$. The sampling fraction reduces to $\mathbb{N}_{k,1}^{**}(\boldsymbol{v})/\mathbb{N}_{k,0}(\boldsymbol{v})$.

For intuition about the second equality, note that $\mathbb{N}_{k,1}^{**}(\boldsymbol{v})$ can be re-expressed as $\mathbb{N}_{k,1}^{**} \times f_{k,**}^{\mathcal{D}3}$, which is equivalent to $\mathbb{N}_{k,1} \times f_{k,1}^{\mathcal{D}2}$ because $\mathbb{N}_{k,1}^{**}$ is chosen by the investigator (so we can drop the superscript) and because $f_{k,**}^{\mathcal{D}3} = f_{k,1}^{\mathcal{D}3}$.

$$\alpha_{k,2}^{\mathcal{D}3}(r, \boldsymbol{a_k})$$
$$= \frac{\mathbb{N}_{k,2}(r, \boldsymbol{a_k})}{\mathbb{N}_{k,1}(r, \boldsymbol{a_k})}$$
$$= \frac{\mathbb{N}_{k,2}(r)}{\mathbb{N}_{k,1}(r)} \times \frac{f_{k,2}^{\mathcal{D}3}}{f_{k,1}^{\mathcal{D}3}}$$
$$= \frac{\mathbb{N}_{k,2}(r)}{\mathbb{N}_{k,1}(r)} \times \frac{P(\boldsymbol{a_k} | Q_k^{\dagger}=0, Q_k^{\ddagger}=1, T=1, k)}{P(\boldsymbol{a_k} | Q_k^{\dagger}=0, Q_k^{\ddagger}=1, R=r, k)}$$

<u>G-Computation Estimator</u>

$$\mu_k^{\mathcal{D}3}(r)$$

$$= \sum_{\boldsymbol{a},\boldsymbol{n}} E\left(Y_{k+J}\middle|Q_k = 1, R = r, \boldsymbol{n_k}, \boldsymbol{a_k}, k\right) P\left(\boldsymbol{n_k}\middle|Q_k^{\dagger} = 0, Q_k^{\ddagger} = 1, R = r, \boldsymbol{a_k}, k\right) P\left(\boldsymbol{a_k}\middle|Q_k^{\dagger} = 0, Q_k^{\ddagger} = 1, T = 1, k\right)$$

$$= E_{\boldsymbol{a}}\left[\!\left[ E_{\boldsymbol{n}}\left( E\left[Y_{k+J}\middle|Q_k = 1, R = r, \boldsymbol{n_k}, \boldsymbol{a_k}, k\right]\middle|Q_k^{\dagger} = 0, Q_k^{\ddagger} = 1, R = r, \boldsymbol{a_k}, k\right) \middle| Q_k^{\dagger} = 0, Q_k^{\ddagger} = 1, T = 1, k \right]\!\right]$$

where $E_{\boldsymbol{n}}(\cdot)$ is over $P\left(\boldsymbol{n_k}\middle|Q_k^{\dagger} = 0, Q_k^{\ddagger} = 1, R = r, \boldsymbol{a_k}, k\right)$ and $E_{\boldsymbol{a}}[\![\cdot]\!]$ is over $P\left(\boldsymbol{a_k}\middle|Q_k^{\dagger} = 0, Q_k^{\ddagger} = 1, T = 1, k\right)$

By definition of an iterated expectation

<u>Weighting Estimator</u>

$$\mu_k^{\mathcal{D}2}(r)$$

$$= \sum_{\boldsymbol{a},\boldsymbol{n}} E\left(Y_{k+J}\middle|Q_k = 1, R = r, \boldsymbol{n_k}, \boldsymbol{a_k}, k\right) P\left(\boldsymbol{n_k}\middle|Q_k^{\dagger} = 0, Q_k^{\ddagger} = 1, R = r, \boldsymbol{a_k}, k\right) P\left(\boldsymbol{a_k}\middle|Q_k^{\dagger} = 0, Q_k^{\ddagger} = 1, T = 1, k\right)$$

$$= \sum_{\boldsymbol{a},\boldsymbol{n}} E\left(Y_{k+J}\middle|Q_k = 1, R = r, \boldsymbol{n_k}, \boldsymbol{a_k}, k\right) P(\boldsymbol{n_k}|Q_k = 1, R = r, \boldsymbol{a_k}, k) P(\boldsymbol{a_k}|Q_k = 1, R = r, k)$$

$$\times \frac{P(\boldsymbol{n_k}|Q_k^{\dagger} = 0, Q_k^{\ddagger} = 1, R = r, \boldsymbol{a_k}, k)}{P(\boldsymbol{n_k}|Q_k = 1, R = r, \boldsymbol{a_k}, k)} \times \frac{P(\boldsymbol{a_k}|Q_k^{\dagger} = 0, Q_k^{\ddagger} = 1, T = 1, k)}{P(\boldsymbol{a_k}|Q_k = 1, R = r, k)}$$

$$= \sum_{\boldsymbol{a},\boldsymbol{n}} E\left(Y_{k+J}\middle|Q_k = 1, R = r, \boldsymbol{n_k}, \boldsymbol{a_k}, k\right) P(\boldsymbol{n_k}|Q_k = 1, R = r, \boldsymbol{a_k}, k) P(\boldsymbol{a_k}|Q_k = 1, R = r, k)$$

$$\times \frac{P(Q_k^{\dagger}=0|Q_k^{\ddagger}=1,R=r,\boldsymbol{n_k},\boldsymbol{a_k},k)}{P(Q_k^{\dagger}=1|Q_k^{\ddagger}=1,R=r,\boldsymbol{n_k},\boldsymbol{a_k},k)} \times \frac{P\left(Q_k^{\dagger}=1,Q_k^{\ddagger}=1,R=r,k\right)}{P\left(Q_k^{\dagger}=0,Q_k^{\ddagger}=1,R=r,k\right)} \times \frac{P\left(\boldsymbol{a_k}|Q_k^{\dagger}=0,Q_k^{\ddagger}=1,T=1,k\right)}{P\left(\boldsymbol{a_k}|Q_k^{\dagger}=0,Q_k^{\ddagger}=1,R=r,k\right)}$$

$$= \sum_{\boldsymbol{a},\boldsymbol{n}} E\left(Y_{k+J}\middle|Q_k = 1, R = r, \boldsymbol{n_k}, \boldsymbol{a_k}, k\right) P(\boldsymbol{n_k}|Q_k = 1, R = r, \boldsymbol{a_k}, k) P(\boldsymbol{a_k}|Q_k = 1, R = r, k)$$

$$\times \frac{P(Q_k^{\dagger}=0|Q_k^{\ddagger}=1,R=r,\boldsymbol{n_k},\boldsymbol{a_k},k)}{P(Q_k^{\dagger}=1|Q_k^{\ddagger}=1,R=r,\boldsymbol{n_k},\boldsymbol{a_k},k)} \times \frac{P(Q_k^{\dagger}=1|Q_k^{\ddagger}=1,R=r,k)}{P(Q_k^{\dagger}=0|Q_k^{\ddagger}=1,R=r,k)} \times \frac{P(T=1|Q_k^{\dagger}=0,Q_k^{\ddagger}=1,\boldsymbol{a_k},k)}{P(R=r|Q_k^{\dagger}=0,Q_k^{\ddagger}=1,\boldsymbol{a_k},k)} \times \frac{P(R=r|Q_k^{\dagger}=0,Q_k^{\ddagger}=1,k)}{P(T=1|Q_k^{\dagger}=0,Q_k^{\ddagger}=1,k)}$$

$$= E[Y_{k+J} \times \omega_{r,Q_k=1,k,}^{\mathcal{D}3} | Q_k = 1, R = r, k]$$

where

$$\omega_{r,Q_k=1,k}^{\mathcal{D}3} = \frac{P(Q_k^{\dagger}=0|Q_k^{\ddagger}=1,R=r,\boldsymbol{n_k},\boldsymbol{a_k},k)}{P(Q_k^{\dagger}=1|Q_k^{\ddagger}=1,R=r,\boldsymbol{n_k},\boldsymbol{a_k},k)} \times \frac{P(Q_k^{\dagger}=1|Q_k^{\ddagger}=1,R=r,k)}{P(Q_k^{\dagger}=0|Q_k^{\ddagger}=1,R=r,k)} \times \frac{P(T=1|Q_k^{\dagger}=0,Q_k^{\ddagger}=1,\boldsymbol{a_k},k)}{P(R=r|Q_k^{\dagger}=0,Q_k^{\ddagger}=1,\boldsymbol{a_k},k)} \times \frac{P(R=r|Q_k^{\dagger}=0,Q_k^{\ddagger}=1,k)}{P(T=1|Q_k^{\dagger}=0,Q_k^{\ddagger}=1,k)}$$

<u>Aggregated G-Computation Estimator (over calendar time $k$)</u>

$$\tau^{\mathcal{D}2}(r) = \sum_k \mu_k^{\mathcal{D}2}(r)\, P(k|Q_k^{\dagger} = 0, Q_k^{\ddagger} = 1, T = 1)$$

$$= \sum_{\boldsymbol{a},\boldsymbol{n},k} E\left(Y_{k+J}\middle|Q_k = 1, R = r, \boldsymbol{n_k}, \boldsymbol{a_k}, k\right) P\left(\boldsymbol{n_k}\middle|Q_k^{\dagger} = 0, Q_k^{\ddagger} = 1, R = r, \boldsymbol{a_k}, k\right) P\left(\boldsymbol{a_k}\middle|Q_k^{\dagger} = 0, Q_k^{\ddagger} = 1, T = 1, k\right)$$

$$\times P(k|Q_k^{\dagger} = 0, Q_k^{\ddagger} = 1, T = 1)$$

$$= \sum_{\boldsymbol{a},\boldsymbol{n},k} E\left(Y_{k+J}\middle|Q_k = 1, R = r, \boldsymbol{n_k}, \boldsymbol{a_k}, k\right) P\left(\boldsymbol{n_k}\middle|Q_k^{\dagger} = 0, Q_k^{\ddagger} = 1, R = r, \boldsymbol{a_k}, k\right) P\left(\boldsymbol{a_k}, k\middle|Q_k^{\dagger} = 0, Q_k^{\ddagger} = 1, T = 1\right)$$

$$= E_{\boldsymbol{a},k}\left[\!\left[ E_{\boldsymbol{n}}\left( E\left[Y_{k+J}\middle|Q_k = 1, R = r, \boldsymbol{n_k}, \boldsymbol{a_k}, k\right]\middle|Q_k^{\dagger} = 0, Q_k^{\ddagger} = 1, R = r, \boldsymbol{a_k}, k\right) \middle| Q_k^{\dagger} = 0, Q_k^{\ddagger} = 1, T = 1 \right]\!\right]$$

where $E_{\boldsymbol{n}}(\cdot)$ is over $P\left(\boldsymbol{n_k}\middle|Q_k^{\dagger} = 0, Q_k^{\ddagger} = 1, T = 1, \boldsymbol{a_k}, k\right)$ and $E_{\boldsymbol{a},k}[\![\cdot]\!]$ is over $P\left(\boldsymbol{a_k}, k\middle|Q_k^{\dagger} = 0, Q_k^{\ddagger} = 1, T = 1\right)$

By definition of an iterated expectation

<u>Aggregated Weighting Estimator (over calendar time $k$)</u>

$$\tau^{\mathcal{D}3}(r) = \sum_k \mu_k^{\mathcal{D}3}(r)\, P(k|Q_k^{\dagger} = 0, Q_k^{\ddagger} = 1, T = 1)$$

$$= \sum_{\boldsymbol{a},\boldsymbol{n},k} E\left(Y_{k+J}\middle|Q_k = 1, R = r, \boldsymbol{n_k}, \boldsymbol{a_k}, k\right) P\left(\boldsymbol{n_k}\middle|Q_k^{\dagger} = 0, Q_k^{\ddagger} = 1, R = r, \boldsymbol{a_k}, k\right) P\left(\boldsymbol{a_k}\middle|Q_k^{\dagger} = 0, Q_k^{\ddagger} = 1, T = 1, k\right)$$

$$\times P(k|Q_k^{\dagger} = 0, Q_k^{\ddagger} = 1, T = 1)$$

$$= \sum_{\boldsymbol{a},\boldsymbol{n},k} E\left(Y_{k+J}\middle|Q_k = 1, R = r, \boldsymbol{n_k}, \boldsymbol{a_k}, k\right) P(\boldsymbol{n_k}|Q_k = 1, R = r, \boldsymbol{a_k}, k) P(\boldsymbol{a_k}, k|Q_k = 1, R = r)$$

$$\times \frac{P(\boldsymbol{n_k}|Q_k^{\dagger}=0,Q_k^{\ddagger}=1,R=r,\boldsymbol{a_k},k)}{P(\boldsymbol{n_k}|Q_k=1,R=r,\boldsymbol{a_k},k)} \times \frac{P(\boldsymbol{a_k},k|Q_k^{\dagger}=0,Q_k^{\ddagger}=1,T=1)}{P(\boldsymbol{a_k},k|Q_k=1,R=r)}$$

$$= \sum_{\boldsymbol{a},\boldsymbol{n},k} E\left(Y_{k+J}\middle|Q_k = 1, R = r, \boldsymbol{n_k}, \boldsymbol{a_k}, k\right) P(\boldsymbol{n_k}|Q_k = 1, R = r, \boldsymbol{a_k}, k) P(\boldsymbol{a_k}, k|Q_k = 1, R = r)$$

$$\times \frac{P(Q_k^{\dagger}=0|Q_k^{\ddagger}=1,R=r,\boldsymbol{n_k},\boldsymbol{a_k},k)}{P(Q_k^{\dagger}=1|Q_k^{\ddagger}=1,R=r,\boldsymbol{n_k},\boldsymbol{a_k},k)} \times \frac{P(Q_k^{\dagger}=1|Q_k^{\ddagger}=1,R=r)}{P(Q_k^{\dagger}=0|Q_k^{\ddagger}=1,R=r)} \times \frac{P(T=1|Q_k^{\dagger}=0,Q_k^{\ddagger}=1,\boldsymbol{a_k},k)}{P(R=r|Q_k^{\dagger}=0,Q_k^{\ddagger}=1,\boldsymbol{a_k},k)} \times \frac{P(R=r|Q_k^{\dagger}=0,Q_k^{\ddagger}=1)}{P(T=1|Q_k^{\dagger}=0,Q_k^{\ddagger}=1)}$$

$$= E[Y_{k+J} \times \omega_{r,Q_k=1,k,}^{\mathcal{D}3-pool} | Q_k = 1, R = r]$$

where

$$\omega_{r,Q_k=1,k}^{\mathcal{D}3-pool} = \frac{P(Q_k^{\dagger}=0|Q_k^{\ddagger}=1,R=r,\boldsymbol{n_k},\boldsymbol{a_k},k)}{P(Q_k^{\dagger}=1|Q_k^{\ddagger}=1,R=r,\boldsymbol{n_k},\boldsymbol{a_k},k)} \times \frac{P(Q_k^{\dagger}=1|Q_k^{\ddagger}=1,R=r)}{P(Q_k^{\dagger}=0|Q_k^{\ddagger}=1,R=r)} \times \frac{P(T=1|Q_k^{\dagger}=0,Q_k^{\ddagger}=1,\boldsymbol{a_k},k)}{P(R=r|Q_k^{\dagger}=0,Q_k^{\ddagger}=1,\boldsymbol{a_k},k)} \times \frac{P(R=r|Q_k^{\dagger}=0,Q_k^{\ddagger}=1)}{P(T=1|Q_k^{\dagger}=0,Q_k^{\ddagger}=1)}$$

*DESIGN 4*

<u>Lemma 1.</u>

$$P\left(\boldsymbol{n_k} \middle| Q_k^{G_k} = 1, R = r, \boldsymbol{a_k}, k\right)$$

$$= P\left(\boldsymbol{n_k} \middle| Q_k^{\wr\,G_k} = 1, Q_k^{\dagger\,G_k} = 1, Q_k^{\ddagger} = 1, R = r, \boldsymbol{a_k}, k\right)$$

$$= \frac{P(Q_k^{\wr\,G_k}=1|Q_k^{\dagger\,G_k}=1,Q^{\ddagger}=1,R=r,\boldsymbol{n_k},\boldsymbol{a_k},k)P(Q_k^{\dagger\,G_k}=1|Q_k^{\ddagger}=1,R=r,\boldsymbol{n_k},\boldsymbol{a_k},k)P(Q_k^{\ddagger}=1,R=r,\boldsymbol{n_k},\boldsymbol{a_k},k)}{\sum_{\boldsymbol{n}} P(Q_k^{\wr\,G_k}=1|Q_k^{\dagger\,G_k}=1,Q_k^{\ddagger}=1,R=r,\boldsymbol{n_k},\boldsymbol{a_k},k)P(Q_k^{\dagger\,G_k}=1|Q_k^{\ddagger}=1,R=r,\boldsymbol{n_k},\boldsymbol{a_k},k)P(Q_k^{\ddagger}=1,R=r,\boldsymbol{n_k},\boldsymbol{a_k},k)}$$

$$= \frac{P(Q_k^{\wr\,G_k}=1|Q_k^{\ddagger}=1,R=r,\boldsymbol{n_k},\boldsymbol{a_k},k)P(Q_k^{\dagger\,G_k}=1|Q_k^{\ddagger}=1,R=r,\boldsymbol{n_k},\boldsymbol{a_k},k)P(Q_k^{\ddagger}=1,R=r,\boldsymbol{n_k},\boldsymbol{a_k},k)}{\sum_{\boldsymbol{n}} P(Q_k^{\wr\,G_k}=1|Q_k^{\ddagger}=1,R=r,\boldsymbol{n_k},\boldsymbol{a_k},k)P(Q_k^{\dagger\,G_k}=1|Q_k^{\ddagger}=1,R=r,\boldsymbol{n_k},\boldsymbol{a_k},k)P(Q_k^{\ddagger}=1,R=r,\boldsymbol{n_k},\boldsymbol{a_k},k)}$$

$$= \frac{P(Q_k^{\wr\,G_k}=1|Q_k^{\dagger}=1,Q_k^{\ddagger}=1,R=r,\boldsymbol{n_k},\boldsymbol{a_k},k)P(Q_k^{\dagger\,G_k}=1|Q_k^{\ddagger}=1,R=r,\boldsymbol{n_k},\boldsymbol{a_k},k)P(Q_k^{\ddagger}=1,R=r,\boldsymbol{n_k},\boldsymbol{a_k},k)}{\sum_{\boldsymbol{n}} P(Q_k^{\wr\,G_k}=1|Q_k^{\dagger}=1,Q_k^{\ddagger}=1,R=r,\boldsymbol{n_k},\boldsymbol{a_k},k)P(Q_k^{\dagger\,G_k}=1|Q_k^{\ddagger}=1,R=r,\boldsymbol{n_k},\boldsymbol{a_k},k)P(Q_k^{\ddagger}=1,R=r,\boldsymbol{n_k},\boldsymbol{a_k},k)}$$

$$= \frac{P(Q_k^{\wr}=1|Q_k^{\dagger}=1,Q_k^{\ddagger}=1,R=r,\boldsymbol{n_k},\boldsymbol{a_k},k)P(Q_k^{\dagger\,G_k}=1|Q_k^{\ddagger}=1,R=r,\boldsymbol{n_k},\boldsymbol{a_k},k)P(Q_k^{\ddagger}=1,R=r,\boldsymbol{n_k},\boldsymbol{a_k},k)}{\sum_{\boldsymbol{n}} P(Q_k^{\wr}=1|Q_k^{\dagger}=1,Q_k^{\ddagger}=1,R=r,\boldsymbol{n_k},\boldsymbol{a_k},k)P(Q_k^{\dagger\,G_k}=1|Q_k^{\ddagger}=1,R=r,\boldsymbol{n_k},\boldsymbol{a_k},k)P(Q_k^{\ddagger}=1,R=r,\boldsymbol{n_k},\boldsymbol{a_k},k)}$$

$$= \frac{P(Q_k^{\wr}=1|Q_k^{\dagger}=1,Q_k^{\ddagger}=1,R=r,\boldsymbol{n_k},\boldsymbol{a_k},k)\mathscr{g}_k(\cdot)P(Q_k^{\ddagger}=1,R=r,\boldsymbol{n_k},\boldsymbol{a_k},k)}{\sum_{\boldsymbol{n}} P(Q_k^{\wr}=1|Q_k^{\dagger}=1,Q_k^{\ddagger}=1,R=r,\boldsymbol{n_k},\boldsymbol{a_k},k)\mathscr{g}_k(\cdot)P(Q_k^{\ddagger}=1,R=r,\boldsymbol{n_k},\boldsymbol{a_k},k)}$$

The first equality holds by definition of $Q_k$, the third and sixth by definition of $G_k$ as a random assignment of $\boldsymbol{W}^{\dagger}$ from the distribution $\mathscr{g}_k(\cdot)$ given some (but not all) of the elements in $(Q_k^{\ddagger} = 1, R = r, \boldsymbol{n_k}, \boldsymbol{a_k}, k)$ and definition of $Q_k^{\dagger} = I(G_k = \boldsymbol{w}^{\dagger} \in \mathbb{w}^{\dagger})$, the fourth by exchangeability (18) and positivity (12), and the fifth by consistency. Note that $\mathscr{g}_k(\cdot) = P(Q_k^{\dagger\,G_k} = 1 | Q_k^{\ddagger} = 1, R = r, \boldsymbol{n_k}, \boldsymbol{a_k}, k)$ the conditional distribution of $Q_k^{\dagger}$ under the intervention $G_k$.

<u>Lemma 2.</u>

For any arbitrarily defined subgroup $S = s$

$$P\left(\boldsymbol{a_k} \middle| Q_k^{G_k} = 1, S = s, k\right)$$

$$= P\left(\boldsymbol{a_k} \middle| Q_k^{\wr\,G_k} = 1, Q_k^{\dagger\,G_k} = 1, Q_k^{\ddagger} = 1, S = s, k\right)$$

$$= \frac{\sum_{r,\boldsymbol{n}} P(Q_k^{\wr\,G_k}=1|Q_k^{\dagger\,G_k}=1,Q_k^{\ddagger}=1,S=s,R=r,\boldsymbol{n_k},\boldsymbol{a_k},k)P(Q_k^{\dagger\,G_k}=1|Q_k^{\ddagger}=1,S=s,R=r,\boldsymbol{n_k},\boldsymbol{a_k},k)P(Q_k^{\ddagger}=1,S=s,R=r,\boldsymbol{n_k},\boldsymbol{a_k},k)}{\sum_{r,\boldsymbol{n},\boldsymbol{a}} P(Q_k^{\wr\,G_k}=1|Q_k^{\dagger\,G_k}=1,Q_k^{\ddagger}=1,S=s,R=r,\boldsymbol{n_k},\boldsymbol{a_k},k)P(Q_k^{\dagger\,G_k}=1|Q_k^{\ddagger}=1,S=s,R=r,\boldsymbol{n_k},\boldsymbol{a_k},k)P(Q_k^{\ddagger}=1,S=s,R=r,\boldsymbol{n_k},\boldsymbol{a_k},k)}$$

$$= \frac{\sum_{r,\boldsymbol{n}} P(Q_k^{\wr\,G_k}=1|Q_k^{\ddagger}=1,S=s,R=r,\boldsymbol{n_k},\boldsymbol{a_k},k)P(Q_k^{\dagger\,G_k}=1|Q_k^{\ddagger}=1,S=s,R=r,\boldsymbol{n_k},\boldsymbol{a_k},k)P(Q_k^{\ddagger}=1,S=s,R=r,\boldsymbol{n_k},\boldsymbol{a_k},k)}{\sum_{r,\boldsymbol{n},\boldsymbol{a}} P(Q_k^{\wr\,G_k}=1|Q_k^{\ddagger}=1,S=s,R=r,\boldsymbol{n_k},\boldsymbol{a_k},k)P(Q_k^{\dagger\,G_k}=1|Q_k^{\ddagger}=1,S=s,R=r,\boldsymbol{n_k},\boldsymbol{a_k},k)P(Q_k^{\ddagger}=1,S=s,R=r,\boldsymbol{n_k},\boldsymbol{a_k},k)}$$

$$= \frac{\sum_{r,\boldsymbol{n}} P(Q_k^{\wr\,G_k}=1|Q_k^{\dagger}=1,Q_k^{\ddagger}=1,S=s,R=r,\boldsymbol{n_k},\boldsymbol{a_k},k)P(Q_k^{\dagger\,G_k}=1|Q_k^{\ddagger}=1,S=s,R=r,\boldsymbol{n_k},\boldsymbol{a_k},k)P(Q_k^{\ddagger}=1,S=s,R=r,\boldsymbol{n_k},\boldsymbol{a_k},k)}{\sum_{r,\boldsymbol{n},\boldsymbol{a}} P(Q_k^{\wr\,G_k}=1|Q_k^{\dagger}=1,Q_k^{\ddagger}=1,S=s,R=r,\boldsymbol{n_k},\boldsymbol{a_k},k)P(Q_k^{\dagger\,G_k}=1|Q_k^{\ddagger}=1,S=s,R=r,\boldsymbol{n_k},\boldsymbol{a_k},k)P(Q_k^{\ddagger}=1,S=s,R=r,\boldsymbol{n_k},\boldsymbol{a_k},k)}$$

$$= \frac{\sum_{r,\boldsymbol{n}} P(Q_k^{\wr}=1|Q_k^{\dagger}=1,Q_k^{\ddagger}=1,S=s,R=r,\boldsymbol{n_k},\boldsymbol{a_k},k)P(Q_k^{\dagger\,G_k}=1|Q_k^{\ddagger}=1,S=s,R=r,\boldsymbol{n_k},\boldsymbol{a_k},k)P(Q_k^{\ddagger}=1,S=s,R=r,\boldsymbol{n_k},\boldsymbol{a_k},k)}{\sum_{r,\boldsymbol{n},\boldsymbol{a}} P(Q_k^{\wr}=1|Q_k^{\dagger}=1,Q_k^{\ddagger}=1,S=s,R=r,,\boldsymbol{n_k},\boldsymbol{a_k},k)P(Q_k^{\dagger\,G_k}=1|Q_k^{\ddagger}=1,S=s,R=r,\boldsymbol{n_k},\boldsymbol{a_k},k)P(Q_k^{\ddagger}=1,S=s,R=r,\boldsymbol{n_k},\boldsymbol{a_k},k)}$$

$$= \frac{\sum_{r,\boldsymbol{n}} P(Q_k^{\wr}=1|Q_k^{\dagger}=1,Q_k^{\ddagger}=1,S=s,R=r,\boldsymbol{n_k},\boldsymbol{a_k},k)\mathscr{g}_k(\cdot)P(Q_k^{\ddagger}=1,S=s,R=r,\boldsymbol{n_k},\boldsymbol{a_k},k)}{\sum_{r,\boldsymbol{n},\boldsymbol{a}} P(Q_k^{\wr}=1|Q_k^{\dagger}=1,Q_k^{\ddagger}=1,S=s,R=r,\boldsymbol{n_k},\boldsymbol{a_k},k)\mathscr{g}_k(\cdot)P(Q^{\ddagger}=1,S=s,R=r,\boldsymbol{n_k},\boldsymbol{a_k},k)}$$

$$= \frac{\sum_{\boldsymbol{n}} P(Q_k^{\wr}=1|Q_k^{\dagger}=1,Q_k^{\ddagger}=1,S=s,\boldsymbol{n_k},\boldsymbol{a_k},k)\sum_r\{\mathscr{g}_k(\cdot)P(r|Q^{\ddagger}=1,S=s,\boldsymbol{n_k},\boldsymbol{a_k},k)\}P(Q_k^{\ddagger}=1,S=s,\boldsymbol{n_k},\boldsymbol{a_k},k)}{\sum_{\boldsymbol{n},\boldsymbol{a}} P(Q_k^{\wr}=1|Q_k^{\dagger}=1,Q_k{}^{\ddagger}=1,S=s,\boldsymbol{n_k},\boldsymbol{a_k},k)\sum_r\{\mathscr{g}_k(\cdot)P(r|Q^{\ddagger}=1,S=s,\boldsymbol{n_k},\boldsymbol{a_k},k)\}P(Q_k^{\ddagger}=1,S=s,\boldsymbol{n_k},\boldsymbol{a_k},k)}$$

The first equality holds by definition of $Q_k$, the third and sixth by definition of $G_k$ as a random assignment of $\boldsymbol{W_k^{\dagger}}$ from the distribution $\mathscr{g}_k(\cdot)$ given some (but not all) of the elements in $(Q_k^{\ddagger} = 1, R = r, \boldsymbol{n_k}, \boldsymbol{a_k}, k)$ and definition of $Q_k^{\dagger} = I(G_k = \boldsymbol{w}^{\dagger} \in \mathbb{w}^{\dagger})$, the fourth by exchangeability (18) and positivity (12), and the fifth by consistency. Note that $\mathscr{q}_k(\cdot) = P({Q_k^{\dagger}}^{G_k} = 1|Q_k^{\ddagger} = 1, R = r, \boldsymbol{n_k}, \boldsymbol{a_k}, k)$ the conditional distribution of $Q_k^{\dagger}$ under the intervention $G_k$.

<u>Corollary A of Lemma 2.</u>

$$P\big(\boldsymbol{a_k}\big|Q_k^{G_k} = 1, R = r, k\big)$$

$$= \frac{\sum_{\boldsymbol{n}} P(Q_k^{\wr}=1|Q_k^{\dagger}=1,Q_k^{\ddagger}=1,R=r,\boldsymbol{n_k},\boldsymbol{a_k},k)\mathscr{q}_k(\cdot)P(Q_k^{\ddagger}=1,R=r,\boldsymbol{n_k},\boldsymbol{a_k},k)}{\sum_{\boldsymbol{n},\boldsymbol{a}} P(Q_k^{\wr}=1|Q_k^{\dagger}=1,Q_k^{\ddagger}=1,R=r,\boldsymbol{n_k},\boldsymbol{a_k},k)\mathscr{q}_k(\cdot)P(Q_k^{\ddagger}=1,R=r,\boldsymbol{n_k},\boldsymbol{a_k},k)}$$

This holds by simply choosing a specific social group, denoted by $R = r'$, as the subgroup $S = s$ and noting that $P(r|Q_k^{\ddagger} = 1, R = r', \boldsymbol{n_k}, \boldsymbol{a_k})$ equals 1 when $r = r'$ and equals 0 when $r \neq r'$.

<u>Corollary B of Lemma 2.</u>

$$P\big(\boldsymbol{a_k}\big|Q_k^{G_k} = 1, T = 1, k\big)$$

$$= \frac{\sum_{\boldsymbol{n}} P(Q_k^{\wr}=1|Q_k^{\dagger}=1,Q_k^{\ddagger}=1,T=1,\boldsymbol{n_k},\boldsymbol{a_k},k)\sum_r\{\mathscr{q}_k(\cdot)P(r|Q_k^{\ddagger}=1,T=1,\boldsymbol{n_k},\boldsymbol{a_k},k)\}P(Q_k^{\ddagger}=1,T=1,\boldsymbol{n_k},\boldsymbol{a_k},k)}{\sum_{\boldsymbol{n},\boldsymbol{a}} P(Q_k^{\wr}=1|Q_k^{\dagger}=1,Q_k^{\ddagger}=1,T=1,\boldsymbol{n_k},\boldsymbol{a_k},k)\sum_r\{\mathscr{q}_k(\cdot)P(r|Q_k^{\ddagger}=1,T=1,\boldsymbol{n_k},\boldsymbol{a_k},k)\}P(Q_k^{\ddagger}=1,T=1,\boldsymbol{n_k},\boldsymbol{a_k},k)}$$

This holds by simply choosing the standard population, denoted by $T = 1$, as the subgroup $S = s$.

<u>Corollary C of Lemma 3.</u>

$$P\big(\boldsymbol{a_k}, k\big|Q_k^{G_k} = 1, T = 1\big)$$

$$= \frac{\sum_{\boldsymbol{n}} P(Q_k^{\wr}=1|Q_k^{\dagger}=1,Q_k^{\ddagger}=1,T=1,\boldsymbol{n_k},\boldsymbol{a_k},k)\sum_r\{\mathscr{q}_k(\cdot)P(r|Q_k^{\ddagger}=1,T=1,\boldsymbol{n_k},\boldsymbol{a_k},k)\}P(Q_k^{\ddagger}=1,T=1,\boldsymbol{n_k},\boldsymbol{a_k},k)}{\sum_{\boldsymbol{n},\boldsymbol{a},k} P(Q_k^{\wr}=1|Q_k^{\dagger}=1,Q_k^{\ddagger}=1,T=1,\boldsymbol{n_k},\boldsymbol{a_k},k)\sum_r\{\mathscr{q}_k(\cdot)P(r|Q_k^{\ddagger}=1,T=1,\boldsymbol{n_k},\boldsymbol{a_k},k)\}P(Q_k^{\ddagger}=1,T=1,\boldsymbol{n_k},\boldsymbol{a_k},k)}$$

This is straightforward to show using similar steps used in the derivation of Lemma 2.

<u>Lemma 4.</u>

$$P\big(Y_{k+J}^{G_k} = y\big|Q_k^{G_k} = 1, R = r, \boldsymbol{n_k}, \boldsymbol{a_k}, k\big)P\big(\boldsymbol{n_k}\big|Q_k^{G_k} = 1, R = r, \boldsymbol{a_k}, k\big)P\big(\boldsymbol{a_k}\big|Q_k^{G_k} = 1, T = 1, k\big)$$

$$= P\left(Y_{k+J}^{G_k} = y\middle|{Q_k^{\wr}}^{G_k} = 1, {Q_k^{\dagger}}^{G_k} = 1, Q_k^{\ddagger} = 1, R = r, \boldsymbol{n_k}, \boldsymbol{a_k}, k\right)$$

$$\times \frac{P(Q_k^{\wr}=1|Q_k^{\dagger}=1,Q_k^{\ddagger}=1,R=r,\boldsymbol{n_k},\boldsymbol{a_k},k)\mathscr{q}_k(\cdot)P(Q_k^{\ddagger}=1,R=r,\boldsymbol{n_k},\boldsymbol{a_k},k)}{\sum_{\boldsymbol{n}} P(Q_k^{\wr}=1|Q_k^{\dagger}=1,Q_k^{\ddagger}=1,R=r,\boldsymbol{n_k},\boldsymbol{a_k},k)\mathscr{q}_k(\cdot)P(Q_k^{\ddagger}=1,R=r,\boldsymbol{n_k},\boldsymbol{a_k},k)}$$

$$\times \frac{\sum_{\boldsymbol{n}} P(Q_k^{\wr}=1|Q_k^{\dagger}=1,Q_k^{\ddagger}=1,T=1,\boldsymbol{n_k},\boldsymbol{a_k},k)\sum_r\{\mathscr{q}_k(\cdot)P(r|Q_k^{\ddagger}=1,T=1,\boldsymbol{n_k},\boldsymbol{a_k},k)\}P(Q_k^{\ddagger}=1,T=1,\boldsymbol{n_k},\boldsymbol{a_k},k)}{\sum_{\boldsymbol{n},\boldsymbol{a}} P(Q_k^{\wr}=1|Q_k^{\dagger}=1,Q_k^{\ddagger}=1,T=1,\boldsymbol{n_k},\boldsymbol{a_k},k)\sum_r\{\mathscr{q}_k(\cdot)P(r|Q_k^{\ddagger}=1,T=1,\boldsymbol{n_k},\boldsymbol{a_k},k)\}P(Q_k^{\ddagger}=1,T=1,\boldsymbol{n_k},\boldsymbol{a_k},k)}$$

$$= P\left(Y_{k+J}^{G_k} = y\middle|{Q_k^{\wr}}^{G_k} = 1, {Q_k^{\dagger}}^{G_k} = 1, Q^{\ddagger} = 1, R = r, \boldsymbol{n_k}, \boldsymbol{a_k}, k\right)$$

$$\times \frac{P({Q_k^{\wr}}^{G_k}=1|{Q_k^{\dagger}}^{G_k}=1,Q_k^{\ddagger}=1,R=r,\boldsymbol{n_k},\boldsymbol{a_k},k)\mathscr{q}_k(\cdot)P(Q_k^{\ddagger}=1,R=r,\boldsymbol{n_k},\boldsymbol{a_k},k)}{\sum_{\boldsymbol{n}} P(Q_k^{\wr}=1|Q_k^{\dagger}=1,Q_k^{\ddagger}=1,R=r,\boldsymbol{n_k},\boldsymbol{a_k},k)\mathscr{q}_k(\cdot)P(Q_k^{\ddagger}=1,R=r,\boldsymbol{n_k},\boldsymbol{a_k},k)}$$

$$\times \frac{\sum_{\boldsymbol{n}} P(Q_k^{\wr}=1|Q_k^{\dagger}=1,Q_k^{\ddagger}=1,T=1,\boldsymbol{n_k},\boldsymbol{a_k},k)\sum_r\{\mathscr{q}_k(\cdot)P(r|Q_k^{\ddagger}=1,T=1,\boldsymbol{n_k},\boldsymbol{a_k},k)\}P(Q_k^{\ddagger}=1,T=1,\boldsymbol{n_k},\boldsymbol{a_k},k)}{\sum_{\boldsymbol{n},\boldsymbol{a}} P(Q_k^{\wr}=1|Q_k^{\dagger}=1,Q_k^{\ddagger}=1,T=1,\boldsymbol{n_k},\boldsymbol{a_k},k)\sum_r\{\mathscr{q}_k(\cdot)P(r|Q_k^{\ddagger}=1,T=1,\boldsymbol{n_k},\boldsymbol{a_k},k)\}P(Q_k^{\ddagger}=1,T=1,\boldsymbol{n_k},\boldsymbol{a_k},k)}$$

$$= P\left(Y_{k+J}^{G_k} = y, {Q_k^{\wr}}^{G_k} = 1\middle|{Q_k^{\dagger}}^{G_k} = 1, Q_k^{\ddagger} = 1, R = r, \boldsymbol{n_k}, \boldsymbol{a_k}, k\right)$$

$$\times \frac{\mathscr{q}_k(\cdot)P(Q_k^{\ddagger}=1,R=r,\boldsymbol{n_k},\boldsymbol{a_k},k)}{\sum_{\boldsymbol{n}} P(Q_k^{\wr}=1|Q_k^{\dagger}=1,Q_k^{\ddagger}=1,R=r,\boldsymbol{n_k},\boldsymbol{a_k},k)\mathscr{q}_k(\cdot)P(Q_k^{\ddagger}=1,R=r,\boldsymbol{n_k},\boldsymbol{a_k},k)}$$

$$\times \frac{\sum_{\boldsymbol{n}} P(Q_k^{\wr}=1|Q_k^{\dagger}=1,Q_k^{\ddagger}=1,T=1,\boldsymbol{n_k},\boldsymbol{a_k},k)\sum_r\{\mathscr{q}_k(\cdot)P(r|Q_k^{\ddagger}=1,T=1,\boldsymbol{n_k},\boldsymbol{a_k},k)\}P(Q_k^{\ddagger}=1,T=1,\boldsymbol{n_k},\boldsymbol{a_k},k)}{\sum_{\boldsymbol{n},\boldsymbol{a}} P(Q_k^{\wr}=1|Q_k^{\dagger}=1,Q_k^{\ddagger}=1,T=1,\boldsymbol{n_k},\boldsymbol{a_k},k)\sum_r\{\mathscr{q}_k(\cdot)P(r|Q^{\ddagger}=1,T=1,\boldsymbol{n_k},\boldsymbol{a_k},k)\}P(Q_k^{\ddagger}=1,T=1,\boldsymbol{n_k},\boldsymbol{a_k},k)}$$

$$= P\big(Y_{k+J}^{G_k} = y, {Q_k^{\wr}}^{G_k} = 1\big|Q^{\ddagger} = 1, R = r, \boldsymbol{n_k}, \boldsymbol{a_k}, k\big)$$

$$\times \frac{\mathscr{q}_k(\cdot)P(Q_k^{\ddagger}=1,R=r,\boldsymbol{n_k},\boldsymbol{a_k},k)}{\sum_{\boldsymbol{n}} P(Q_k^{\wr}=1|Q_k^{\dagger}=1,Q_k^{\ddagger}=1,R=r,\boldsymbol{n_k},\boldsymbol{a_k},k)\mathscr{q}_k(\cdot)P(Q_k^{\ddagger}=1,R=r,\boldsymbol{n_k},\boldsymbol{a_k},k)}$$

$$\times \frac{\sum_{\boldsymbol{n}} P(Q_k^{\wr}=1|Q_k^{\dagger}=1,Q_k^{\ddagger}=1,T=1,\boldsymbol{n_k},\boldsymbol{a_k},k)\sum_r\{\mathscr{q}_k(\cdot)P(r|Q_k^{\ddagger}=1,T=1,\boldsymbol{n_k},\boldsymbol{a_k},k)\}P(Q_k^{\ddagger}=1,T=1,\boldsymbol{n_k},\boldsymbol{a_k},k)}{\sum_{\boldsymbol{n},\boldsymbol{a}} P(Q_k^{\wr}=1|Q_k^{\dagger}=1,Q_k^{\ddagger}=1,T=1,\boldsymbol{n_k},\boldsymbol{a_k},k)\sum_r\{\mathscr{q}_k(\cdot)P(r|Q_k^{\ddagger}=1,T=1,\boldsymbol{n_k},\boldsymbol{a_k},k)\}P(Q_k^{\ddagger}=1,T=1,\boldsymbol{n_k},\boldsymbol{a_k},k)}$$

$$= P\left(Y_{k+J}^{G_k} = y, Q_k^{\wr\, G_k} = 1 \middle| Q_k^{\dagger} = 1, Q_k^{\ddagger} = 1, R = r, \boldsymbol{n_k}, \boldsymbol{a_k}, k\right)$$

$$\times \frac{\mathscr{g}_k(\cdot)P(Q_k^{\ddagger}=1,R=r,\boldsymbol{n_k},\boldsymbol{a_k},k)}{\sum_{\boldsymbol{n}} P(Q_k^{\wr}=1|Q_k^{\dagger}=1,Q_k^{\ddagger}=1,R=r,\boldsymbol{n_k},\boldsymbol{a_k},k)\mathscr{g}_k(\cdot)P(Q_k^{\ddagger}=1,R=r,\boldsymbol{n_k},\boldsymbol{a_k},k)}$$

$$\times \frac{\sum_{\boldsymbol{n}} P(Q_k^{\wr}=1|Q_k^{\dagger}=1,Q_k^{\ddagger}=1,T=1,\boldsymbol{n_k},\boldsymbol{a_k},k)\sum_r\{\mathscr{g}_k(\cdot)P(r|Q_k^{\ddagger}=1,T=1,\boldsymbol{n_k},\boldsymbol{a_k},k)\}P(Q_k^{\ddagger}=1,T=1,\boldsymbol{n_k},\boldsymbol{a_k},k)}{\sum_{\boldsymbol{n},\boldsymbol{a}} P(Q_k^{\wr}=1|Q_k^{\dagger}=1,Q_k^{\ddagger}=1,T=1,\boldsymbol{n_k},\boldsymbol{a_k},k)\sum_r\{\mathscr{g}_k(\cdot)P(r|Q_k^{\ddagger}=1,T=1,\boldsymbol{n_k},\boldsymbol{a_k},k)\}P(Q_k^{\ddagger}=1,T=1,\boldsymbol{n_k},\boldsymbol{a_k},k)}$$

$$= P\left(Y_{k+J} = y, Q_k^{\wr} = 1 \middle| Q_k^{\dagger} = 1, Q_k^{\ddagger} = 1, R = r, \boldsymbol{n_k}, \boldsymbol{a_k}, k\right)$$

$$\times \frac{\mathscr{g}_k(\cdot)P(Q_k^{\ddagger}=1,R=r,\boldsymbol{n_k},\boldsymbol{a_k},k)}{\sum_{\boldsymbol{n}} P(Q_k^{\wr}=1|Q_k^{\dagger}=1,Q_k^{\ddagger}=1,R=r,\boldsymbol{n_k},\boldsymbol{a_k},k)\mathscr{g}_k(\cdot)P(Q_k^{\ddagger}=1,R=r,\boldsymbol{n_k},\boldsymbol{a_k},k)}$$

$$\times \frac{\sum_{\boldsymbol{n}} P(Q_k^{\wr}=1|Q_k^{\dagger}=1,Q_k^{\ddagger}=1,T=1,\boldsymbol{n_k},\boldsymbol{a_k},k)\sum_r\{\mathscr{g}_k(\cdot)P(r|Q_k^{\ddagger}=1,T=1,\boldsymbol{n_k},\boldsymbol{a_k},k)\}P(Q_k^{\ddagger}=1,T=1,\boldsymbol{n_k},\boldsymbol{a_k},k)}{\sum_{\boldsymbol{n},\boldsymbol{a}} P(Q_k^{\wr}=1|Q_k^{\dagger}=1,Q_k^{\ddagger}=1,T=1,\boldsymbol{n_k},\boldsymbol{a_k},k)\sum_r\{\mathscr{g}_k(\cdot)P(r|Q_k^{\ddagger}=1,T=1,\boldsymbol{n_k},\boldsymbol{a_k},k)\}P(Q_k^{\ddagger}=1,T=1,\boldsymbol{n_k},\boldsymbol{a_k},k)}$$

$$= P\left(Y_{k+J} = y \middle| Q_k^{\wr} = 1, Q_k^{\dagger} = 1, Q_k^{\ddagger} = 1, R = r, \boldsymbol{n_k}, \boldsymbol{a_k}, k\right)$$

$$\times \frac{P(Q_k^{\wr}=1|Q_k^{\dagger}=1,Q_k^{\ddagger}=1,R=r,\boldsymbol{n_k},\boldsymbol{a_k},k)\mathscr{g}_k(\cdot)P(Q_k^{\ddagger}=1,R=r,\boldsymbol{n_k},\boldsymbol{a_k},k)}{\sum_{\boldsymbol{n}} P(Q_k^{\wr}=1|Q_k^{\dagger}=1,Q_k^{\ddagger}=1,R=r,\boldsymbol{n_k},\boldsymbol{a_k},k)\mathscr{g}_k(\cdot)P(Q_k^{\ddagger}=1,R=r,\boldsymbol{n_k},\boldsymbol{a_k},k)}$$

$$\times \frac{\sum_{\boldsymbol{n}} P(Q_k^{\wr}=1|Q_k^{\dagger}=1,Q_k^{\ddagger}=1,T=1,\boldsymbol{n_k},\boldsymbol{a_k},k)\sum_r\{\mathscr{g}_k(\cdot)P(r|Q_k^{\ddagger}=1,T=1,\boldsymbol{n_k},\boldsymbol{a_k},k)\}P(Q_k^{\ddagger}=1,T=1,\boldsymbol{n_k},\boldsymbol{a_k},k)}{\sum_{\boldsymbol{n},\boldsymbol{a}} P(Q_k^{\wr}=1|Q_k^{\dagger}=1,Q_k^{\ddagger}=1,T=1,\boldsymbol{n_k},\boldsymbol{a_k},k)\sum_r\{\mathscr{g}_k(\cdot)P(r|Q_k^{\ddagger}=1,T=1,\boldsymbol{n_k},\boldsymbol{a_k},k)\}P(Q_k^{\ddagger}=1,T=1,\boldsymbol{n_k},\boldsymbol{a_k},k)}$$

$$= P\left(Y_{k+J} = y \middle| Q_k^{\wr} = 1, Q_k^{\dagger} = 1, Q_k^{\ddagger} = 1, R = r, \boldsymbol{n_k}, \boldsymbol{a_k}, k\right)$$

$$\times \frac{P(Q_k^{\wr}=1|Q_k^{\dagger}=1,Q_k^{\ddagger}=1,R=r,\boldsymbol{n_k},\boldsymbol{a_k},k)\mathscr{g}_k(\cdot)}{\sum_{\boldsymbol{n}} P(Q_k^{\wr}=1|Q_k^{\dagger}=1,Q_k^{\ddagger}=1,R=r,\boldsymbol{n_k},\boldsymbol{a_k},k)\mathscr{g}_k(\cdot)P(\boldsymbol{n_k}|Q_k^{\ddagger}=1,R=r,\boldsymbol{a_k},k)} \times P(\boldsymbol{n_k}|Q_k^{\ddagger} = 1, R = r, \boldsymbol{a_k}, k)$$

$$\times \frac{\sum_{\boldsymbol{n}} P(Q_k^{\wr}=1|Q_k^{\dagger}=1,Q_k^{\ddagger}=1,T=1,\boldsymbol{n_k},\boldsymbol{a_k},k)\sum_r\{\mathscr{g}_k(\cdot)P(r|Q_k^{\ddagger}=1,T=1,\boldsymbol{n_k},\boldsymbol{a_k},k)\}P(\boldsymbol{n_k}|Q_k^{\ddagger}=1,T=1,\boldsymbol{a_k},k)}{\sum_{\boldsymbol{n},\boldsymbol{a}} P(Q_k^{\wr}=1|Q_k^{\dagger}=1,Q_k^{\ddagger}=1,T=1,\boldsymbol{n_k},\boldsymbol{a_k},k)\sum_r\left\{\mathscr{g}_k(\cdot)P(r|Q_k^{\ddagger}=1,T=1,\boldsymbol{n_k},\boldsymbol{a_k},k)\right\}P(\boldsymbol{n_k},\boldsymbol{a_k}|Q_k^{\ddagger}=1,T=1,k)} \times P(\boldsymbol{a_k}|Q^{\ddagger} = 1, T = 1, k)$$

The first equality holds by by design of the target study and under overlap (3) in the main text (see footnote 21), the second by definition of $Q_k$, Lemma 1, Lemma 2, and Corollary B of Lemma 2, the third by exchangeability (18) and positivity (12) and consistency, the fifth by definition of $G_k$ as a random assignment of $\boldsymbol{W_k^{\dagger}}$ from the distribution $\mathscr{g}_k(\cdot)$ given some (but not all) of the elements in $(Q_k^{\ddagger} = 1, R = r, \boldsymbol{n_k}, \boldsymbol{a_k}, k)$ and definition of $Q_k^{\dagger} = I(G_k = \boldsymbol{w}^{\dagger} \in \mathbb{w}^{\dagger})$, the sixth by exchangeability (18) and positivity (12), and the seventh by consistency.

This result establishes that, under the identification assumptions:

$$P\left(Y_{k+J}^{G_k} = y \middle| Q_k^{G_k} = 1, R = r, \boldsymbol{n_k}, \boldsymbol{a_k}, k\right) = P\left(Y_{k+J}^{G_k} = y \middle| Q_k = 1, R = r, \boldsymbol{n_k}, \boldsymbol{a_k}, k\right)$$

Joint distribution of $(Y_{k+J}^{G_k}, \boldsymbol{A_k}, \boldsymbol{N_k})$ given $Q_k^{G_k} = 1$, and $R = r$ in the Counterfactual Eligible Source Population

$f_{k,***}^{\mathcal{D}4} = P\left(Y_{k+J}^{G_k} = y \middle| Q_k^{G_k} = 1, R = r, \boldsymbol{n_k}, \boldsymbol{a_k}, k\right) P(\boldsymbol{n_k}|Q_k^{G_k} = 1, R = r, \boldsymbol{a_k}, k) P\left(\boldsymbol{a_k} \middle| Q_k^{G_k} = 1, R = r, k\right)$

Joint distribution of $(Y_{k+J}^{G_k}, \boldsymbol{A_k}, \boldsymbol{N_k})$ given $Q_k^{G_k} = 1$, and $R = r$ in the Available Eligible Source Population

$f_{k,0}^{\mathcal{D}4} = P\left(Y_{k+J}^{G_k} = y \middle| Q_k^{G_k} = 1, R = r, \boldsymbol{n_k}, \boldsymbol{a_k}, k\right) P(\boldsymbol{n_k}|Q_k = 1, R = r, \boldsymbol{a_k}, k) P(\boldsymbol{a_k}|Q_k = 1, R = r, k)$

Joint distribution of $(Y_{k+J}^{G_k}, \boldsymbol{A_k}, \boldsymbol{N_k})$ given $Q_k^{G_k} = 1$, and $R = r$ after Stage One of Sampling

$f_{k,1}^{\mathcal{D}4} = P\left(Y_{k+J}^{G_k} = y \middle| Q_k^{G_k} = 1, R = r, \boldsymbol{n_k}, \boldsymbol{a_k}, k\right) P(\boldsymbol{n_k}|Q_k^{G_k} = 1, R = r, \boldsymbol{a_k}, k) P\left(\boldsymbol{a_k} \middle| Q_k^{G_k} = 1, R = r, k\right)$

Joint distribution of $(Y_{k+J}^{G_k}, \boldsymbol{A_k}, \boldsymbol{N_k})$ given $Q_k^{G_k} = 1$, and $R = r$ after Stage Two of Sampling

$f_{k,2}^{\mathcal{D}4} = P\left(Y_{k+J}^{G_k} = y \middle| Q_k^{G_k} = 1, R = r, \boldsymbol{n_k}, \boldsymbol{a_k}, k\right) P(\boldsymbol{n_k}|Q_k^{G_k} = 1, R = r, \boldsymbol{a_k}, k) P\left(\boldsymbol{a_k} \middle| Q_k^{G_k} = 1, T = 1, k\right)$

These joint distribution holds by design of the target study and under overlap (3) (see footnote 21) and innocuous sampling (4), exchangeability (18) or (19), positivity (12), and consistency as in the main text. See also Lemmas 1-4.

Mean of $Y_{k+J}$ given $R = r$ in the target study

$$\mu_k^{\mathcal{D}4}(r)$$

$$= \sum_{\boldsymbol{a},\boldsymbol{n}} E\left(Y_{k+J}^{G_k} \middle| Q_k^{G_k} = 1, R = r, \boldsymbol{n_k}, \boldsymbol{a_k}, k\right) P(\boldsymbol{n_k} | Q_k^{G_k} = 1, R = r, \boldsymbol{a_k}, k) P\left(\boldsymbol{a_k} \middle| Q_k^{G_k} = 1, T = 1, k\right)$$

$$= E\left(Y_{k+J} \middle| Q_k^{\wr} = 1, Q_k^{\dagger} = 1, Q_k^{\ddagger} = 1, R = r, \boldsymbol{n_k}, \boldsymbol{a_k}, k\right)$$

$$\times \frac{P(Q_k^{\wr}=1|Q_k^{\dagger}=1,Q_k^{\ddagger}=1,R=r,\boldsymbol{n_k},\boldsymbol{a_k},k) q_k(\cdot)}{\sum_{\boldsymbol{n}} P(Q_k^{\wr}=1|Q_k^{\dagger}=1,Q_k^{\ddagger}=1,R=r,\boldsymbol{n_k},\boldsymbol{a_k},k) q_k(\cdot) P(\boldsymbol{n_k}|Q_k^{\ddagger}=1,R=r,\boldsymbol{a_k},k)} \times P(\boldsymbol{n_k} | Q_k^{\ddagger} = 1, R = r, \boldsymbol{a_k}, k)$$

$$\times \frac{\sum_{\boldsymbol{n}} P(Q_k^{\wr}=1|Q_k^{\dagger}=1,Q_k^{\ddagger}=1,T=1,\boldsymbol{n_k},\boldsymbol{a_k},k) \sum_r \{q_k(\cdot) P(r|Q_k^{\ddagger}=1,T=1,\boldsymbol{n_k},\boldsymbol{a_k},k)\} P(\boldsymbol{n_k}|Q_k^{\ddagger}=1,T=1,\boldsymbol{a_k},k)}{\sum_{\boldsymbol{n},\boldsymbol{a}} P(Q_k^{\wr}=1|Q_k^{\dagger}=1,Q_k^{\ddagger}=1,T=1,\boldsymbol{n_k},\boldsymbol{a_k},k) \sum_r \left\{q_k(\cdot) P(r|Q_k^{\ddagger}=1,T=1,\boldsymbol{n_k},\boldsymbol{a_k},k)\right\} P(\boldsymbol{n_k},\boldsymbol{a_k}|Q_k^{\ddagger}=1,T=1,k)} \times P(\boldsymbol{a_k} | Q^{\ddagger} = 1, T = 1, k)$$

By definition of a mean and under exchangeability (18), positivity (12), and consistency and the definition of $G_k$.

Sampling Fractions

$$\alpha_{k,1}^{\mathcal{D}4}(r, \boldsymbol{a_k}, \boldsymbol{n_k})$$

$$= \frac{\mathbb{N}_{k,1}(r,\boldsymbol{a_k},\boldsymbol{n_k})}{\mathbb{N}_{k,0}(r,\boldsymbol{a_k},\boldsymbol{n_k})}$$

$$= \frac{\mathbb{N}_{k,1}(r)}{\mathbb{N}_{k,0}(r)} \times \frac{f_{k,1}^{\mathcal{D}4}}{f_{k,0}^{\mathcal{D}4}}$$

$$= \frac{\mathbb{N}_{k,1}(r)}{\mathbb{N}_{k,0}(r)} \times \frac{P\left(\boldsymbol{n_k}|Q_k^{G_k}=1,R=r,\boldsymbol{a_k},k\right)}{P(\boldsymbol{n_k}|Q_k=1,R=r,\boldsymbol{a_k},k)} \times \frac{P\left(\boldsymbol{a_k}|Q_k^{G_k}=1,R=r,k\right)}{P(\boldsymbol{a_k}|Q_k=1,R=r,k)}$$

$$= \frac{\frac{\sum_{\boldsymbol{n}} P(Q_k^{\wr}=1|Q_k^{\dagger}=1,Q_k^{\ddagger}=1,S=s,\boldsymbol{n_k},\boldsymbol{a_k},k) \sum_r \{q_k(\cdot) P(r|Q_k^{\ddagger}=1,S=s,\boldsymbol{n_k},\boldsymbol{a_k},k)\} P(Q_k^{\ddagger}=1,S=s,\boldsymbol{n_k},\boldsymbol{a_k},k)}{\sum_{\boldsymbol{n},\boldsymbol{a}} P(Q_k^{\wr}=1|Q_k^{\dagger}=1,Q_k^{\ddagger}=1,S=s,\boldsymbol{n_k},\boldsymbol{a_k},k) \sum_r \{q_k(\cdot) P(r|Q_k^{\ddagger}=1,S=s,\boldsymbol{n_k},\boldsymbol{a_k},k)\} P(Q_k^{\ddagger}=1,S=s,\boldsymbol{n_k},\boldsymbol{a_k},k)}}{P(\boldsymbol{n_k}|Q_k=1,R=r,\boldsymbol{a_k},k)} \times \frac{\frac{\sum_{\boldsymbol{n}} P(Q_k^{\wr}=1|Q_k^{\dagger}=1,Q_k^{\ddagger}=1,R=r,\boldsymbol{n_k},\boldsymbol{a_k},k) q_k(\cdot) P(Q_k^{\ddagger}=1,R=r,\boldsymbol{n_k},\boldsymbol{a_k},k)}{\sum_{\boldsymbol{n},\boldsymbol{a}} P(Q_k^{\wr}=1|Q_k^{\dagger}=1,Q_k^{\ddagger}=1,R=r,\boldsymbol{n_k},\boldsymbol{a_k},k) q_k(\cdot) P(Q_k^{\ddagger}=1,R=r,\boldsymbol{n_k},\boldsymbol{a_k},k)}}{P(\boldsymbol{a_k}|Q_k=1,R=r,k)}$$

$$\alpha_{k,2}^{\mathcal{D}4}(r, \boldsymbol{n_k}, \boldsymbol{a_k})$$

$$= \frac{\mathbb{N}_{k,2}(r,\boldsymbol{n_k},\boldsymbol{a_k})}{\mathbb{N}_{k,1}(r,\boldsymbol{n_k},\boldsymbol{a_k})}$$

$$= \frac{\mathbb{N}_{k,2}(r)}{\mathbb{N}_{k,1}(r)} \times \frac{f_{k,2}^{\mathcal{D}4}}{f_{k,1}^{\mathcal{D}4}}$$

$$= \frac{\mathbb{N}_{k,2}(r)}{\mathbb{N}_{k,1}(r)} \times \frac{P\left(\boldsymbol{a_k}|Q_k^{G_k}=1,T=1,k\right)}{P\left(\boldsymbol{a_k}|Q_k^{G_k}=1,R=r,k\right)}$$

$$= \frac{\mathbb{N}_{k,2}(r)}{\mathbb{N}_{k,1}(r)} \times \frac{\frac{\sum_{\boldsymbol{n}} P(Q_k^{\wr}=1|Q_k^{\dagger}=1,Q_k^{\ddagger}=1,T=1,\boldsymbol{n_k},\boldsymbol{a_k},k) \sum_r \{q_k(\cdot) P(r|Q_k^{\ddagger}=1,T=1,\boldsymbol{n_k},\boldsymbol{a_k},k)\} P(Q_k^{\ddagger}=1,T=1,\boldsymbol{n_k},\boldsymbol{a_k},k)}{\sum_{\boldsymbol{n},\boldsymbol{a}} P(Q_k^{\wr}=1|Q_k^{\dagger}=1,Q_k^{\ddagger}=1,T=1,\boldsymbol{n_k},\boldsymbol{a_k},k) \sum_r \{q_k(\cdot) P(r|Q_k^{\ddagger}=1,T=1,\boldsymbol{n_k},\boldsymbol{a_k},k)\} P(Q_k^{\ddagger}=1,T=1,\boldsymbol{n_k},\boldsymbol{a_k},k)}}{\frac{\sum_{\boldsymbol{n}} P(Q_k^{\wr}=1|Q_k^{\dagger}=1,Q_k^{\ddagger}=1,R=r,\boldsymbol{n_k},\boldsymbol{a_k},k) q_k(\cdot) P(Q_k^{\ddagger}=1,R=r,\boldsymbol{n_k},\boldsymbol{a_k},k)}{\sum_{\boldsymbol{n},\boldsymbol{a}} P(Q_k^{\wr}=1|Q_k^{\dagger}=1,Q_k^{\ddagger}=1,R=r,\boldsymbol{n_k},\boldsymbol{a_k},k) q_k(\cdot) P(Q_k^{\ddagger}=1,R=r,\boldsymbol{n_k},\boldsymbol{a_k},k)}}$$

G-Computation Estimator

First, define:

$$\phi_k$$

$$= \frac{P(Q_k^{\wr}=1|Q_k^{\dagger}=1,Q_k^{\ddagger}=1,R=r,\boldsymbol{n_k},\boldsymbol{a_k},k) q_k(\cdot)}{\sum_{\boldsymbol{n}} P(Q_k^{\wr}=1|Q_k^{\dagger}=1,Q_k^{\ddagger}=1,R=r,\boldsymbol{n_k},\boldsymbol{a_k},k) q_k(\cdot) P(\boldsymbol{n_k}|Q_k^{\ddagger}=1,R=r,\boldsymbol{a_k},k)}$$

$$= \frac{q_k(\cdot) P(Q_k^{\wr}=1|Q_k^{\dagger}=1,Q_k^{\ddagger}=1,R=r,\boldsymbol{n_k},\boldsymbol{a_k},k)}{E_{\boldsymbol{n}}[q_k(\cdot) P(Q_k^{\wr}=1|Q_k^{\dagger}=1,Q_k^{\ddagger}=1,R=r,\boldsymbol{n_k},\boldsymbol{a_k},k)|Q_k^{\ddagger}=1,R=r,\boldsymbol{a_k},k]}$$ where $E_{\boldsymbol{n}}[\cdot]$ is over $P(\boldsymbol{n_k} | Q_k^{\ddagger} = 1, R = r, \boldsymbol{a_k}, k)$

and also:

$$\theta_k$$

$$= \frac{\sum_{\boldsymbol{n}} P(Q_k^{\wr}=1|Q_k^{\dagger}=1,Q_k^{\ddagger}=1,T=1,\boldsymbol{n_k},\boldsymbol{a_k},k) \sum_r \{q_k(\cdot) P(r|Q_k^{\ddagger}=1,T=1,\boldsymbol{n_k},\boldsymbol{a_k},k)\} P(\boldsymbol{n_k}|Q_k^{\ddagger}=1,T=1,k,\boldsymbol{a_k})}{\sum_{\boldsymbol{n},\boldsymbol{a}} P(Q_k^{\wr}=1|Q_k^{\dagger}=1,Q_k^{\ddagger}=1,T=1,\boldsymbol{n_k},\boldsymbol{a_k},k) \sum_r \left\{q_k(\cdot) P(r|Q_k^{\ddagger}=1,T=1,\boldsymbol{n_k},\boldsymbol{a_k},k)\right\} P(\boldsymbol{n_k},\boldsymbol{a_k}|Q_k^{\ddagger}=1,T=1,k)}$$

$$= \frac{E_{\boldsymbol{n}}[E_r\{q_k(\cdot)|Q_k^{\ddagger}=1,T=1,\boldsymbol{n_k},\boldsymbol{a_k},k\} P(Q_k^{\wr}=1|Q_k^{\dagger}=1,Q_k^{\ddagger}=1,R=r,\boldsymbol{n_k},\boldsymbol{a_k},k)|Q_k^{\ddagger}=1T=1,\boldsymbol{a_k},k]}{E_{(\boldsymbol{n},\boldsymbol{a})}[E_r\{q_k(\cdot)|Q_k^{\ddagger}=1,T=1,\boldsymbol{n_k},\boldsymbol{a_k},k\} P(Q_k^{\wr}=1|Q_k^{\dagger}=1,Q_k^{\ddagger}=1,R=r,\boldsymbol{n_k},\boldsymbol{a_k},k)|Q_k^{\ddagger}=1,T=1,k]}$$ where $\boldsymbol{E}_r\{\cdot\}$ is over $P(r | Q_k^{\ddagger} = 1, T = 1, \boldsymbol{n_k}, \boldsymbol{a_k}, k)$

and $E_{\boldsymbol{n}}[\cdot]$ is over $P(\boldsymbol{n_k} | Q_k^{\ddagger} = 1, T = 1, k, \boldsymbol{a_k})$ and $E_{(\boldsymbol{n},\boldsymbol{a})}[\cdot]$ is over $P(\boldsymbol{n_k}, \boldsymbol{a_k} | Q_k^{\ddagger} = 1, T = 1, k)$

Therefore:

$$
\begin{aligned}
&\mu_k^{\mathcal{D}4}(r) \\
&= \textstyle\sum_{\boldsymbol{a},\boldsymbol{n}} E\left[Y_{k+J} \middle| Q_k = 1, R = r, \boldsymbol{n_k}, \boldsymbol{a_k}, k\right] \phi_k P\left(\boldsymbol{n_k} \middle| Q_k^{\ddagger} = 1, R = r, \boldsymbol{a_k}, k\right) \theta_k P\left(\boldsymbol{a_k} \middle| Q_k^{\ddagger} = 1, T = 1, k\right) \\
&= \textstyle\sum_{\boldsymbol{a}} E_{\boldsymbol{n}}\left(\phi_k E\left[Y_{k+J} \middle| Q_k = 1, R = r, \boldsymbol{n_k}, \boldsymbol{a_k}, k\right] \middle| Q_k^{\ddagger} = 1, R = r, \boldsymbol{a_k}, k\right) \theta_k P\left(\boldsymbol{a_k} \middle| Q_k^{\ddagger} = 1, T = 1, k\right) \\
&= E_{\boldsymbol{a}}\left[\!\left[ \theta_k E_{\boldsymbol{n}}\left(\phi_k E\left[Y_{k+J} \middle| Q_k = 1, R = r, \boldsymbol{n_k}, \boldsymbol{a_k}, k\right] \middle| Q_k^{\ddagger} = 1, R = r, \boldsymbol{a_k}, k\right) | Q_k^{\ddagger} = 1, T = 1, k \right]\!\right]
\end{aligned}
$$

where $E_{\boldsymbol{n}}(\cdot)$ is over $P\left(\boldsymbol{n_k} \middle| Q_k^{G_k} = 1, R = r, \boldsymbol{a_k}, k\right)$ which is identified by $\phi_k P\left(\boldsymbol{n_k} \middle| Q_k^{\ddagger} = 1, R = r, \boldsymbol{a_k}, k\right)$,
and $E_{\boldsymbol{a}}[\![\cdot]\!]$ is over $P\left(\boldsymbol{a_k} \middle| Q_k^{G_k} = 1, T = 1, k\right)$ which is identified by $\theta_k P\left(\boldsymbol{a_k} \middle| Q_k^{\ddagger} = 1, T = 1, k\right)$

The second and third equalities follow by definition of an iterated expectation.

<u>Weighting Estimator</u>

$$
\begin{aligned}
&\mu_k^{\mathcal{D}4}(r) \\
&= \textstyle\sum_{\boldsymbol{a},\boldsymbol{n}} E\left[Y_{k+J} \middle| Q_k = 1, R = r, \boldsymbol{n_k}, \boldsymbol{a_k}, k\right] \phi_k P\left(\boldsymbol{n_k} \middle| Q_k^{\ddagger} = 1, R = r, \boldsymbol{a_k}, k\right) \theta_k P\left(\boldsymbol{a_k} \middle| Q_k^{\ddagger} = 1, T = 1, k\right) \\
&= \textstyle\sum_{\boldsymbol{a},\boldsymbol{n}} E\left[Y_{k+J} \middle| Q_k = 1, R = r, \boldsymbol{n_k}, \boldsymbol{a_k}, k\right] P(\boldsymbol{n_k} | Q_k = 1, R = r, \boldsymbol{a_k}, k) P(\boldsymbol{a_k} | Q_k = 1, R = r, k) \\
&\quad \times \phi_k \times \theta_k \times \frac{P\left(\boldsymbol{n_k} \middle| Q_k^{\ddagger} = 1, R = r, \boldsymbol{a_k}, k\right)}{P(\boldsymbol{n_k} | Q_k = 1, R = r, \boldsymbol{a_k}, k)} \times \frac{P\left(\boldsymbol{a_k} \middle| Q_k^{\ddagger} = 1, T = 1, k\right)}{P(\boldsymbol{a_k} | Q_k = 1, R = r, k)} \\
&= \textstyle\sum_{\boldsymbol{a},\boldsymbol{n}} E\left[Y_{k+J} \middle| Q_k = 1, R = r, \boldsymbol{n_k}, \boldsymbol{a_k}, k\right] P(\boldsymbol{n_k} | Q_k = 1, R = r, \boldsymbol{a_k}, k) P(\boldsymbol{a_k} | Q_k = 1, R = r, k) \\
&\quad \times \phi_k \times \theta_k \times \frac{P(Q_k^{\wr}=1 | Q_k^{\dagger}=1, Q_k^{\ddagger}=1, R=r, k)}{P(Q_k^{\wr}=1 | Q_k^{\dagger}=1, Q_k^{\ddagger}=1, R=r, \boldsymbol{n_k}, \boldsymbol{a_k}, k)} \times \frac{P(Q_k^{\dagger}=1 | Q_k^{\ddagger}=1, R=r, k)}{P(Q_k^{\dagger}=1 | Q_k^{\ddagger}=1, R=r, \boldsymbol{n_k}, \boldsymbol{a_k}, k)} \times \frac{P(T=1 | Q_k^{\ddagger}=1, \boldsymbol{a_k}, k)}{P(R=r | Q_k^{\ddagger}=1, \boldsymbol{a_k}, k)} \times \frac{P(R=r | Q_k^{\ddagger}=1, k)}{P(T=1 | Q_k^{\ddagger}=1, k)} \\
&= E[Y_{k+J} \times \omega_{r, Q_k=1, k,}^{\mathcal{D}4} | Q_k = 1, R = r, k]
\end{aligned}
$$

where

$$
\omega_{r, Q_k=1, k}^{\mathcal{D}4} = \phi_k \times \theta_k \times \frac{P(Q_k^{\wr}=1 | Q_k^{\dagger}=1, Q_k^{\ddagger}=1, R=r, k)}{P(Q_k^{\wr}=1 | Q_k^{\dagger}=1, Q_k^{\ddagger}=1, R=r, \boldsymbol{n_k}, \boldsymbol{a_k}, k)} \times \frac{P(Q_k^{\dagger}=1 | Q_k^{\ddagger}=1, R=r, k)}{P(Q_k^{\dagger}=1 | Q_k^{\ddagger}=1, R=r, \boldsymbol{n_k}, \boldsymbol{a_k}, k)} \times \frac{P(T=1 | Q_k^{\ddagger}=1, \boldsymbol{a_k}, k)}{P(R=r | Q_k^{\ddagger}=1, \boldsymbol{a_k}, k)} \times \frac{P(R=r | Q_k^{\ddagger}=1, k)}{P(T=1 | Q_k^{\ddagger}=1, k)}
$$

<u>Aggregated G-Computation Estimator</u>

First, define:

$$
\begin{aligned}
&\theta_k^{pool} \\
&= \frac{\sum_{\boldsymbol{n}} P(Q^{\wr}=1 | Q^{\dagger}=1, Q^{\ddagger}=1, T=1, \boldsymbol{n_k}, \boldsymbol{a_k}, k) \sum_r \{ \text{ꝗ}_k(\cdot) P(r | Q^{\ddagger}=1, T=1, \boldsymbol{n_k}, \boldsymbol{a_k}, k) \} P(\boldsymbol{n_k} | Q_k^{\ddagger}=1, T=1, \boldsymbol{a_k}, k)}{\sum_{\boldsymbol{n},\boldsymbol{a},k} P(Q^{\wr}=1 | Q^{\dagger}=1, Q^{\ddagger}=1, T=1, \boldsymbol{n_k}, \boldsymbol{a_k}, k) \sum_r \{ \text{ꝗ}_k(\cdot) P(r | Q^{\ddagger}=1, T=1, \boldsymbol{n_k}, \boldsymbol{a_k}, k) \} P(\boldsymbol{n_k}, \boldsymbol{a_k}, k | Q_k^{\ddagger}=1, T=1)} \\
&= \frac{\boldsymbol{E_n}[E_r\{ \text{ꝗ}_k(\cdot) | Q^{\ddagger}=1, T=1, \boldsymbol{n_k}, \boldsymbol{a_k}, k \} P(Q^{\wr}=1 | Q^{\dagger}=1, Q^{\ddagger}=1, T=1, \boldsymbol{n_k}, \boldsymbol{a_k}, k) | Q_k^{\ddagger}=1, T=1, \boldsymbol{a_k}, k]}{E_{(\boldsymbol{n},\boldsymbol{a},k)}[E_r\{ \text{ꝗ}_k(\cdot) | Q^{\ddagger}=1, T=1, \boldsymbol{n_k}, \boldsymbol{a_k}, k \} P(Q^{\wr}=1 | Q^{\dagger}=1, Q^{\ddagger}=1, T=1, \boldsymbol{n_k}, \boldsymbol{a_k}, k) | Q_k^{\ddagger}=1, T=1]}
\end{aligned}
$$

where $E_r\{\cdot\}$ is over $P(r | Q_k^{\ddagger} = 1, T = 1, \boldsymbol{n_k}, \boldsymbol{a_k}, k)$

and where $E_{\boldsymbol{n}}[\cdot]$ is over $P(\boldsymbol{n_k} | Q_k^{\ddagger} = 1, T = 1, \boldsymbol{a_k}, k)$ and $E_{(\boldsymbol{n},\boldsymbol{a},k)}[\cdot]$ is over $P(\boldsymbol{n_k}, \boldsymbol{a_k}, k | Q_k^{\ddagger} = 1, T = 1)$

Note also that:

$$
\begin{aligned}
&P\left(\boldsymbol{a_k}, k \middle| Q_k^{G_k} = 1, T = 1\right) \\
&= \frac{\sum_{\boldsymbol{n}} P(Q^{\wr}=1 | Q^{\dagger}=1, Q^{\ddagger}=1, T=1, \boldsymbol{n_k}, \boldsymbol{a_k}, k) \sum_r \{ \text{ꝗ}_k(\cdot) P(r | Q^{\ddagger}=1, T=1, \boldsymbol{n_k}, \boldsymbol{a_k}, k) \} P(Q_k^{\ddagger}=1, T=1, \boldsymbol{n_k}, \boldsymbol{a_k}, k)}{\sum_{\boldsymbol{n},\boldsymbol{a},k} P(Q^{\wr}=1 | Q^{\dagger}=1, Q^{\ddagger}=1, T=1, \boldsymbol{n_k}, \boldsymbol{a_k}, k) \sum_r \{ \text{ꝗ}_k(\cdot) P(r | Q^{\ddagger}=1, T=1, \boldsymbol{n_k}, \boldsymbol{a_k}, k) \} P(Q_k^{\ddagger}=1, T=1, \boldsymbol{n_k}, \boldsymbol{a_k}, k)} \\
&= \theta_k^{pool} \times P(\boldsymbol{a_k}, k | Q_k^{\ddagger} = 1, T = 1)
\end{aligned}
$$

Now,

$$
\begin{aligned}
&\tau^{\mathcal{D}4}(r) = \textstyle\sum_k \mu_k^{\mathcal{D}4}(r) P(k | Q_k^{G_k} = 1, T = 1) \\
&= \textstyle\sum_{\boldsymbol{a},\boldsymbol{n},k} E\left(Y_{k+J}^{G_k} \middle| Q_k^{G_k} = 1, R = r, \boldsymbol{n_k}, \boldsymbol{a_k}, k\right) P\left(\boldsymbol{n_k} \middle| Q_k^{G_k} = 1, R = r, \boldsymbol{a_k}, k\right) P\left(\boldsymbol{a_k} \middle| Q_k^{G_k} = 1, T = 1, k\right) \\
&\quad \times P(k | Q_k^{G_k} = 1, T = 1) \\
&= \textstyle\sum_{\boldsymbol{a},\boldsymbol{n},k} E\left(Y_{k+J}^{G_k} \middle| Q_k^{G_k} = 1, R = r, \boldsymbol{n_k}, \boldsymbol{a_k}, k\right) P\left(\boldsymbol{n_k} \middle| Q_k^{G_k} = 1, R = r, \boldsymbol{a_k}, k\right) P\left(\boldsymbol{a_k}, k \middle| Q_k^{G_k} = 1, T = 1\right) \\
&= \textstyle\sum_{\boldsymbol{a},\boldsymbol{n},k} E\left[Y_{k+J} \middle| Q_k = 1, R = r, \boldsymbol{n_k}, \boldsymbol{a_k}, k\right] \phi_k P\left(\boldsymbol{n_k} \middle| Q_k^{\ddagger} = 1, R = r, \boldsymbol{a_k}, k\right) \theta_k^{pool} P\left(\boldsymbol{a_k}, k \middle| Q_k^{\ddagger} = 1, T = 1\right)
\end{aligned}
$$

$= \sum_{\boldsymbol{a},k} E_{\boldsymbol{n}}\left(\phi_k E\left[Y_{k+J} \middle| Q_k = 1, R = r, \boldsymbol{n_k}, \boldsymbol{a_k}, k\right] \middle| Q_k^{\ddagger} = 1, R = r, \boldsymbol{a_k}, k\right) \theta_k^{pool} P\left(\boldsymbol{a_k}, k \middle| Q_k^{\ddagger} = 1, T = 1\right)$

$= E_{\boldsymbol{a},k}\left[\!\left[ \theta_k^{pool} E_{\boldsymbol{n}}\left(\phi_k E\left[Y_{k+J} \middle| Q_k = 1, R = r, \boldsymbol{n_k}, \boldsymbol{a_k}, k\right] \middle| Q_k^{\ddagger} = 1, R = r, \boldsymbol{a_k}, k\right) | Q_k^{\ddagger} = 1, T = 1 \right]\!\right]$

where $E_{\boldsymbol{n}}(\cdot)$ is over $P\left(\boldsymbol{n_k} \middle| Q_k^{G_k} = 1, R = r, \boldsymbol{a_k}, k\right)$ which is identified by $\phi_k P\left(\boldsymbol{n_k} \middle| Q_k^{\ddagger} = 1, R = r, \boldsymbol{a_k}, k\right)$,
and $E_{\boldsymbol{a},k}[\![\cdot]\!]$ is over $P\left(\boldsymbol{a_k}, k \middle| Q_k^{G_k} = 1, T = 1\right)$ which is identified by $\theta_k^{pool} P\left(\boldsymbol{a_k}, k \middle| Q_k^{\ddagger} = 1, T = 1\right)$

<u>Aggregated Weighting Estimator</u>

$\tau^{\mathcal{D}4}(r) = \sum_k \mu_k^{\mathcal{D}4}(r) P(k | Q_k^{G_k} = 1, T = 1)$

$= \sum_{\boldsymbol{a},\boldsymbol{n},k} E\left(Y_{k+J}^{G_k} \middle| Q_k^{G_k} = 1, R = r, \boldsymbol{n_k}, \boldsymbol{a_k}, k\right) P\left(\boldsymbol{n_k} \middle| Q_k^{G_k} = 1, R = r, \boldsymbol{a_k}, k\right) P\left(\boldsymbol{a_k} \middle| Q_k^{G_k} = 1, T = 1, k\right)$

$\times P(k | Q_k^{G_k} = 1, T = 1)$

$= \sum_{\boldsymbol{a},\boldsymbol{n},k} E\left(Y_{k+J}^{G_k} \middle| Q_k^{G_k} = 1, R = r, \boldsymbol{n_k}, \boldsymbol{a_k}, k\right) P\left(\boldsymbol{n_k} \middle| Q_k^{G_k} = 1, R = r, \boldsymbol{a_k}, k\right) P\left(\boldsymbol{a_k}, k \middle| Q_k^{G_k} = 1, T = 1\right)$

$= \sum_{\boldsymbol{a},\boldsymbol{n},k} E\left[Y_{k+J} \middle| Q_k = 1, R = r, \boldsymbol{n_k}, \boldsymbol{a_k}, k\right] \phi_k P\left(\boldsymbol{n_k} \middle| Q_k^{\ddagger} = 1, R = r, \boldsymbol{a_k}, k\right) \theta_k^{pool} P\left(\boldsymbol{a_k}, k \middle| Q_k^{\ddagger} = 1, T = 1\right)$

$= \sum_{\boldsymbol{a},\boldsymbol{n},k} E\left[Y_{k+J} \middle| Q_k = 1, R = r, \boldsymbol{n_k}, \boldsymbol{a_k}, k\right] P(\boldsymbol{n_k} | Q_k = 1, R = r, \boldsymbol{a_k}, k) P(\boldsymbol{a_k}, k | Q_k = 1, R = r)$

$\times \phi_k \times \theta_k^{pool} \times \frac{P\left(\boldsymbol{n_k} \middle| Q_k^{\ddagger} = 1, R = r, \boldsymbol{a_k}, k\right)}{P(\boldsymbol{n_k} | Q_k = 1, R = r, \boldsymbol{a_k}, k)} \times \frac{P\left(\boldsymbol{a_k}, k \middle| Q_k^{\ddagger} = 1, T = 1\right)}{P(\boldsymbol{a_k}, k | Q_k = 1, R = r)}$

$= \sum_{\boldsymbol{a},\boldsymbol{n},k} E\left[Y_{k+J} \middle| Q_k = 1, R = r, \boldsymbol{n_k}, \boldsymbol{a_k}, k\right] P(\boldsymbol{n_k} | Q_k = 1, R = r, \boldsymbol{a_k}, k) P(\boldsymbol{a_k}, k | Q_k = 1, R = r)$

$\times \phi_k \times \theta_k^{pool} \times \frac{P(Q_k^{\wr}=1 | Q_k^{\dagger}=1, Q_k^{\ddagger}=1, R=r)}{P(Q_k^{\wr}=1 | Q_k^{\dagger}=1, Q_k^{\ddagger}=1, R=r, \boldsymbol{n_k}, \boldsymbol{a_k}, k)} \times \frac{P(Q_k^{\dagger}=1 | Q_k^{\ddagger}=1, R=r)}{P(Q_k^{\dagger}=1 | Q_k^{\ddagger}=1, R=r, \boldsymbol{n_k}, \boldsymbol{a_k}, k)} \times \frac{P(T=1 | Q_k^{\ddagger}=1, \boldsymbol{a_k}, k)}{P(R=r | Q_k^{\ddagger}=1, \boldsymbol{a_k}, k)} \times \frac{P(R=r | Q_k^{\ddagger}=1)}{P(T=1 | Q_k^{\ddagger}=1)}$

$= E[Y_{k+J} \times \omega_{r,Q_k=1,k,}^{\mathcal{D}4-pool} | Q_k = 1, R = r]$

where

$\omega_{r,Q_k=1,k}^{\mathcal{D}4-pool} = \phi_k \times \theta_k^{pool} \times \frac{P(Q_k^{\wr}=1 | Q_k^{\dagger}=1, Q_k^{\ddagger}=1, R=r)}{P(Q_k^{\wr}=1 | Q_k^{\dagger}=1, Q_k^{\ddagger}=1, R=r, \boldsymbol{n_k}, \boldsymbol{a_k}, k)} \times \frac{P(Q_k^{\dagger}=1 | Q^{\ddagger}=1, R=r)}{P(Q_k^{\dagger}=1 | Q_k^{\ddagger}=1, R=r, \boldsymbol{n_k}, \boldsymbol{a_k}, k)} \times \frac{P(T=1 | Q_k^{\ddagger}=1, \boldsymbol{a_k}, k)}{P(R=r | Q_k^{\ddagger}=1, \boldsymbol{a_k}, k)} \times \frac{P(R=r | Q_k^{\ddagger}=1)}{P(T=1 | Q_k^{\ddagger}=1)}$

*Remark 3.*

Under Design 4a, when there are no other eligibility variables downstream from $\boldsymbol{W_k^{\dagger}}$ (i.e., when $\boldsymbol{W_k^{\wr}} = \varnothing$) and under an allocation strategy that assigns the eligibility variables $\boldsymbol{W_k^{\dagger}}$ at random (i.e., when $g_k(\cdot) = P(\boldsymbol{W_k^{\dagger}} | Q_k^{\ddagger} = 1, k)$), all results of Design 2 (generalization) apply, i.e., $f_{k,\ell}^{\mathcal{D}4a} \Rightarrow f_{k,\ell}^{\mathcal{D}2}$, implying that $\tau_k^{\mathcal{D}4a}(r) \Rightarrow \tau_k^{\mathcal{D}2}$, $\omega_{r,Q_k=1,k}^{\mathcal{D}4a} \Rightarrow \omega_{r,Q_k=1,k}^{\mathcal{D}2}$, and $\omega_{r,Q_k=1,k}^{\mathcal{D}4a-pool} \Rightarrow \omega_{r,Q_k=1,k}^{\mathcal{D}2-pool}$ and also the G-Computation estimators of Design 4 reduce to those of Design 2. To see why this holds, consider that (i) the terms involving $Q_k^{\wr}$ disappear from all expressions, and that (ii) $q_k(\cdot)$, the implied distribution of partial eligibility $Q_k^{\dagger} = 1$ under the intervention, being independent of $(\boldsymbol{A_k}, \boldsymbol{N_k}, R)$ given $k$, drops out of all expressions through cancellation. Thus, under these constraints, i.e., when $\boldsymbol{W_k^{\wr}} = \varnothing$ and $q_k(\cdot) = P(\boldsymbol{W_k^{\dagger}} | Q_k^{\ddagger} = 1, k)$, Design 4a allows for generalization (through hypothetical intervention) when $\boldsymbol{W_k^{\dagger}}$ affects the outcome $Y_{k+J}$ while Design 2 does not. This difference is due to the fact that Design 4a relies on exchangeability (19), positivity (12), and consistency, whereas Design 2 instead relies on independence (11) and positivity (12). Next, we show that simplifications with different choices of $g(\cdot)$ when $\boldsymbol{W_k^{\wr}} = \varnothing$ do not reduce Design 4 to Design 2.

*Remark 4.*

Under Design 4b, when there are no other eligibility variables downstream from $\boldsymbol{W_k^{\dagger}}$ (i.e., when $\boldsymbol{W_k^{\wr}} = \varnothing$) and under an allocation strategy that assigns the eligibility variables $\boldsymbol{W_k^{\dagger}}$ at random given the social group $R = r$ and the allowables $\boldsymbol{A_k}$ (i.e., when $g_k(\cdot) = P(\boldsymbol{w_k^{\dagger}} | Q_k^{\ddagger} = 1, R = r, \boldsymbol{a_k}, k)$), we obtain the following results:

$P\left(\boldsymbol{a_k} \middle| Q_k^{G_k} = 1, T = 1, k\right)$

$= \frac{\sum_{\boldsymbol{n}} \sum_r \{P(G_k = \boldsymbol{w_k^{\dagger}} \in \mathbb{w}_k^{\dagger} | Q_k^{\ddagger}=1, R=r, \boldsymbol{a_k}, k) P(r | Q_k^{\ddagger}=1, T=1, \boldsymbol{n_k}, \boldsymbol{a_k}, k)\} P(Q_k^{\ddagger}=1, T=1, \boldsymbol{n_k}, \boldsymbol{a_k}, k)}{\sum_{\boldsymbol{n},\boldsymbol{a}} \sum_r \{P(G_k = \boldsymbol{w_k^{\dagger}} \in \mathbb{w}_k^{\dagger} | Q_k^{\ddagger}=1, R=r, \boldsymbol{a_k}, k) P(r | Q_k^{\ddagger}=1, T=1, \boldsymbol{n_k}, \boldsymbol{a_k}, k)\} P(Q_k^{\ddagger}=1, T=1, \boldsymbol{n_k}, \boldsymbol{a_k}, k)}$

$= \frac{\sum_{\boldsymbol{n}} \sum_r \{P(G_k = \boldsymbol{w_k^{\dagger}} \in \mathbb{w}_k^{\dagger} | Q_k^{\ddagger}=1, R=r, \boldsymbol{n_k}, \boldsymbol{a_k}, k) P(r | Q_k^{\ddagger}=1, T=1, \boldsymbol{n_k}, \boldsymbol{a_k}, k)\} P(Q_k^{\ddagger}=1, T=1, \boldsymbol{n_k}, \boldsymbol{a_k}, k)}{\sum_{\boldsymbol{n},\boldsymbol{a}} \sum_r \{P(G_k = \boldsymbol{w_k^{\dagger}} \in \mathbb{w}_k^{\dagger} | Q_k^{\ddagger}=1, R=r, \boldsymbol{n_k}, \boldsymbol{a_k}, k) P(r | Q_k^{\ddagger}=1, T=1, \boldsymbol{n_k}, \boldsymbol{a_k}, k)\} P(Q_k^{\ddagger}=1, T=1, \boldsymbol{n_k}, \boldsymbol{a_k}, k)}$ by definition of $G_k$

$= \frac{P(G_k = \boldsymbol{w}_k^{\dagger} \in \mathbb{w}_k^{\dagger} | Q_k^{\ddagger} = 1, T = 1, \boldsymbol{a_k}) P(Q_k^{\ddagger} = 1, T = 1, \boldsymbol{a_k}, k)}{\sum_{\boldsymbol{a}} P(G_k = \boldsymbol{w}_k^{\dagger} \in \mathbb{w}_k^{\dagger} | Q_k^{\ddagger} = 1, T = 1, \boldsymbol{a_k}, k) P(Q_k^{\ddagger} = 1, T = 1, \boldsymbol{a_k}, k)}$

$= \frac{P(Q_k^{\dagger} = 1 | Q_k^{\ddagger} = 1, T = 1, \boldsymbol{a_k}, k) P(Q_k^{\ddagger} = 1, T = 1, \boldsymbol{a_k}, k)}{P(Q_k^{\dagger} = 1 | Q_k^{\ddagger} = 1, T = 1, k) P(Q_k^{\ddagger} = 1, T = 1, k)}$

$= P(\boldsymbol{a_k} | Q_k = 1, T = 1, k)$

Furthermore,

$P\left(\boldsymbol{n_k} \middle| Q_k^{G_k} = 1, R = r, \boldsymbol{a_k}, k\right)$

$= P\left(\boldsymbol{n_k} \middle| Q_k^{\dagger^{G_k}} = 1, Q_k^{\ddagger} = 1, R = r, \boldsymbol{a_k}, k\right)$

$= P\left(\boldsymbol{n_k} \middle| Q_k^{\ddagger} = 1, R = r, \boldsymbol{a_k}, k\right)$ by definition of $G_k$

Therefore,

$\mu_k^{\mathcal{D}4b}(r)$
$= \sum_{\boldsymbol{a},\boldsymbol{n}} E\left(Y_{k+J}^{G_k} \middle| Q_k^{G_k} = 1, R = r, \boldsymbol{n_k}, \boldsymbol{a_k}, k\right) P\left(\boldsymbol{n_k} \middle| Q_k^{G_k} = 1, R = r, \boldsymbol{a_k}, k\right) P\left(\boldsymbol{a_k} \middle| Q_k^{G_k} = 1, R = r, k\right)$
$= \sum_{\boldsymbol{a},\boldsymbol{n}} E\left(Y_{k+J}^{G_k} \middle| Q_k^{\dagger^{G_k}} = 1, Q_k^{\ddagger} = 1, \boldsymbol{n_k}, \boldsymbol{a_k}, k\right) P\left(\boldsymbol{n_k} \middle| Q_k^{\ddagger} = 1, R = r, \boldsymbol{a_k}, k\right) P(\boldsymbol{a_k} | Q_k = 1, T = 1, k)$
$= \sum_{\boldsymbol{a},\boldsymbol{n}} E\left(Y_{k+J}^{G_k} \middle| Q_k^{\ddagger} = 1, R = r, \boldsymbol{n_k}, \boldsymbol{a_k}, k\right) P\left(\boldsymbol{n_k} \middle| Q_k^{\ddagger} = 1, R = r, \boldsymbol{a_k}, k\right) P(\boldsymbol{a_k} | Q_k = 1, T = 1, k)$
$= \sum_{\boldsymbol{a},\boldsymbol{n}} E\left(Y_{k+J}^{G_k} \middle| Q_k^{\dagger} = 1, Q_k^{\ddagger} = 1, R = r, \boldsymbol{n_k}, \boldsymbol{a_k}, k\right) P\left(\boldsymbol{n_k} \middle| Q_k^{\ddagger} = 1, R = r, \boldsymbol{a_k}, k\right) P(\boldsymbol{a_k} | Q_k = 1, T = 1, k)$
$= \sum_{\boldsymbol{a},\boldsymbol{n}} E\left(Y_{k+J} \middle| Q_k = 1, R = r, \boldsymbol{n_k}, \boldsymbol{a_k}, k\right) P\left(\boldsymbol{n_k} \middle| Q_k^{\ddagger} = 1, R = r, \boldsymbol{a_k}, k\right) P(\boldsymbol{a_k} | Q_k = 1, T = 1, k)$

It can be shown that with this identifying expression for $\tau_k^{\mathcal{D}4b}(r)$, we obtain the following estimators for Design 4b:

<u>G-Computation Estimator</u>

$\mu_k^{\mathcal{D}4b}(r)$
$= E_{\boldsymbol{a}} \llbracket E_{\boldsymbol{n}}\left(E\left[Y_{k+J} \middle| Q_k = 1, R = r, \boldsymbol{a_k}, k\right] \middle| Q_k^{\ddagger} = 1, R = r, \boldsymbol{a_k}, k\right) | Q_k = 1, T = 1, k \rrbracket$

where $E_{\boldsymbol{n}}(\cdot)$ is over $P\left(\boldsymbol{n_k} \middle| Q_k^{\ddagger} = 1, R = r, \boldsymbol{a_k}, k\right)$ and $E_{\boldsymbol{a}}\llbracket\cdot\rrbracket$ is over $P(\boldsymbol{a_k} | Q_k = 1, T = 1, k)$

By definition of an iterated expectation

<u>Weighting Estimator</u>

$\mu_k^{\mathcal{D}4b}(r)$
$= E[Y_{k+J} \times \omega_{r,Q_k=1,k}^{\mathcal{D}4b} | Q_k = 1, R = r, k]$

where

$\omega_{r,Q_k=1,k}^{\mathcal{D}4b} = \frac{P(Q_k^{\dagger} = 1 | Q_k^{\ddagger} = 1, R = r, \boldsymbol{a_k}, k)}{P(Q_k^{\dagger} = 1 | Q_k^{\ddagger} = 1, R = r, \boldsymbol{n_k}, \boldsymbol{a_k}, k)} \times \frac{P(T = 1 | Q_k = 1, \boldsymbol{a_k}, k)}{P(R = r | Q_k = 1, \boldsymbol{a_k}, k)} \times \frac{P(R = r | Q_k = 1, k)}{P(T = 1 | Q_k = 1, k)}$

<u>Aggregated G-Computation Estimator (over calendar time $k$)</u>

$\tau^{\mathcal{D}4b}(r) = \sum_k \mu_k^{\mathcal{D}4b}(r) \, P(k | Q_k^{\ddagger} = 1, T = 1)$
$= E_{\boldsymbol{a},k} \llbracket E_{\boldsymbol{n}}\left(E\left[Y_{k+J} \middle| Q_k = 1, R = r, \boldsymbol{a_k}, k\right] \middle| Q_k^{\ddagger} = 1, R = r, \boldsymbol{a_k}, k\right) | Q_k = 1, T = 1 \rrbracket$

where $E_{\boldsymbol{n}}(\cdot)$ is over $P\left(\boldsymbol{n_k} \middle| Q_k^{\ddagger} = 1, R = r, \boldsymbol{a_k}, k\right)$ and $E_{\boldsymbol{a},k}\llbracket\cdot\rrbracket$ is over $P(\boldsymbol{a_k}, k | Q_k = 1, T = 1)$

By definition of an iterated expectation

<u>Aggregated Weighting Estimator (over calendar time $k$)</u>

$\tau^{\mathcal{D}4b}(r) = \sum_k \mu_k^{\mathcal{D}4b}(r) \, P(k | Q_k^{\ddagger} = 1, T = 1)$
$= \sum_{\boldsymbol{a},\boldsymbol{n}} E\left(Y_{k+J} \middle| Q_k = 1, R = r, \boldsymbol{n_k}, \boldsymbol{a_k}, k\right) P(\boldsymbol{n_k} | Q_k = 1, R = r, \boldsymbol{a_k}, k) P(\boldsymbol{a_k}, k | Q_k = 1, R = r)$
$\times \frac{P(Q_k^{\dagger} = 1 | Q_k^{\ddagger} = 1, R = r, \boldsymbol{a_k}, k)}{P(Q_k^{\dagger} = 1 | Q_k^{\ddagger} = 1, R = r, \boldsymbol{n_k}, \boldsymbol{a_k}, k)} \times \frac{P(T = 1 | Q_k = 1, \boldsymbol{a_k}, k)}{P(R = r | Q_k = 1, \boldsymbol{a_k}, k)} \times \frac{P(R = r | Q_k = 1)}{P(T = 1 | Q_k = 1)}$
$= E[Y_{k+J} \times \omega_{r,Q_k=1,k}^{\mathcal{D}4b-pool} | Q_k = 1, R = r]$

where

$$\omega_{r,Q_k=1,k}^{\mathcal{D}4b-pool} = \frac{P(Q_k^{\dagger}=1|Q_k^{\ddagger}=1,R=r,\boldsymbol{a_k},k)}{P(Q_k^{\dagger}=1|Q_k^{\ddagger}=1,R=r,\boldsymbol{n_k},\boldsymbol{a_k},k)} \times \frac{P(T=1|Q_k=1,\boldsymbol{a_k},k)}{P(R=r|Q_k=1,\boldsymbol{a_k},k)} \times \frac{P(R=r|Q_k=1)}{P(T=1|Q_k=1)}$$

The estimators for Design 4b are not the same as those of Design 2. This is because, under Design 4b, $P\left(\boldsymbol{a_k}\middle|Q_k^{G_k}=1,T=1,k\right)$ reduces to $P(\boldsymbol{a_k}|Q_k=1,T=1,k)$. This means that the estimators for Design 4b cannot be used to generalize (as in Design 2), whereas those for Design 4a can.

*Remark 5.*

Under Design 4c, when there are no other eligibility variables downstream from $\boldsymbol{W_k^{\dagger}}$ (i.e., when $\boldsymbol{W_k^{\wr}} = \oslash$) and under an allocation strategy that assigns the eligibility variables $\boldsymbol{W_k^{\dagger}}$ at random given the allowables $\boldsymbol{A_k}$ and non-allowables $\boldsymbol{N_k}$ among the standard population (i.e., when $g_k(\cdot) = P(\boldsymbol{w_k^{\dagger}}|Q_k^{\ddagger}=1,T=1,\boldsymbol{n_k},\boldsymbol{a_k},k)$), we obtain the results:

$$P\left(\boldsymbol{a_k}\middle|Q_k^{G_k}=1,T=1,k\right)$$

$$= \frac{\sum_{\boldsymbol{n}}\sum_r\{P(G_k=\boldsymbol{w_k^{\dagger}}\in\mathbb{w}_k^{\dagger}|Q_k^{\ddagger}=1,T=1,\boldsymbol{n_k},\boldsymbol{a_k},k)P(r|Q_k^{\ddagger}=1,T=1,\boldsymbol{n_k},\boldsymbol{a_k},k)\}P(Q_k^{\ddagger}=1,T=1,\boldsymbol{n_k},\boldsymbol{a_k},k)}{\sum_{\boldsymbol{n},\boldsymbol{a}}\sum_r\{P(G_k=\boldsymbol{w_k^{\dagger}}\in\mathbb{w}_k^{\dagger}|Q_k^{\ddagger}=1,T=1,\boldsymbol{n_k},\boldsymbol{a_k},k)P(r|Q_k^{\ddagger}=1,T=1,\boldsymbol{n_k},\boldsymbol{a_k},k)\}P(Q_k^{\ddagger}=1,T=1,\boldsymbol{n_k},\boldsymbol{a_k},k)} \text{ by definition of } G_k$$

$$= \frac{\sum_{\boldsymbol{n}} P(G_k=\boldsymbol{w_k^{\dagger}}\in\mathbb{w}_k^{\dagger}|Q_k^{\ddagger}=1,T=1,\boldsymbol{n_k},\boldsymbol{a_k},k)P(Q_k^{\ddagger}=1,T=1,\boldsymbol{n_k},\boldsymbol{a_k},k)}{\sum_{\boldsymbol{n},\boldsymbol{a}} P(G_k=\boldsymbol{w_k^{\dagger}}\in\mathbb{w}_k^{\dagger}|Q_k^{\ddagger}=1,T=1,\boldsymbol{n_k},\boldsymbol{a_k},k)P(Q_k^{\ddagger}=1,T=1,\boldsymbol{n_k},\boldsymbol{a_k},k)}$$

$$= \frac{P(Q_k^{\dagger}=1|Q_k^{\ddagger}=1,T=1,\boldsymbol{a_k})P(Q_k^{\ddagger}=1,T=1,\boldsymbol{a_k},k)}{P(Q_k^{\dagger}=1|Q_k^{\ddagger}=1,T=1,k)P(Q_k^{\ddagger}=1,T=1,k)}$$

$$= P(\boldsymbol{a_k}|Q_k=1,T=1,k)$$

Furthermore,

$$P\left(\boldsymbol{n_k}\middle|Q_k^{G_k}=1,R=r,\boldsymbol{a_k},k\right)$$

$$= \frac{P(G_k=\boldsymbol{w_k^{\dagger}}\in\mathbb{w}_k^{\dagger}|Q_k^{\ddagger}=1,T=1,\boldsymbol{n_k},\boldsymbol{a_k},k)P(Q_k^{\ddagger}=1,R=r,\boldsymbol{n_k},\boldsymbol{a_k},k)}{\sum_{\boldsymbol{n}} P(G_k=\boldsymbol{w_k^{\dagger}}\in\mathbb{w}_k^{\dagger}|Q_k^{\ddagger}=1,T=1,\boldsymbol{n_k},\boldsymbol{a_k},k)P(Q_k^{\ddagger}=1,R=r,\boldsymbol{n_k},\boldsymbol{a_k},k)} \text{ by definition of } G_k$$

$$= \frac{P(Q_k^{\dagger}=1|Q_k^{\ddagger}=1,T=1,\boldsymbol{n_k},\boldsymbol{a_k},k)}{\sum_{\boldsymbol{n}} P(Q_k^{\dagger}=1|Q_k^{\ddagger}=1,T=1,\boldsymbol{n_k},\boldsymbol{a_k},k)P(\boldsymbol{n_k}|Q_k^{\ddagger}=1,R=r,\boldsymbol{a_k},k)} \times P(\boldsymbol{n_k}|Q_k^{\ddagger}=1,R=r,\boldsymbol{a_k},k)$$

Therefore,

$$\begin{aligned}
&\mu_k^{\mathcal{D}4c}(r) \\
&= \textstyle\sum_{\boldsymbol{a},\boldsymbol{n}} E\left(Y_{k+J}^{G_k}\middle|Q_k^{G_k}=1,R=r,\boldsymbol{n_k},\boldsymbol{a_k},k\right)P\left(\boldsymbol{n_k}\middle|Q_k^{G_k}=1,R=r,\boldsymbol{a_k},k\right)P\left(\boldsymbol{a_k}\middle|Q_k^{G_k}=1,R=r,k\right) \\
&= \textstyle\sum_{\boldsymbol{a},\boldsymbol{n}} E\left(Y_{k+J}^{G_k}\middle|Q_k^{\dagger G_k}=1,Q_k^{\ddagger}=1,\right)\phi_k^c P\left(\boldsymbol{n_k}\middle|Q_k^{\ddagger}=1,R=r,\boldsymbol{a_k},k\right)P(\boldsymbol{a_k}|Q_k=1,T=1,k) \\
&= \textstyle\sum_{\boldsymbol{a},\boldsymbol{n}} E\left(Y_{k+J}^{G_k}\middle|Q_k^{\ddagger}=1,R=r,\boldsymbol{n_k},\boldsymbol{a_k},k\right)\phi_k^c P\left(\boldsymbol{n_k}\middle|Q_k^{\ddagger}=1,R=r,\boldsymbol{a_k},k\right)P(\boldsymbol{a_k}|Q_k=1,T=1,k) \\
&= \textstyle\sum_{\boldsymbol{a},\boldsymbol{n}} E\left(Y_{k+J}^{G_k}\middle|Q_k^{\dagger}=1,Q_k^{\ddagger}=1,R=r,\boldsymbol{n_k},\boldsymbol{a_k},k\right)\phi_k^c P\left(\boldsymbol{n_k}\middle|Q_k^{\ddagger}=1,R=r,\boldsymbol{a_k},k\right)P(\boldsymbol{a_k}|Q_k=1,T=1,k) \\
&= \textstyle\sum_{\boldsymbol{a},\boldsymbol{n}} E\left(Y_{k+J}^{G_k}\middle|Q_k^{\dagger}=1,Q_k^{\ddagger}=1,R=r,\boldsymbol{n_k},\boldsymbol{a_k},k\right)\phi_k^c P\left(\boldsymbol{n_k}\middle|Q_k^{\ddagger}=1,R=r,\boldsymbol{a_k},k\right)P(\boldsymbol{a_k}|Q_k=1,T=1,k) \\
&= \textstyle\sum_{\boldsymbol{a},\boldsymbol{n}} E\left(Y_{k+J}\middle|Q_k=1,R=r,\boldsymbol{n_k},\boldsymbol{a_k},k\right)\phi_k^c P\left(\boldsymbol{n_k}\middle|Q_k^{\ddagger}=1,R=r,\boldsymbol{a_k},k\right)P(\boldsymbol{a_k}|Q_k=1,T=1,k)
\end{aligned}$$

where $\phi_k^c = \frac{P(Q_k^{\dagger}=1|Q_k^{\ddagger}=1,T=1,\boldsymbol{n_k},\boldsymbol{a_k},k)}{\sum_{\boldsymbol{n}} P(Q_k^{\dagger}=1|Q_k^{\ddagger}=1,T=1,\boldsymbol{n_k},\boldsymbol{a_k},k)P(\boldsymbol{n_k}|Q_k^{\ddagger}=1,R=r,\boldsymbol{a_k},k)}$

$= \frac{P(Q_k^{\dagger}=1|Q_k^{\ddagger}=1,T=1,\boldsymbol{n_k},\boldsymbol{a_k},k)}{\boldsymbol{E_n}[P(Q_k^{\dagger}=1|Q_k^{\ddagger}=1,T=1,\boldsymbol{n_k},\boldsymbol{a_k},k)|Q_k^{\ddagger}=1,R=r,\boldsymbol{a_k},k]}$

with $E_{\boldsymbol{n}}[\cdot]$ over the distribution $P(\boldsymbol{n_k}|Q_k^{\ddagger}=1,R=r,\boldsymbol{a_k},k)$

It can be shown that with this identifying expression for $\tau_k^{\mathcal{D}4c}(r)$, we obtain the following estimators for Design 4c:

<u>G-Computation Estimator</u>

$$\mu_k^{\mathcal{D}4c}(r)$$
$$= E_{\boldsymbol{a}}\llbracket E_{\boldsymbol{n}}\left(\phi_k^c E\left[Y_{k+J}\middle|Q_k = 1, R = r, \boldsymbol{n_k}, \boldsymbol{a_k}, k\right]\middle|Q_k^{\ddagger} = 1, R = r, \boldsymbol{a_k}, k\right)|Q_k = 1, T = 1, k\rrbracket$$

where $E_{\boldsymbol{n}}(\cdot)$ is over $P\left(\boldsymbol{n_k}\middle|Q_k^{G_k} = 1, R = r, \boldsymbol{a_k}, k\right)$ which is identified by $\phi_k^c P\left(\boldsymbol{n_k}\middle|Q_k^{\ddagger} = 1, R = r, \boldsymbol{a_k}, k\right)$,
and $E_{\boldsymbol{a}}\llbracket\cdot\rrbracket$ is over $P(\boldsymbol{a_k}|Q_k = 1, T = 1, k)$ and $\phi_k^c$ defined as above for the identifying expression for $\mu_k^{\mathcal{D}4c}(r)$

<u>Weighting Estimator</u>

$$\mu_k^{\mathcal{D}4c}(r)$$
$$= E[Y_{k+J} \times \omega_{r,Q_k=1,k,}^{\mathcal{D}4c}|Q_k = 1, R = r, k]$$

where

$$\omega_{r,Q_k=1,k}^{\mathcal{D}4} = \phi_k^c \times \frac{P(Q_k^{\dagger}=1|Q_k^{\ddagger}=1,R=r,\boldsymbol{a_k},k)}{P(Q_k^{\dagger}=1|Q_k^{\ddagger}=1,R=r,\boldsymbol{n_k},\boldsymbol{a_k},k)} \times \frac{P(T=1|Q_k=1,\boldsymbol{a_k},k)}{P(R=r|Q_k=1,\boldsymbol{a_k},k)} \times \frac{P(R=r|Q_k=1,k)}{P(T=1|Q_k=1,k)}$$

with $\phi_k^c$ defined as above for the identifying expression for $\mu_k^{\mathcal{D}4c}(r)$

<u>Aggregated G-Computation Estimator</u>

$$\tau^{\mathcal{D}4c}(r) = \textstyle\sum_k \mu_k^{\mathcal{D}4c}(r)\, P(k|Q_k^{G_k} = 1, T = 1)$$
$$= E_{\boldsymbol{a}}\llbracket E_{\boldsymbol{n}}\left(\phi_k^c E\left[Y_{k+J}\middle|Q_k = 1, R = r, \boldsymbol{n_k}, \boldsymbol{a_k}, k\right]\middle|Q_k^{\ddagger} = 1, R = r, \boldsymbol{a_k}, k\right)|Q_k = 1, T = 1\rrbracket$$

where $E_{\boldsymbol{n}}(\cdot)$ is over $P\left(\boldsymbol{n_k}\middle|Q_k^{G_k} = 1, R = r, \boldsymbol{a_k}, k\right)$ which is identified by $\phi_k^c P\left(\boldsymbol{n_k}\middle|Q_k^{\ddagger} = 1, R = r, \boldsymbol{a_k}, k\right)$,
and $E_{\boldsymbol{a}}\llbracket\cdot\rrbracket$ is over $P(\boldsymbol{a_k}, k|Q_k = 1, T = 1)$

with $\phi_k^c$ defined as above for the identifying expression for $\mu_k^{\mathcal{D}4c}(r)$

<u>Aggregated Weighting Estimator</u>

$$\tau^{\mathcal{D}4c}(r) = \textstyle\sum_k \mu_k^{\mathcal{D}4c}(r)\, P(k|Q_k^{G_k} = 1, T = 1)$$
$$= E[Y_{k+J} \times \omega_{r,Q_k=1,k,}^{\mathcal{D}4c-pool}|Q_k = 1, R = r]$$

where

$$\omega_{r,Q_k=1,k}^{\mathcal{D}4-pool} = \phi_k^c \times \frac{P(Q_k^{\dagger}=1|Q_k^{\ddagger}=1,R=r,\boldsymbol{a_k},k)}{P(Q_k^{\dagger}=1|Q_k^{\ddagger}=1,R=r,\boldsymbol{n_k},\boldsymbol{a_k},k)} \times \frac{P(T=1|Q_k=1,\boldsymbol{a_k},k)}{P(R=r|Q_k=1,\boldsymbol{a_k},k)} \times \frac{P(R=r|Q_k=1)}{P(T=1|Q_k=1)}$$

with $\phi_k^c$ defined as above for the identifying expression for $\mu_k^{\mathcal{D}4c}(r)$

The estimators for Design 4c are not the same as those of Design 2. This is because, under Design 4c, $P\left(\boldsymbol{n_k}\middle|Q_k^{G_k} = 1, R = r, \boldsymbol{a_k}, k\right)$ reduces to $\phi_k^c P\left(\boldsymbol{n_k}\middle|Q_k^{\ddagger} = 1, R = r, \boldsymbol{a_k}, k\right)$ and $P\left(\boldsymbol{a_k}\middle|Q_k^{G_k} = 1, T = 1, k\right)$ reduces to $P(\boldsymbol{a_k}|Q_k = 1, T = 1, k)$. This means that the estimators for Design 4c cannot be used to generalize (as in Design 2), whereas those for Design 4a can.

*Remark 6.*

Under Design 4c, when there are no other eligibility variables downstream from $\boldsymbol{W_k^{\dagger}}$ (i.e., when $\boldsymbol{W_k^{\wr}} = \emptyset$) and under an allocation strategy that assigns the eligibility variables $\boldsymbol{W_k^{\dagger}}$ at random given the allowables $\boldsymbol{A_k}$ and non-allowables $\boldsymbol{N_k}$ among the standard population (i.e., when $\mathscr{g}_k(\cdot) = P(\boldsymbol{w_k^{\dagger}}|Q_k^{\ddagger} = 1, T = 1, \boldsymbol{n_k}, \boldsymbol{a_k}, k)$), when a particular social group $R = r'$ is chosen as the standard population $T = 1$, $P\left(\boldsymbol{n_k}\middle|Q_k^{G_k} = 1, R = r', \boldsymbol{a_k}, k\right)$ reduces to $P(\boldsymbol{n_k}|Q_k = 1, R = r', \boldsymbol{a_k}, k)$ and $P\left(\boldsymbol{a_k}\middle|Q_k^{G_k} = 1, T = 1, k\right)$ reduces to $P(\boldsymbol{a_k}|Q_k = 1, R = r', k)$. Thus, for Design 4c under these constraints, the mean outcome for the social group $R = r'$ is fully descriptive among $Q_k = 1$.
To see that $P\left(\boldsymbol{n_k}\middle|Q_k^{G_k} = 1, R = r', \boldsymbol{a_k}, k\right)$ reduces to $P(\boldsymbol{n_k}|Q_k = 1, R = r', \boldsymbol{a_k}, k)$ consider that:

$P\left(\boldsymbol{n_k} \middle| Q_k^{G_k} = 1, R = r', \boldsymbol{a_k}, k\right)$

$= \frac{P(Q_k^{\dagger}=1|Q_k^{\ddagger}=1,R=r',\boldsymbol{n_k},\boldsymbol{a_k},k)P(\boldsymbol{n_k}|Q_k^{\ddagger}=1,R=r',\boldsymbol{a_k},k)}{\sum_{\boldsymbol{n}} P(Q_k^{\dagger}=1|Q_k^{\ddagger}=1,R=r',\boldsymbol{n_k},\boldsymbol{a_k},k)P(\boldsymbol{n_k}|Q_k^{\ddagger}=1,R=r',\boldsymbol{a_k},k)}$ (after replacing $R = r$ with $R = r'$ and $T = 1$ with $R = r'$)

$= \frac{P(Q_k^{\dagger}=1,Q_k^{\ddagger}=1,R=r',\boldsymbol{n_k},\boldsymbol{a_k},k)}{P(Q_k^{\dagger}=1,Q_k^{\ddagger}=1,R=r',\boldsymbol{a_k},k)}$

$= P(\boldsymbol{n_k}|Q_k = 1, R = r', \boldsymbol{a_k}, k)$

SAMPLE CODE

Note: updated code may be made available at the first author's website: https://github.com/jwjackson/targetstudy/